\documentclass{article}

\usepackage{arxiv}

\usepackage[utf8]{inputenc}
\usepackage[T1]{fontenc}
\usepackage{hyperref}
\usepackage{url}
\usepackage{booktabs}
\usepackage{amsfonts}
\usepackage{nicefrac}
\usepackage{microtype}
\usepackage{graphicx}
\usepackage{doi}
\usepackage{amssymb}
\usepackage{amsmath}
\usepackage{bm}
\usepackage{multirow}
\usepackage{subcaption}
\usepackage{nomencl}
\usepackage{tikz, pgfplots}
\usepackage{ifthen}
\usepackage{etoolbox}
\usepackage{pgfgantt}
\usepackage{comment}
\usepackage{adjustbox}
\usepackage{siunitx}    
\usetikzlibrary{positioning, shapes.geometric, arrows, fit}

\renewcommand{\nomgroup}[1]{%
  \item[\bfseries
  \ifstrequal{#1}{A}{Roman Symbols}{%
  \ifstrequal{#1}{G}{Greek Symbols}{%
  \ifstrequal{#1}{S}{Subscripts}{%
  \ifstrequal{#1}{P}{Superscripts}{%
  \ifstrequal{#1}{Z}{Abbreviations}{}}}}}%
]}

\title{Investigation of Aeroacoustics and In-flight Particle Transport in Thermal Spray Supersonic Jets}

\author{
  \href{https://orcid.org/0009-0007-3357-0359}{\includegraphics[scale=0.06]{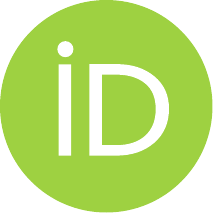}\hspace{1mm}D.~Rahmat Samii} \\
  Mechanical, Industrial and Aerospace Engineering\\
  Concordia University\\
  \texttt{dara.rahmatsamii@mail.concordia.ca} \\
  \And
  \href{https://orcid.org/0000-0003-4655-4351}{\includegraphics[scale=0.06]{orcid.pdf}\hspace{1mm}M.~Tembely} \\
  Mechanical, Industrial and Aerospace Engineering\\
  Concordia University\\
  \texttt{moussa.tembely@concordia.ca} \\
}

\renewcommand{\shorttitle}{Aeroacoustics and In-flight Particle Transport in Supersonic Jets}

\hypersetup{
  pdftitle={Investigation of Aeroacoustics and In-flight Particle Transport in Thermal Spray Supersonic Jets},
  pdfauthor={D. Rahmat Samii, M. Tembely},
  pdfkeywords={Thermal spray, Aeroacoustics, CFD, Supersonic jet, Lagrangian particle tracking},
}

\begin{document}
\maketitle

\begin{abstract}
The acoustic signature of thermal spray processes is known to vary with changes in operating conditions, which also influence particle in-flight velocity and distribution. Building on this idea, the present work first develops an analytical model that links chamber/nozzle parameters to far-field acoustic levels using gas-dynamics relations and simplified acoustic power propagation; the model is then calibrated to reduce systematic error associated with neglected turbulence effects and to improve agreement across operating conditions. In addition, a numerical framework is implemented to complement the analytical model and to resolve supersonic jet flow and in-flight particle transport. The second part of the study uses unsteady compressible simulations with hybrid turbulence modeling such as Unsteady Reynolds-Averaged Navier-Stokes(URANS) and Delayed Detached Eddy Simulation(DDES) to capture the development of the shock-containing jet and the associated near-field pressure fluctuations. Far-field sound is predicted using the Ffowcs Williams–Hawkings acoustic analogy, while a Lagrangian approach tracks particles injected at the nozzle exit to quantify velocity evolution, radial spreading, and downstream flux distributions. The influence of operating conditions (e.g. chamber pressure and temperature) is assessed, and predictions are evaluated against published microphone spectra and particle-flux measurements. Overall, the combined analytical and numerical approach captures how changes in nozzle operating conditions affect jet unsteadiness and mixing, leading to measurable shifts in acoustic level and spectral content. These results suggest that aeroacoustic signatures could be used as a non-intrusive pathway to monitor and potentially control thermal spray operating conditions.
\end{abstract}


\makenomenclature

\section{Introduction}

It is well known in the thermal spray community that the sound of the torch is a significant indicator of spraying conditions. A skilled thermal spray specialist can often determine the overall quality of a coating solely based on the sound produced by the operating torch. In this study, the computational fluid dynamics(CFD) package OpenFOAM together with the Ffowcs Williams–Hawkings analogy, coupled with a Lagrangian particle-tracking method, is employed to simulate cold spray operation and capture pressure-induced noise fluctuations. Furthermore, the particle distribution downstream of the nozzle exit is analyzed. A complete CFD setup and configuration that can be used in future research to accurately model in-flight particle distributions, predict the generated acoustic signals, and establish a link between the sound characteristics and particle behavior is described. In \autoref{sec:literature}, a brief overview of the existing literature on the CFD simulation of jet aeroacoustic noise is provided. In \autoref{sec:analytical}, an analytical model is developed, linking various design parameters of the nozzle and operating conditions to pressure fluctuations. In \autoref{sec:numerical}, the CFD governing equations, boundary conditions, discretization techniques, and simulation procedures are discussed. The methods and results of model verification and validation are detailed in \autoref{sec:valid}. Finally, the CFD results are presented and examined in \autoref{sec:results}.

\section{Literature Review}\label{sec:literature}


Lighthill's groundbreaking work in 1952, which developed the acoustic analogy approach by reformulating the Navier-Stokes equations to separate sound creation from propagation, laid the theoretical groundwork for aeroacoustic prediction~\cite{Lighthill1952}. In 1969, Ffowcs Williams and Hawkings expanded this innovative framework to include solid boundaries and moving surfaces~\cite{FfowcsWilliamsHawkings1969}, resulting in what is still the most popular approach for far-field noise prediction in contemporary computational aeroacoustics. As noted by Petrosino and Barbarino, since these early advancements in the 1950s, research into jet-noise generation has continued as more advanced engine designs require faster and more accurate simulation and prediction techniques~\cite{Petrosino2023SemiEmpiricalJetNoise}.


Early computational attempts to predict jet noise relied heavily on Reynolds-Averaged Navier-Stokes (RANS) methods due to their computational efficiency. Nevertheless, these methods turned out to be essentially insufficient for capturing the unsteady turbulent structures responsible for noise generation. Tyacke et al.~\cite{Tyacke2016PredictiveLES} demonstrated that low-order RANS methods are inadequate for complex jets, as they inherently average out the temporal fluctuations crucial for acoustic radiation. This limitation prompted a paradigm shift toward Large-Eddy Simulation (LES), which directly resolves the energy-containing turbulent scales while modeling solely the smallest, dissipative scales.

The transition to LES marked a fundamental improvement in predictive capability. Brès and Lele describe how advances in meshing, numerical schemes, and subgrid modeling now allow LES to represent nozzle-shape variations and capture turbulent boundary layers at the nozzle exit, significantly improving quantitative accuracy in far-field sound pressure and spectral shape~\cite{Bres2019JetNoiseLES}. The superiority of LES over RANS was definitively demonstrated by Murugu et al., who compared URANS, LES, and DES turbulence models for a Mach 0.8 chevron nozzle jet, finding that LES and DES predictions agreed well with experimental measurements, whereas URANS substantially under-predicted the noise~\cite{sym14101975}.


A critical consideration in LES implementation is the choice between structured and unstructured grid strategies. Fosso Pouangué et al. conducted a comparison of two LES methodologies for a Mach 0.9 cold jet, contrasting block-structured grids with low-dissipative finite-volume schemes against fully unstructured tetrahedral grids employing low-dissipative Taylor-Galerkin finite-element schemes~\cite{fosso2015subsonic}. Both methods used LES to calculate acoustic sources and the FW-H analogy to propagate noise, showing that unstructured grids are just as accurate as structured methods at handling intricate noise-reduction devices like chevrons and dual-stream nozzles. This validation was essential for extending LES to industrially relevant configurations with geometric complexity.


Despite LES's success, the requirement to resolve thin near-wall layers makes pure LES computationally unfeasible for complex installations and high Reynolds-number jets. Due to this difficulty, hybrid RANS/LES techniques have been developed, in which LES resolves the free shear layers where the majority of noise is produced while RANS is used in the near-wall region~\cite{Tyacke2016PredictiveLES}. Further advances include Wall-Modeled LES (WMLES), which Stich et al. describe as applying an analytical wall-stress law to eliminate the need for $y^+ \approx 1$ resolution, enabling better aspect ratios and improved acoustic predictions within practical turnaround times~\cite{Stich2022WallModeledLES}.

For far-field noise prediction, hybrid acoustic methods have become standard practice. The resolved turbulent field is generally enclosed by a permeable boundary, and small-amplitude acoustic perturbations are transmitted analytically via the Ffowcs Williams-Hawkings equation. Positioning the FW-H surface excessively near the turbulent region eliminates source contributions, whilst positioning it too distantly results in numerical errors. The "end-caps" method has been developed to mitigate this issue by using several surfaces and phase averaging to reduce spurious noise~\cite{Bres2019JetNoiseLES}.


The creation of specialized software platforms has made it easier to put these cutting-edge techniques into practice. Commercial CFD software has shown to be very capable. Accordingly, West and Caraeni performed LES of a Mach 0.75 jet from a round nozzle using STAR-CCM+ and achieved agreement with experimental data to within 1-2 dB for both near-field turbulence and far-field noise~\cite{west2015jetnoise}. Similarly, Prasad et al. investigated passive and active flow-control strategies in supersonic jets using STAR-CCM+, demonstrating significant alterations in turbulence development and far-field sound pressure levels~\cite{prasad2019fluidinserts,prasad2019effect}. Dewan has also effectively utilized ANSYS Fluent, leveraging hybrid DDES formulations to capture important flow characteristics while producing accurate acoustic forecasts using the FW-H analogy~\cite{dewan2013supersonic}.

Among specialist high-fidelity solvers, AVBP, created at CERFACS, is distinguished by its sophisticated numerical methods. This unstructured compressible LES solver utilizes the Taylor-Galerkin method TTG4A, delivering third-order spatial precision and fourth-order temporal accuracy~\cite{TTG4A}. The EXEJET project deployed AVBP to create a comprehensive experimental database of dual-stream jets, validating three configurations with varying meshing procedures. The results indicated that regulated tripping tactics near the nozzle exit were essential for precise predictions, with acoustic forecasts at $30^\circ$ and $60^\circ$ closely aligning with experimental data.~\cite{yue2014exejet,barras2012nozzle}.


The democratization of jet aeroacoustic simulation has been significantly advanced by open-source platforms, particularly OpenFOAM and libAcoustics library. This framework has become widely adopted due to its flexibility and extensibility in implementing the Ffowcs Williams-Hawkings acoustic analogy. Epikhin et al. pioneered the integration of libAcoustics with the Quasi-Gas Dynamic (QGDFoam) solver to predict free jet noise at Mach 0.9, showing predominant radiation at $30^\circ$ relative to the jet axis, in alignment with experimental data~\cite{Epikhin2020FreeJetNoiseQGD}. The framework has been successfully extended to low Reynolds number compressible jets~\cite{Epikhin2019LowReJetOpenFOAM}, complex multiphase flows using Eulerian-Lagrangian approaches~\cite{popov2021eulerian,Melnikova2021GasDropletAcoustic}, and various industrial applications including synthetic jet control~\cite{MurilloRincon2023SyntheticJetNoise}, combustion noise in residential furnaces~\cite{Williamson2022CombustionNoiseFurnaces}, and under-expanded jets in aerospace contexts~\cite{Wang2023UnderexpandedJetOpenFOAM}.

The success of libAcoustics shows that open-source implementations can provide the transparency and customisation necessary for research applications while achieving accuracy similar to commercial and specialist solvers. This makes OpenFOAM with libAcoustics the perfect framework for the current investigation since it combines the ability to apply cutting-edge numerical techniques for jet noise simulation with verified acoustic prediction capabilities.


To validate the CFD model, six simulations were conducted in total. For aeroacoustic validation, the study by Arkhipov et al. \cite{Arkhipov2025} was employed, in which two sets of microphones captured the generated noise of VRC Nozzle 58 and VRC Nozzle 70 cold spray jets in free jet conditions. They provided SPL as a function of frequency for different chamber pressures and temperatures, but with limited data on particle distribution and behavior. Five simulations were based on the Arkhipov configuration with varying temperature and pressure, three cases at constant 757 K with pressures of 45, 55, and 65 bar, and three cases at constant 45 bar with temperatures of 707 K, 757 K, and 805 K. To validate particle distribution, the study by Allofs et al. \cite{Allofs2023, Allofs2022} was used, in which shadowgraphy was used to monitor the mass flux of different materials at the nozzle exit. One simulation was conducted using the nozzle configuration and operating conditions from this study. The common configuration parameters are shown in \autoref{tab:validation_conditions}.

\begin{table}
\centering
\caption{Experimental conditions used for CFD model validation}
\label{tab:validation_conditions}
\begin{tabular}{lcc}
\toprule
\textbf{Parameter} & \textbf{Arkhipov et al.}\cite{Arkhipov2025} & \textbf{Allofs et al.}\cite{Allofs2023} \\
\midrule
\multicolumn{3}{l}{\textbf{Nozzle Geometry}} \\
Inlet Diameter (mm) & 9.53 & 70.3 \\
Throat Diameter (mm) & 1.73 & 22.13 \\
Exit Diameter (mm) & 5.08 & 30.0 \\
Converging Length (mm) & 44.45 & 63 \\
Diverging Length (mm) & 153.16 & 49 \\
\midrule
\multicolumn{3}{l}{\textbf{Operating Conditions}} \\
Material & Cu & Al$_2$O$_3$ \\
Inlet Temperature (K) & 707, 757, 807 & 374.7 \\
Inlet Pressure (MPa) & 4.5, 5.5, 6.5 & 0.952 \\
Particle Density (kg/m$^3$) & 8800 & 3950 \\
Feed Rate (kg/min) & 0.0148 & 2.73 \\
\midrule
\multicolumn{3}{l}{\textbf{Particle Size Distribution}} \\
Mean Diameter ($\mu$m) & 37 & 20 \\
Standard Deviation ($\mu$m) & 15 & 10 \\
Minimum Diameter ($\mu$m) & 5 & 9 \\
Maximum Diameter ($\mu$m) & 100 & 30 \\
\bottomrule
\end{tabular}
\end{table}

\section{Analytical Model}\label{sec:analytical}

This section presents fundamental acoustic calculations for jet nozzle systems, emphasizing the relationship between nozzle chamber conditions and far-field sound propagation. Spherical wave propagation is assumed, and conservation of acoustic power is applied to derive the sound pressure level at a designated distance from the nozzle exit.

\begin{figure}[ht]
    \centering
    \includegraphics[width=0.5\linewidth]{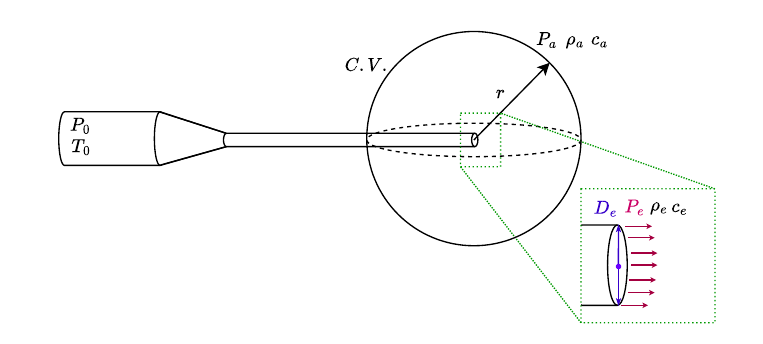}
    \caption{Schematic of Analytical model}
    \label{fig:acousticSchematic}
\end{figure}

\subsection{Derivation}

A nozzle with exit diameter $D_e$ exhausting a high-speed jet into a quiescent medium is considered, as illustrated in \autoref{fig:acousticSchematic}. To simplify the model, several key assumptions are introduced.

The acoustic field is assumed to exhibit spherical symmetry, with no dependence on the polar angle $\theta$ or the azimuthal angle $\phi$. This assumption is considered reasonable in the far field, where the nozzle behaves as a point source. Furthermore, the nozzle walls are assumed not to interfere with acoustic wave propagation, and the exit diameter is considered negligible compared to the observation distance, such that $D_e \ll r$. The principle of acoustic power conservation is applied, requiring the following equation~\cite{langley1971acoustic}:

\begin{equation}
    W_{in} - W_{out} = 0,
\end{equation}

where $W_{in}$ and $W_{out}$ denote the acoustic power entering and leaving a control surface, respectively.

The acoustic power generated at the nozzle exit is regarded as being related to the sound intensity and the exit area. For a plane wave at the nozzle exit, the connection between sound intensity and pressure is described by the following~\cite{morgan1961use}:

\begin{equation}
\label{eq:intensity_exit}
I_e = \frac{P_e^2}{\rho_e c_e}
\end{equation}

In \autoref{eq:intensity_exit}, $I_e$ denotes the sound intensity at the nozzle exit [$\mathrm{W/m^2}$]. The variable $P_e$ represents the pressure at the exit [$\mathrm{Pa}$], $\rho_e$ denotes the fluid density at the nozzle exit [$\mathrm{kg/m^3}$], and $c_e$ represents the speed of sound at the nozzle exit [$\mathrm{m/s}$].

The acoustic power input is defined as the total acoustic energy flux passing through the nozzle exit plane and is expressed as the product of the sound intensity and the nozzle exit area.

\begin{equation}
\label{eq:power_in}
W_{in} = I_e \cdot A_e = \frac{P_e^2}{\rho_e c_e} \cdot \frac{\pi D_e^2}{4}
\end{equation}

Here, $W_{in}$ denotes the acoustic power input [$W$], $A_e$ represents the nozzle exit area [$m^2$], and $D_e$ is the nozzle exit diameter [$m$].

In the far field, the assumption of spherical wave propagation becomes valid as the nozzle dimensions become negligible compared to the observation distance. Under this condition, the acoustic power radiates uniformly through a spherical surface of radius $r$ centered at the nozzle exit. The total acoustic power output is expressed as:

\begin{equation}
\label{eq:power_out}
W_{out} = I(r) \cdot 4\pi r^2
\end{equation}

In \autoref{eq:power_out}, $W_{out}$ represents the acoustic power output [$W$], $I(r)$ denotes the sound intensity at distance $r$ [$W/m^2$], and $r$ is the distance from the nozzle exit to the observation point [$m$].

The conservation of acoustic power principle requires that the power input at the nozzle exit equals the power output in the far field, assuming no dissipative losses:

\begin{equation}
\label{eq:power_conservation}
W_{in} = W_{out}
\end{equation}

Substituting \autoref{eq:power_in} and \autoref{eq:power_out} into \autoref{eq:power_conservation} yields the following:

\begin{equation}
\label{eq:power_balance}
\frac{P_e^2}{\rho_e c_e} \cdot \frac{\pi D_e^2}{4} = I(r) \cdot 4\pi r^2
\end{equation}

Solving \autoref{eq:power_balance} for the far-field sound intensity provides the relationship governing sound propagation from the nozzle:

\begin{equation}
\label{eq:intensity_farfield}
I(r) = \frac{P_e^2 D_e^2}{16 \rho_e c_e r^2}
\end{equation}

\autoref{eq:intensity_farfield} demonstrates that the sound intensity decreases as $1/r^2$, which is the characteristic signature of spherical wave propagation from a point source.

The relationship between sound intensity and root-mean-square pressure in the far field depends critically on the local medium properties, which may differ from the nozzle exit conditions. This relationship is governed by the acoustic impedance of the medium~\cite{kinsler2000fundamentals}:

\begin{equation}
\label{eq:pressure_intensity}
P_{rms}(r) = \sqrt{I(r) \cdot \rho_a c_a}
\end{equation}

In \autoref{eq:pressure_intensity}, $P_{rms}(r)$ represents the root-mean-square sound pressure at distance $r$ [$Pa$], $\rho_a$ denotes the density of the ambient medium [$kg/m^3$], and $c_a$ represents the speed of sound in the ambient medium [$m/s$].

Substituting \autoref{eq:intensity_farfield} into \autoref{eq:pressure_intensity} yields the explicit expression for far-field pressure:

\begin{equation}
\label{eq:pressure_farfield}
P_{rms}(r) = \sqrt{\frac{P_e^2 D_e^2}{16 \rho_e c_e r^2} \cdot \rho_a c_a} = \frac{P_e D_e}{4r} \sqrt{\frac{\rho_a c_a}{\rho_e c_e}}
\end{equation}

It is common to use Sound Pressure Level(SPL) instead of $P_{rms}$ as it expresses sound intensity on a logarithmic scale (in decibels [$dB$]), which better matches how the human ear perceives loudness~\cite{fletcher1933loudness} and is defined as~\cite{ansi2013s}:

\begin{equation}
\label{eq:spl_definition}
SPL = 20 \log_{10}\left(\frac{P_{rms}}{P_{ref}}\right)
\end{equation}

$P_{ref}$ represents the reference pressure equal to $2 \times 10^{-5}$ pascals, which corresponds to the threshold of human hearing.

Substituting \autoref{eq:pressure_farfield} into \autoref{eq:spl_definition} provides the complete expression for the sound pressure level as a function of the nozzle parameters and the observation distance.

\begin{equation}
\label{eq:spl_complete}
SPL(r) = 20 \log_{10}\left(\frac{P_e D_e}{4r P_{ref}} \sqrt{\frac{\rho_a c_a}{\rho_e c_e}}\right)
\end{equation}

The acoustic impedance terms can be expressed in terms of gas properties using the relationships $c = \sqrt{\gamma R T}$ and the ideal gas law $\rho = p/(RT)$. For the ambient medium, in which $R$ is the universal gas constant, $T$ is the temperature $[K]$ and $\gamma$ is heat capacity ratio,  the acoustic impedance becomes:

\begin{equation}
\label{eq:ambient_impedance}
\rho_a c_a = \frac{P_a}{R_a T_a} \sqrt{\gamma_a R_a T_a}  = P_a \sqrt{\frac{\gamma_a}{R_a T_a}}
\end{equation}

Similarly, for the exit conditions:

\begin{equation}
\label{eq:exit_impedance}
\rho_e c_e = \frac{P_e}{R_e T_e} \sqrt{\gamma_e R_e T_e} = P_e \sqrt{\frac{\gamma_e}{R_e T_e}}
\end{equation}

Substituting \autoref{eq:ambient_impedance} and \autoref{eq:exit_impedance} relationships into \autoref{eq:pressure_farfield}:

\begin{equation}
P_{rms} = \frac{P_e D_e}{4r} \sqrt{\frac{P_a \sqrt{\gamma_a/(R_a T_a)}}{P_e \sqrt{\gamma_e/(R_e T_e)}}} = \frac{P_e D_e}{4r} \sqrt{\frac{P_a}{P_e}} \left(\frac{\gamma_a R_e T_e}{\gamma_e R_a T_a}\right)^{1/4}
\end{equation}

For applications involving the same gas composition at exit and ambient conditions ($\gamma_a = \gamma_e = \gamma$ and $R_a = R_e = R$), the impedance ratio simplifies: 

\begin{equation}
\label{eq:pressure_gas_dynamic}
P_{rms} = \frac{P_e D_e}{4r} \sqrt{\frac{P_a}{P_e}} \left(\frac{T_e}{T_a}\right)^{1/4} = \frac{1}{4} \left(\frac{r}{D_e}\right)^{-1} (P_e P_a)^{1/2} \left(\frac{T_e}{T_a}\right)^{1/4}
\end{equation}

Furthermore, the exit pressure and temperature can be expressed in terms of stagnation conditions using the isentropic flow relations from gas dynamics~\cite{john2006gas}:

\begin{equation}
\label{eq:exit_pressure_stagnation}
P_e = P_0 \left(1 + \frac{\gamma-1}{2}Ma_e^2\right)^{-\frac{\gamma}{\gamma-1}}
\end{equation}

\begin{equation}
\label{eq:exit_temperature_stagnation}
T_e = T_0 \left(1 + \frac{\gamma-1}{2}Ma_e^2\right)^{-1}
\end{equation}

Substituting \autoref{eq:exit_pressure_stagnation}  and \autoref{eq:exit_temperature_stagnation} into \autoref{eq:pressure_gas_dynamic}:

\begin{equation}
\label{eq:pressure_complete_stagnation}
P_{rms} = \frac{1}{4} \cdot\left(\frac{r}{D_e}\right)^{-1} \cdot (P_a P_0)^{1/2} \cdot\left(\frac{T_0}{T_a}\right)^{1/4} \cdot \left(1+\frac{\gamma-1}{2}Ma_e^2\right)^{-\left(\frac{\gamma}{2(\gamma-1)}+\frac{1}{4}\right)}
\end{equation}

The exit Mach number $Ma_e$ is implicitly related to the nozzle geometry through the area-Mach relationship~\cite{john2006gas}:

\begin{equation}
\label{eq:area_mach_general}
\frac{A}{A^*} = \frac{1}{Ma}\left[\frac{2}{\gamma+1}\left(1 + \frac{\gamma-1}{2}Ma^2\right)\right]^{\frac{\gamma+1}{2(\gamma-1)}}
\end{equation}

In \autoref{eq:area_mach_general}, $A^*$ represents the sonic throat area where the Mach number equals unity. The area ratio from throat to exit determines the exit Mach number through \autoref{eq:area_mach_general}:

\begin{equation}
\label{eq:exit_area_ratio}
\frac{A_{\text{exit}}}{A_{\text{throat}}} = \left(\frac{D_e}{D_t}\right)^2 = f(Ma_e,\gamma) = \frac{1}{Ma_e}\left[\frac{2}{\gamma+1}\left(1 + \frac{\gamma-1}{2}Ma_e^2\right)\right]^{\frac{\gamma+1}{2(\gamma-1)}}
\end{equation}

So, finally:

\begin{equation}
    P_{rms} = \frac{1}{4} \cdot\left(\frac{r}{D_e}\right)^{-1} \cdot (P_a P_0)^{1/2} \cdot\left(\frac{T_0}{T_a}\right)^{1/4} \cdot \left(1+\frac{\gamma-1}{2}Ma_e^2\right)^{-\left(\frac{\gamma}{2(\gamma-1)}+\frac{1}{4}\right)}
\end{equation}

\begin{equation}
\label{eq:spl_unified_final}
SPL = 20\log_{10}\left(\frac{1}{4P_{ref}} \cdot\left(\frac{r}{D_e}\right)^{-1} \cdot (P_a P_0)^{1/2} \cdot\left(\frac{T_0}{T_a}\right)^{1/4} \cdot \left(1+\frac{\gamma-1}{2}Ma_e^2\right)^{-\left(\frac{\gamma}{2(\gamma-1)}+\frac{1}{4}\right)}\right)
\end{equation}

\begin{equation}
\label{eq:mach_function}
Ma_e = f^{-1}\left(\left(\frac{D_e}{D_t}\right),\gamma\right) = M\left(\left(\frac{D_e}{D_t}\right),\gamma\right)
\end{equation}

The function $f$ in \autoref{eq:exit_area_ratio} maps the exit Mach number $Ma_e$ and heat capacity ratio $\gamma$ to the exit-to-throat area ratio $(A_{exit}/A_{throat}$. The inverse function $f^{-1}$ or $M$ determines the exit Mach number for a given area ratio and $\gamma$. The \autoref{eq:spl_unified_final} can be decomposed into additive contributions by applying logarithmic properties.

\begin{equation}
\label{eq:spl_decomposed_final}
SPL = - \underbrace{20\log_{10}\left(\frac{r}{D_e}\right)}_{\text{Geometry}} + \underbrace{20\log_{10}(\frac{\sqrt{P_a P_0}}{4P_{ref}})}_{\text{Pressure}} + \underbrace{5\log_{10}\left(\frac{T_0}{T_a}\right)}_{\text{Temperature}} + \underbrace{SPL_{\text{comp}}}_{\text{Compressibility}}
\end{equation}

where the compressibility contribution is defined as:

\begin{equation}
\label{eq:spl_compressibility_final}
SPL_{\text{comp}} = -20\left(\frac{\gamma}{2(\gamma-1)}+\frac{1}{4}\right)\log_{10}\left(1+\frac{\gamma-1}{2}M\left(\left(\frac{D_e}{D_t}\right),\gamma\right)^2\right)
\end{equation}

Consequently, five factors determine the SPL at a specific distance from a nozzle:

\begin{equation}
\label{eq:spl_dependency}
SPL = SPL\left(\frac{r}{D_e}, \frac{\sqrt{P_a P_0}}{4P_{ref}}, \frac{T_0}{T_a}, \gamma, \frac{D_e}{D_t}\right)
\end{equation}

\begin{figure}[ht]
    \centering
    \includegraphics[width=\linewidth]{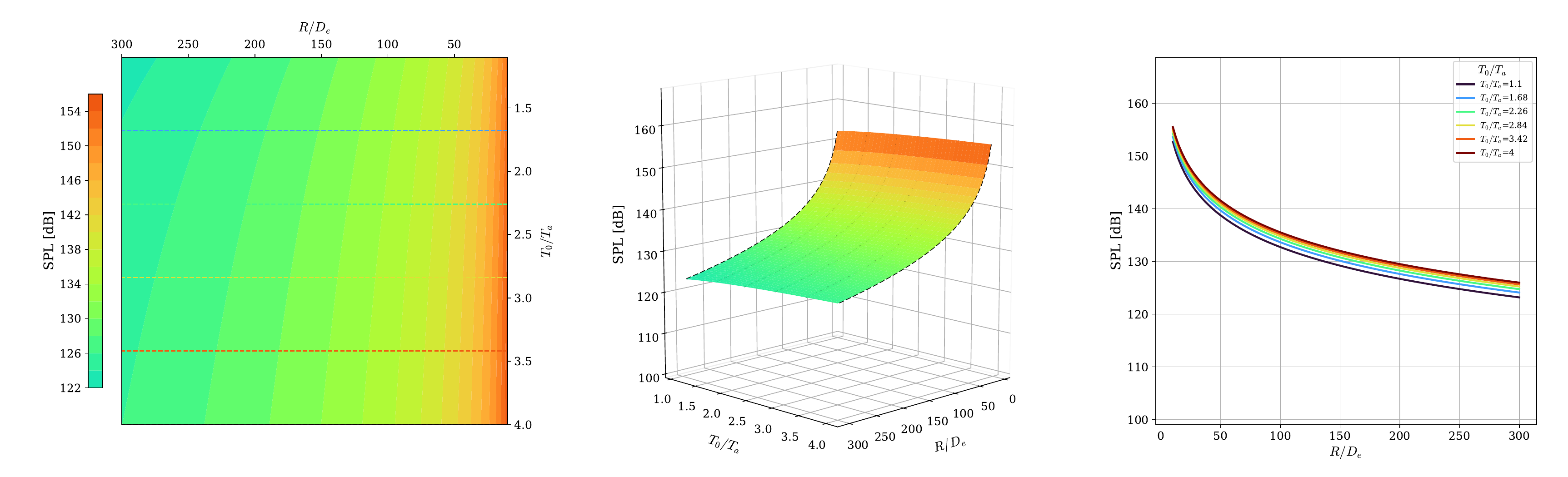}
    \caption{Sensitivity analysis of SPL vs $R/D_e$ and $T_0/Ta$}
    \label{fig:T0Ta}
\end{figure}

\begin{figure}
    \centering
    \includegraphics[width=\linewidth]{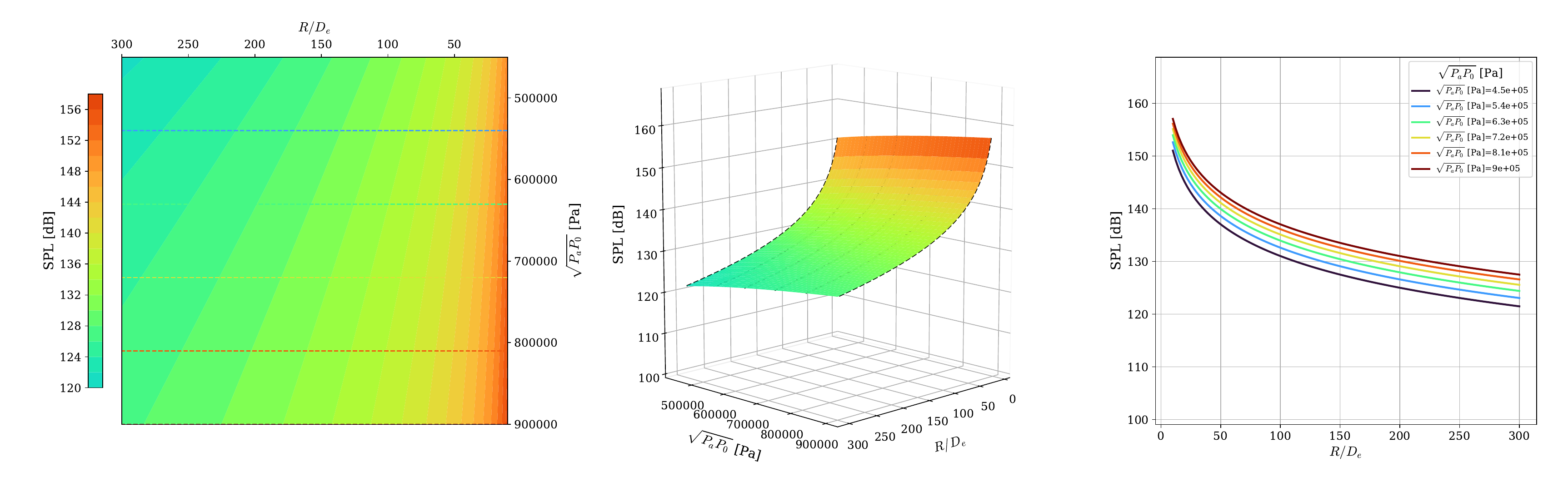}
   \caption{Sensitivity analysis of SPL vs $R/D_e$ and $\sqrt{P_0P_a}$}
    \label{fig:PaP0}
\end{figure}

It is evident that the SPL decreases with increasing radial distance from the nozzle, following the geometric spreading term $-20\log_{10}(r/D_e)$, which corresponds to a 6~dB reduction for each doubling of distance. This trend is illustrated in \autoref{fig:T0Ta} and \autoref{fig:PaP0}. The pressure-dependent term $20\log_{10}\!\left(\frac{\sqrt{P_a P_0}}{4P_{\mathrm{ref}}}\right)$ exhibits a noticeably stronger influence on SPL than the temperature-dependent term $5\log_{10}(T_0/T_a)$. Consequently, changes in chamber or ambient pressure have a more pronounced logarithmic impact on the acoustic field compared to variations in temperature. This behavior is consistent with the validation data from \cite{Arkhipov2025}.

\subsection{Model calibration}

\begin{figure}[htbp]
    \centering
    \begin{subfigure}{0.48\textwidth}
        \centering
        \includegraphics[width=\linewidth]{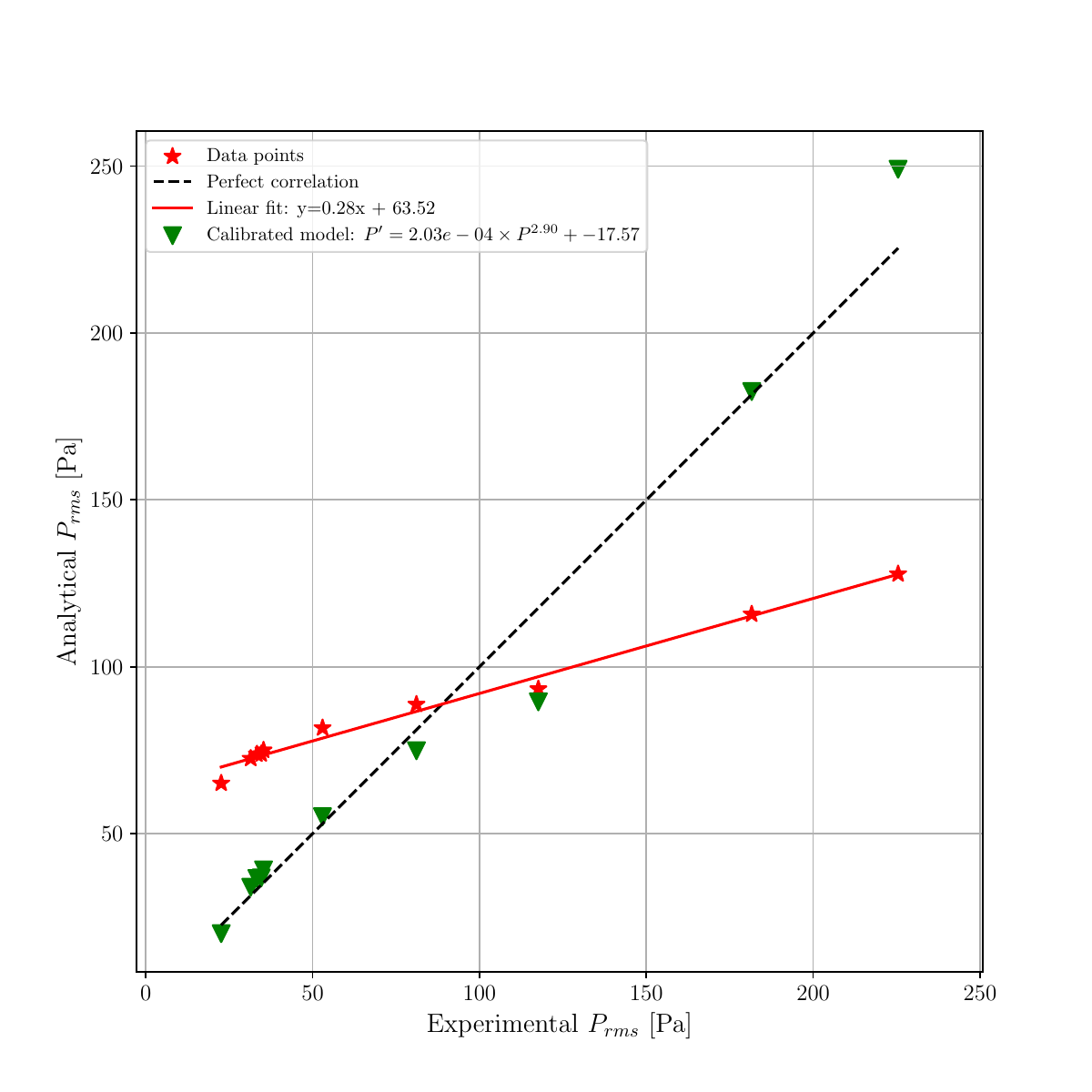}
        \caption{Pressure amplitude validation}
        \label{fig:prms_validation}
    \end{subfigure}
    \hfill
    \begin{subfigure}{0.48\textwidth}
        \centering
        \includegraphics[width=\linewidth]{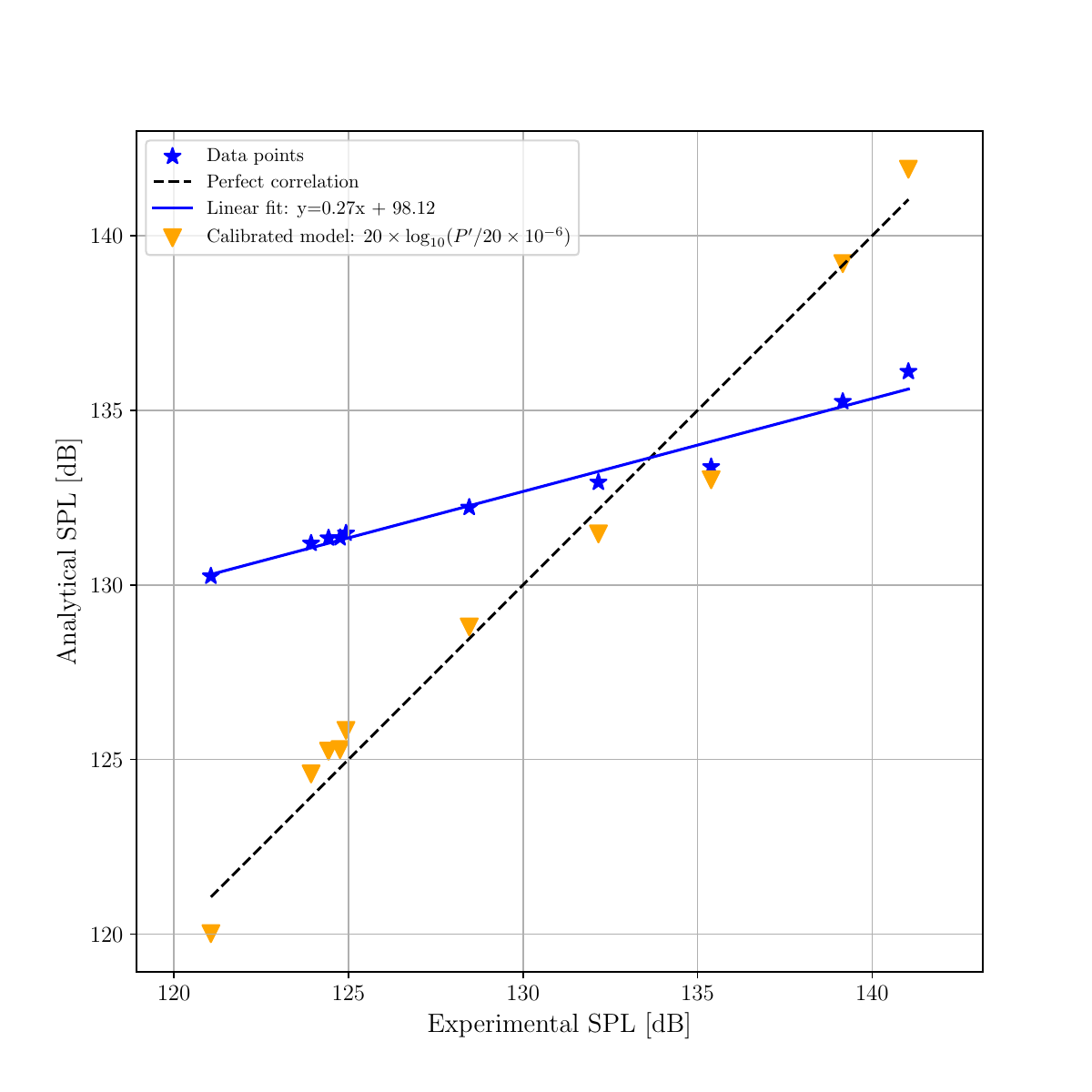}
        \caption{Sound pressure level validation}
        \label{fig:spl_validation}
    \end{subfigure}
    \caption{Comparison of analytical acoustic model predictions against experimental measurements for (a) pressure amplitude P$_{rms}$ and (b) sound pressure level (SPL)}
    \label{fig:acoustic_validation}
\end{figure}

The \autoref{tab:spl_validation_cal} shows the comparison of our analytical model derived in \autoref{eq:pressure_complete_stagnation} and \autoref{eq:spl_decomposed_final}. It is noticeable that the $P_{rms}$ and its corresponding $SPL$ show significant differences between analytical and experimental values. To demonstrate this difference, \autoref{fig:acoustic_validation} is plotted. It is easily observed that the error between the analytical and experimental values follows a trend line, it is possible to add a calibration function to correct the discrepancy.

To calibrate the analytical model to experiments, the pressure was rescaled as follows:
\begin{equation}
    P_{rms}' = A \times P_{rms}^{B} + C
    \label{eq:calibration_function}
\end{equation}
\begin{equation}
    SPL' = 20\times \log_{10}\left(\frac{P_{rms}'}{20 \times 10^{-6}}\right)
    \label{eq:calibrated_spl}
\end{equation}

By fitting the equation, it was found that
\begin{align*}
    A &= 2.03 \times 10^{-4} && B = 2.90 && C = -17.57
\end{align*}

In this way, the new model predicts the $P_{rms}$ and $SPL$ more accurately. The calibrated values in \autoref{tab:spl_validation_cal} demonstrate the improved agreement with experimental data. Additionally, the calibrated model predictions can be seen in \autoref{fig:acoustic_validation} as green triangles for $P_{rms}$ and orange triangles for $SPL$, respectively.

\begin{table}
\centering
\caption{Validation of SPL analytical model against experimental data with calibrated model results}
\label{tab:spl_validation_cal}
\begin{adjustbox}{max width=\textwidth}
\begin{tabular}{lcccccccc}
\toprule
\textbf{Parameter} & \textbf{Value} & \textbf{Experimental} & \textbf{Original} & \textbf{Calibrated} & \textbf{Experimental} & \textbf{Original} & \textbf{Calibrated} \\
& & \textbf{SPL [dB]} & \textbf{SPL [dB]} & \textbf{SPL [dB]} & \textbf{P$_{rms}$ [Pa]} & \textbf{P$_{rms}$ [Pa]} & \textbf{P$_{rms}$ [Pa]} \\
\midrule
\hline
\multicolumn{8}{l}{\textit{Pressure Variation (T$_0$ = 757.15 K, D$_t$ = 1.73 mm)}} \\
35 bar & 3.5 MPa & 121.06 & 130.25 & 120.01 & 22.60 & 65.09 & 20.02 \\
45 bar & 4.5 MPa & 124.76 & 131.35 & 125.28 & 34.60 & 73.88 & 36.73 \\
55 bar & 5.5 MPa & 128.46 & 132.22 & 128.80 & 52.97 & 81.66 & 55.06 \\
65 bar & 6.5 MPa & 132.16 & 132.94 & 131.46 & 81.10 & 88.72 & 74.83 \\
\midrule
\hline

\multicolumn{8}{l}{\textit{Temperature Variation (P$_0$ = 4.5 MPa, D$_t$ = 1.73 mm)}} \\
$434^\circ C$ & 707 K & 123.93 & 131.19 & 124.58 & 31.44 & 72.53 & 33.90 \\
$484^\circ C$ & 757 K & 124.43 & 131.34 & 125.24 & 33.31 & 73.80 & 36.55 \\
$534^\circ C$ & 807 K & 124.93 & 131.48 & 125.83 & 35.28 & 75.00 & 39.14 \\
\midrule
\hline

\multicolumn{8}{l}{\textit{Throat Diameter Variation (P$_0$ = 4.5 MPa, T$_0$ = $434 ^\circ C$)}} \\
2.00 mm & D$_e$/D$_t$ = 2.54 & 135.39 & 133.38 & 133.01 & 117.63 & 93.33 & 89.47 \\
2.26 mm & D$_e$/D$_t$ = 2.25 & 139.16 & 135.25 & 139.20 & 181.56 & 115.75 & 182.44 \\
2.39 mm & D$_e$/D$_t$ = 2.13 & 141.04 & 136.11 & 141.91 & 225.44 & 127.80 & 249.07 \\
\bottomrule
\hline

\end{tabular}
\end{adjustbox}

\footnotesize
\textbf{Constants:} P$_a$ = 101.325 kPa, $\gamma$ = 1.4, R/D$_e$ = 196.85, D$_e$ = 5.08 mm, P$_{ref}$ = 20 $\mu$Pa
\end{table}

Overall, an analytical model has been developed using gas dynamics laws and acoustic equations for jet nozzles. Although the analytical model captures the general trends, several assumptions such as neglecting turbulent flow effects resulted in systematic errors. However, since the errors followed a predictable trend, a power-law calibration function was implemented. This calibration function significantly reduced the error between the analytical model and experimental measurements across various operating conditions, including different chamber pressures, temperatures, and nozzle geometries. The calibrated model demonstrates good agreement with experimental data, making it suitable for practical acoustic predictions in jet nozzle applications.

\section{Numerical Model}\label{sec:numerical}

\begin{figure}[ht]
  \centering
      \resizebox{\linewidth}{!}{
  \input{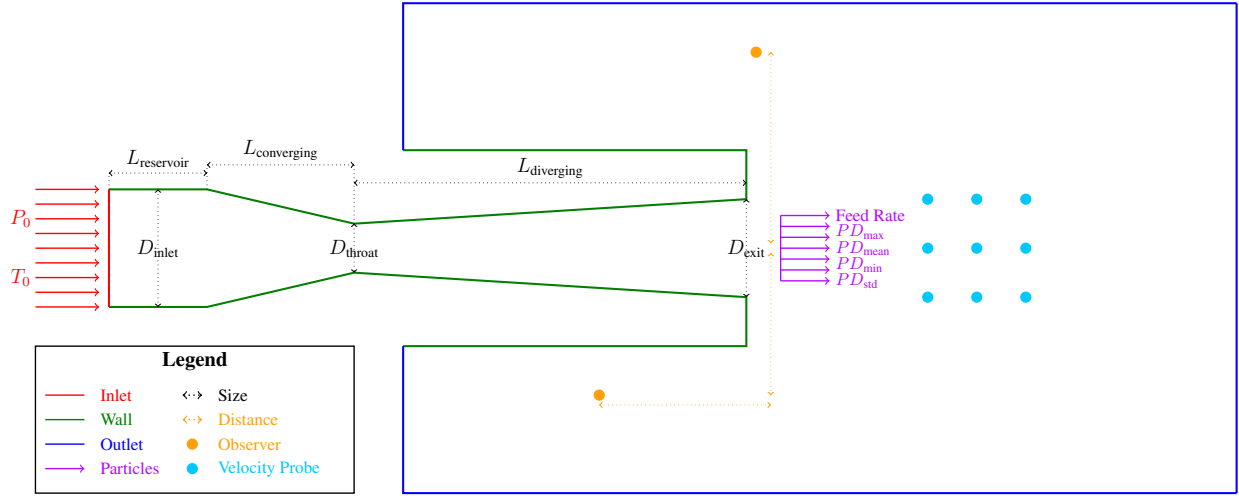}}
  \caption{Schematic of the CFD simulation computational domain, design parameters, inlet conditions, placement of velocity verification probes, particle injection region and acoustic noise observer positions.}
  \label{fig:CFD_schematic}
\end{figure}

\autoref{fig:CFD_schematic} illustrates the computational domain. The nozzle geometry is defined by the section lengths (reservoir, converging, diverging) and by the inlet, throat, and exit diameters. 
The flow is driven by a predetermined stagnation input of total temperature and pressure, and the ambient environment is simulated by a pressure far-field boundary condition. 
An exterior wall expansion of the nozzle is added, and the domain is expanded both upstream and downstream of the nozzle exit to increase the distance between the jet and the far-field barriers in order to reduce boundary effects on the jet development and the radiated acoustics. 
For verification of the flow field, velocity probes are positioned along the jet centerline downstream of the nozzle exit. The recorded velocities are used to plot the energy cascade and to validate the turbulence model configuration used in the simulation.
Two acoustic observers are used to record pressure fluctuations: one located $0.6\,\text{m}$ above the nozzle exit, and another $1.0\,\text{m}$ downstream and $1.0\,\text{m}$ above the centerline. 
The specified mass flow rate, particle density, and a normally distributed particle size determined by its mean, minimum, maximum, and standard deviation are used to govern the injection of particles at the nozzle exit from a circular disc.

\begin{figure}
    \centering
    \includegraphics[width=\linewidth]{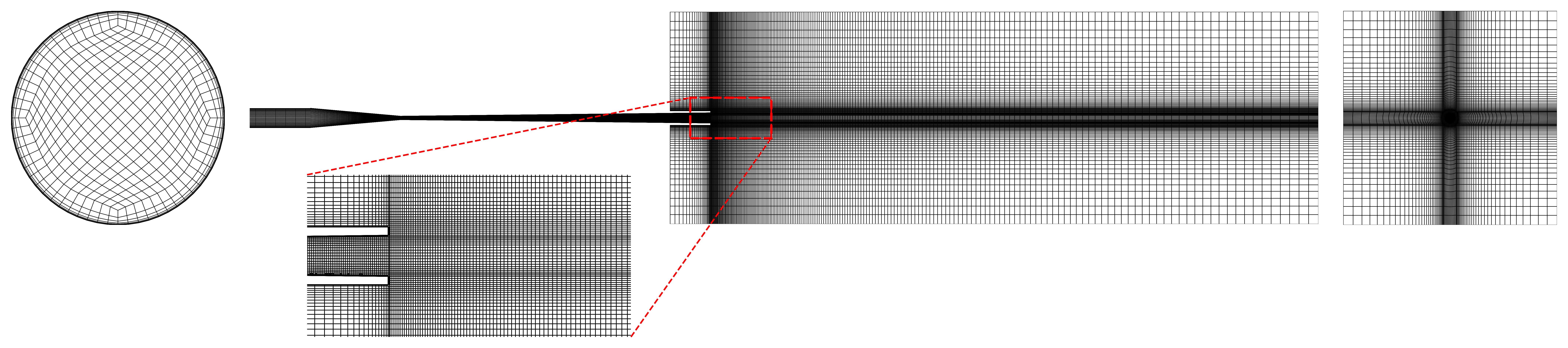}
  \caption{Structured H-topology mesh: cross-section, full domain, and near-nozzle detail.}
  \label{fig:mesh_panels}
\end{figure}

The grid was generated with \texttt{classyBlocks}~\cite{damogranlabs2023classyblocks} using a structured hexahedral multi-block layout. A sample of generated mesh can be seen in \autoref{fig:mesh_panels}. Inside the nozzle, an H topology is used to align the cells in the stream direction. Near the solid walls, the first few layers are contracted to form a boundary layer stack, reducing $y^+$ along the surfaces of the internal and external nozzles. Downstream of the exit, the mesh remains structured but expands smoothly toward the far field so that resolution is concentrated where the shear layer and potential core develop while unnecessary refinement is avoided away from the jet.


The numerical simulation of the high-temperature, high-pressure jet nozzle flow requires solving the complete set of conservation laws governing compressible and turbulent motion, as summarized in \autoref{tab:cfd_summary}. The continuity equation enforces mass conservation throughout the computational domain, while the momentum equations describe the evolution of the velocity field with viscous effects represented through an effective viscosity that combines molecular and turbulence-induced contributions. The energy equation captures the coupling between thermal transport, flow work, and viscous dissipation, with heat flux modeled via Fourier's law using an effective thermal conductivity. The system is closed by the ideal gas equation of state, relating pressure, density, and temperature. Because the flow spans a wide temperature range, viscosity is evaluated using Sutherland's law~\cite{Sutherland01121893}.

\begin{table*}
\centering
\caption{Summary of governing equations for jet nozzle CFD simulation}
\label{tab:cfd_summary}
\begin{adjustbox}{max width=\textwidth}
\begin{tabular}{l>{\hspace{2em}}l >{\hspace{2em}}l}
\toprule
\textbf{Physics} & \textbf{Governing Equation} &  \\
\midrule
\hline
Mass Conservation & $\displaystyle\frac{\partial \rho}{\partial t} + \nabla \cdot (\rho \mathbf{U}) = 0$ &  \\[10pt]
\hline
Momentum Conservation 
& $\displaystyle\frac{\partial (\rho \mathbf{U})}{\partial t} + \nabla \cdot (\rho \mathbf{U}\mathbf{U}) = - \nabla p + \nabla \cdot \boldsymbol{\tau} + \rho \mathbf{g}$ & $\mu_{\mathrm{eff}} = \mu + \mu_t$ \\[8pt]
& $\boldsymbol{\tau} = \mu_{\mathrm{eff}}\!\left(\nabla \mathbf{U} + \nabla \mathbf{U}^{T} - \tfrac{2}{3}(\nabla\!\cdot\!\mathbf{U})\,\mathbf{I}\right)$ & \\[10pt]
\hline
Energy Conservation 
& $\displaystyle\frac{\partial (\rho h)}{\partial t} + \nabla \cdot (\rho \mathbf{U} h) + \frac{\partial (\rho K)}{\partial t} + \nabla \cdot (\rho \mathbf{U} K) - \frac{\partial p}{\partial t}$ & $k_{\mathrm{eff}} = k + k_t$ \\[5pt]
& $\displaystyle = -\,\nabla \cdot \mathbf{q} + \nabla \cdot (\boldsymbol{\tau}\cdot\mathbf{U}) + \rho \mathbf{g}\!\cdot\!\mathbf{U}$ & $k = \mu c_p/Pr$ \\[5pt]
& $\mathbf{q} = -\,k_{\mathrm{eff}}\,\nabla T$ & $k_t = \mu_t c_p/Pr_t$ \\[5pt]
& & $Pr = 0.7$, $Pr_t = 0.85$ \\[10pt]
\hline
Equation of State & $p = \rho R T$ & $R = 287.6$ J/(kg·K) \\[5pt]
& $h = c_p T$ & $\gamma = 1.411$ \\[10pt]
\hline
Transport Properties & $\displaystyle\mu(T) = A_s\,\frac{T^{3/2}}{T + T_s}$ & $A_s = 1.48 \times 10^{-6}$ kg/(m.s.$K^{1/2}$) \\[5pt]
& & $T_s = 116$ K \\[10pt] 
\hline
Turbulence Modeling & \autoref{sec:turbulence} \\[10pt] 
\bottomrule
\end{tabular}
 \end{adjustbox}
\end{table*}


Aeroacoustic modeling requires sophisticated turbulent modeling strategies to resolve unsteady pressure fluctuations across a wide range of temporal and spatial scales. Reynolds-Averaged Navier-Stokes (RANS) and Unsteady RANS (URANS) models, due to their time-averaging procedures, cannot capture the instantaneous turbulent fluctuations that generate acoustic sources in jet flows~\cite{Tyacke2016PredictiveLES, sym14101975}. Large Eddy Simulation (LES) has emerged as the standard for aeroacoustics, explicitly resolving large-scale turbulent structures while modeling only sub-grid scale eddies.

However, high-fidelity LES requires prohibitively fine mesh resolution throughout the domain, particularly near walls where $y^+ \approx 1$ must be maintained. This results in computational costs often requiring billions of grid cells. To address this challenge, Detached Eddy Simulation (DES) is employed, a hybrid approach using RANS in near-wall regions and LES in separated flow regions. This zonal treatment reduces computational expense while preserving resolution of the unsteady turbulent structures critical for acoustic source characterization.

Standard DES formulations, however, can suffer from grid-induced separation when the RANS-to-LES transition occurs prematurely within the attached boundary layer, particularly in regions where the grid spacing becomes comparable to the boundary layer thickness. To enhance simulation stability and prevent this premature mode switching, Delayed Detached Eddy Simulation (DDES) is implemented. The DDES formulation introduces a shielding function that maintains URANS behavior throughout the attached boundary layer until genuine flow separation occurs, at which point the model naturally transitions to LES mode. This approach eliminates grid-induced separation artifacts while ensuring proper resolution of the separated shear layers where acoustic sources are most active.

The $k$-$\omega$ SST (Shear Stress Transport) model serves as the RANS closure~\cite{menter1992improved}, selected for its accurate prediction of flow separation and proper treatment of adverse pressure gradients. Van Driest damping ensures appropriate asymptotic behaviour by lowering the turbulent eddy viscosity ($\nu_t$) close to walls. The mathematical framework of the SST DDES model is described in \autoref{sec:turbulence}.

The computational domain for the simulation consists of four primary boundaries: inlet, nozzle wall, outer nozzle, and outlet.

The inlet boundary conditions employ a ramped approach for both pressure and temperature to ensure numerical stability during the initial transient phase. The pressure inlet utilizes a time-dependent boundary condition that gradually increases from an initial value of 1~MPa to the specified inlet pressure over a defined ramp time. Similarly, the temperature boundary condition ramps from 400~K to the target inlet temperature. The detail of boundary conditions can be seen in \autoref{sec:BC}.

An Euler-Lagrange framework is used to model the particle-laden flow, tracking individual particles in the Lagrangian frame while solving the gas phase in an Eulerian way. The aerodynamic and thermal interactions between phases are captured via two-way momentum and energy coupling~\cite{michaelides2022multiphase}. The governing equations and numerical treatment of the particle phase are summarized in \autoref{tab:particle_summary}.

Particles are introduced at the nozzle exit plane, where they are injected into the high-speed jet and subsequently accelerated by the surrounding flow. Their motion is governed by Newton’s second law while accounting for the physical mechanisms that influence post-exit dynamics. Aerodynamic drag, modeled using a Reynolds-number-dependent spherical drag coefficient, is the dominant force driving particle acceleration. Pressure gradient forces become relevant where the jet experiences rapid expansion and flow deceleration, while the virtual mass force accounts for inertia effects during relative acceleration between the carrier gas and particles. Convective heat transfer between the gas and particles is predicted using a Nusselt number correlation based on the particle Reynolds number.

Particle collisions with the surrounding structures are modeled using a rebound formulation incorporating both normal restitution and tangential friction, with a restitution coefficient $e = 0.97$ representing the nearly elastic nature of metallic particle impacts and a Coulomb friction coefficient $\mu = 0.09$ governing tangential momentum loss. Inter-particle collisions are resolved using a Hertzian spring-slider-dashpot model in which a viscous damping term ($\alpha = 0.12$) dissipates kinetic energy and prevents unrealistically elastic rebounds~\cite{TSUJI1992239, Kuwabara_1987}.

Two-way coupling is enforced by accumulating particle momentum and thermal source terms into the Eulerian control volumes they traverse, allowing the discrete phase to modify the continuous gas flow. Particle trajectories are advanced with an explicit forward Euler scheme.

\begin{table*}
\centering
\caption{Summary of Lagrangian particle tracking model for nozzle CFD simulation}
\label{tab:particle_summary}
\begin{adjustbox}{max width=\textwidth}
\begin{tabular}{ll >{\hspace{1em}}l}
\toprule
\textbf{Physics} & \textbf{Governing Equation} & \textbf{Key Parameters} \\
\midrule
\hline
\multicolumn{3}{l}{\textit{\textbf{Euler-to-Particle Coupling}}} \\
\hline
Particle Motion & $\displaystyle m_p \frac{d\mathbf{u}_p}{dt} = \mathbf{F}_{drag} + \mathbf{F}_{gravity} + \mathbf{F}_{pressure} + \mathbf{F}_{virtual}$ & \\[8pt]
\hline
Drag Force & $\mathbf{F}_{drag} = \frac{1}{2} \rho_g C_D A_p |\mathbf{u}_g - \mathbf{u}_p|(\mathbf{u}_g - \mathbf{u}_p)$ & $C_D = \frac{24}{Re_p}(1 + \frac{1}{6}Re_p^{2/3})$, $Re_p \leq 1000$ \\[5pt]
& & $C_D = 0.424$, $Re_p > 1000$ \\[8pt]
\hline
Other Forces & $\mathbf{F}_{gravity} = m_p \mathbf{g}$, \quad $\mathbf{F}_{pressure} = -V_p \nabla p$ & $C_{vm} = 0.5$, $V_p = \frac{\pi d_p^3}{6}$ \\[5pt]
& $\mathbf{F}_{virtual} = C_{vm} \rho_g V_p \left(\frac{D\mathbf{u}_g}{Dt} - \frac{d\mathbf{u}_p}{dt}\right)$ & \\[8pt]
\hline
\multicolumn{3}{l}{\textit{\textbf{Particle-Wall Interactions}}} \\
\hline
Rebound Model & $u_{p,n}^{\text{after}} = -e \, u_{p,n}^{\text{before}}$ & $e = 0.97$, $\mu = 0.09$ \\[5pt]
& $u_{p,t}^{\text{after}} = u_{p,t}^{\text{before}} - \min(\mu |u_{p,n}^{\text{before}}|, |u_{p,t}^{\text{before}}|) \operatorname{sign}(u_{p,t}^{\text{before}})$ & \\[8pt]
\hline
\multicolumn{3}{l}{\textit{\textbf{Particle-Particle Collisions}}} \\
\hline
Contact Forces & $F_n = k_n \delta_n^{3/2}$, \quad $F_t = \min(k_t \delta_t, \mu F_n)$ & $\delta_{\text{overlap}} = 7 \times 10^{-5}$ m \\[5pt]
& $F_{\text{damp},n} = -\alpha \sqrt{k_n m_{eff}} v_{rel,n}$, \quad $F_{\text{damp},t} = -\alpha \sqrt{k_t m_{eff}} v_{rel,t}$ & $\alpha = 0.12$, $m_{eff} = \frac{m_i m_j}{m_i + m_j}$ \\[8pt]
\hline
\multicolumn{3}{l}{\textit{\textbf{Two-Way Coupling (Particle-to-Euler)}}} \\
\hline
Source Terms & $\displaystyle S_{\text{mom}} = \sum_{p \in \text{cell}} \frac{\mathbf{F}_{\text{drag},p} + \mathbf{F}_{\text{pressure},p}}{V_{\text{cell}}}$ & Units: N/m$^3$
\\[5pt]
\bottomrule
\hline

\end{tabular}
\end{adjustbox}
\end{table*}

Particle trajectories are advanced with an explicit forward Euler scheme. The velocity and position updates read as follows~\cite{elghobashi1994predicting}.
\begin{equation}
\mathbf{u}_p^{\,n+1} = \mathbf{u}_p^{\,n} + \Delta t\,\frac{\sum \mathbf{F}_p}{m_p}, \qquad
\mathbf{x}_p^{\,n+1} = \mathbf{x}_p^{\,n} + \Delta t\,\mathbf{u}_p^{\,n+1}
\end{equation}
The particle time step is restricted by the gas-phase step and a particle Courant constraint based on the local grid scale $\Delta x$,
\begin{equation}
\Delta t_p = \min\!\left(\Delta t_{\text{gas}},\; Co_{\max}\,\frac{\Delta x}{\|\mathbf{u}_p\|}\right), \qquad Co_{\max}=0.1
\end{equation}
which limits particle advection to a fraction of a cell per step and improves stability in regions of strong acceleration and contact.



Since it requires a very fine mesh to resolve acoustic wavelengths from the sources to the observer, a computational domain large enough to contain the observer at an acoustically relevant distance, and a simulation long enough for all wavefronts to reach that observer, directly sampling pressure with an in-domain probe is inefficient and prohibitively expensive. Consequently, acoustic analogies are preferred. The Ffowcs Williams–Hawkings (FW–H) acoustic analogy~\cite{FfowcsWilliamsHawkings1969} reformulates the compressible Navier–Stokes equations as an inhomogeneous wave equation with equivalent sources representing surface motion, unsteady surface loading, and turbulent quadrupole terms, enabling far-field predictions from near-field CFD data without resolving the entire acoustic field on the flow mesh. To capture acoustic noise in the simulation, libAcoustics~\cite{libacoustics, llya_evdokimov_2020_3878439} had been used. The full FH-W formulation is written in \autoref{sec:acoustics}.

Pressure data on the control surface are sampled at $\Delta t = 2\times10^{-5\text{s}}$ to provide sufficient temporal resolution.  The sampling interval yields a sampling frequency of $f_s = 1/\Delta t = 50\,\text{kHz}$ and a Nyquist frequency of $f_N = f_s/2 = 25\,\text{kHz}$.  The sampling rate satisfies the Nyquist requirement with extra margin because the Nyquist frequency is above the audible range of interest (0–20 kHz). This prevents aliasing and allows accurate depiction of the acoustic spectrum across the frequency band of interest.~\cite{shannon2006communication}.


The nozzle flow is advanced with a second-order implicit backward time scheme to maintain stability while resolving broadband unsteadiness. Convective fluxes for momentum and energy use limiter-based, shock-capturing schemes so that internal shocks are resolved without excessive numerical diffusion and second-order accuracy is preserved away from discontinuities~\cite{toro2013riemann}. Gradients are computed with Gauss linear reconstruction and are mildly limited for $\vec{U}$ to suppress overshoots near shocks and strong shear. For the turbulence transports, a bounded upwind scheme is used for $k$ and $\omega$ specifically to preserve numerical stability and positivity, since these scalars are highly sensitive in regions of strong production and dissipation.  For reconstruction of primitive variables, a van-Albada limiter is employed, this limiter preserves smooth peaks and valleys while suppressing false oscillations close to shocks. ~\cite{van1982comparative}. The complete settings are summarized in \autoref{tab:numerical_schemes}.

Numerically, the solver uses the PIMPLE algorithm~\cite{issa1986solution} with two outer correctors and two pressure corrections. Non-orthogonal corrections are limited to one, and pressure limiting factors of 0.5 and 2.0 bound the pressure updates to prevent runaway corrections in the compressible system. The PIMPLE algorithm, combining SIMPLE (Semi-Implicit Method for Pressure-Linked Equations) and PISO (Pressure-Implicit with Splitting of Operators) is known for its robustness and stability in transient simulations.  All linear systems are solved with preconditioned biconjugate gradient stabilized (PBiCGStab)~\cite{van1992bi} preconditioned by Diagonal Incomplete LU (DILU). Convergence tolerances are $10^{-12}$ for pressure and $10^{-15}$ for the remaining transport variables, with a minimum iteration constraint to avoid premature convergence. Field under-relaxation is applied, lower for pressure due to its  sensitivity and slightly higher for other equations to maintain stable coupled iterations~\cite{ferziger2019computational}. A time step of $\Delta t = 10^{-8}\,\text{s}$ was selected after preliminary tests to keep Courant numbers near 0.1. A complete summary of the numerical settings is provided in \autoref{tab:numerical_parameters}.

The simulations began with mesh generation using \texttt{ClassyBlock}, configured to the target geometry and mesh parameters. To improve matrix bandwidth, the mesh was renumbered using Cuthill-McKee algorithm~\cite{cuthill1969reducing} and then decomposed into 382 partitions. The modified solver \texttt{sonicDPMFoam}, a \texttt{sonicFoam} variant with discrete particle tracking (DPM) was used to advance the solution. To enhance stability and avoid large initial pressure/velocity gradients, inlet pressure and temperature were ramped from $1\,bar$ and $400\,K$ to the operating conditions over $t_{\text{ramp}}$. Particle release and acoustic sampling via acoustic analogies began at $t=2\times10^{-3}\,\text{s}$, and the simulation was stopped at $t=4\times10^{-3}\,\text{s}$, yielding $2\times10^{-3}\,\text{s}$ of acoustic data.

Two cases, one with a coarse mesh for URANS and one with a finer mesh for DDES, start at the same time to speed up the simulation as the meshing procedure only requires a few CPUs. SonicFOAM solves the simulation for 0.008 seconds with $\Delta t=10^{-7}$~s as soon as the URANS meshing is completed. Despite the Courant number being roughly 0.9, the simulation converges and runs smoothly with this timestep since its turbulence model is URANS and the mesh is coarser. Once the URANS simulation is complete, the URANS cell values are interpolated to the finer mesh and the results are mapped to the DDES fine mesh. The mesh is decomposed, and the simulation runs for 2000 iterations using the $k$-$\omega$ SST URANS model with a timestep of $\Delta t=10^{-9}$~s to make sure the mapped values are established in the new mesh. Next, to guarantee that the Courant number is roughly 0.1, sonicDPMFoam with the DDES turbulence model is initiated with $\Delta t=10^{-8}$~s. Particles begin to be injected after vortices develop, and the acoustic observers begin recording the aeroacoustic noise. The schematic timetable and procedure of the commands executed in each simulation case are depicted in \autoref{fig:gantt_timeline}.

\begin{figure*}[htbp]
    \centering
    \resizebox{\textwidth}{!}{%
\begin{ganttchart}[
    vgrid,
    hgrid,
    title height=1,
    bar height=0.8,
    group height=.6,
    bar label font=\small\color{black},
    group label font=\bfseries\color{black},
    x unit=1.3cm,  
    y unit chart=0.7cm  
  ]{0}{15} 
  
  \gantttitle{Meshing Parallel}{5}
  \gantttitle{RANS Run}{4}
    \gantttitle{DDES Run}{7}\\ 
  
  
  \ganttbar[bar/.style={fill=orange!40}]{classyBlock}{0}{0} \\
  \ganttbar[bar/.style={fill=orange!40}, ]{BlockMesh}{1}{2} \\
  \ganttbar[bar/.style={fill=orange!40}]{Mesh Renumbering}{3}{3} \\
  \ganttbar[bar/.style={fill=orange!40}]{Decompose}{4}{4} \\

  \ganttnewline[thick, dotted, blue]
  
  \ganttbar[bar/.style={fill=blue!40}]{sonicFoam}{5}{7} \\
  \ganttbar[bar/.style={fill=blue!40}]{Reconstruct}{8}{8} \\

  \ganttnewline[thick, dotted, green]
  \ganttbar[bar/.style={fill=green!40}]{classyBlock}{0}{0} \\
  \ganttbar[bar/.style={fill=green!40}]{BlockMesh}{1}{3} \\
  \ganttbar[bar/.style={fill=green!40}]{Mesh Renumbering}{4}{4} \\

  \ganttnewline[thick, red, dotted]
  
  \ganttbar[bar/.style={fill=red!40}]{mapFields}{10}{10} \\
  \ganttbar[bar/.style={fill=red!40}]{Decompose}{11}{11} \\
  \ganttbar[bar/.style={fill=red!40}]{sonicFoam}{12}{13} \\
  \ganttbar[bar/.style={fill=red!40}]{sonicDPMFoam}{14}{15}

  \ganttlink{elem5}{elem9}
  \ganttlink{elem8}{elem9}




  \ganttvrule{Meshing Finished}{4}
  \ganttvrule{Rans Finished}{8}

\node[anchor=center, text=black, font=\scriptsize] at ($(elem1.center)$) {Coarse Mesh};
\node[anchor=center, text=black, font=\scriptsize] at ($(elem7.center)$) {Fine Mesh};

\node[anchor=center, text=black, font=\scriptsize] at ($(elem3.center)$) {384 cpus};
\node[anchor=center, text=black, font=\scriptsize] at ($(elem10.center)$) {384 cpus};

\node[anchor=center, text=black, font=\scriptsize] at ($(elem9.center)$) {Interpolate};

\node[anchor=center, text=black, font=\scriptsize] at ($(elem4.center)$) {RANS, $T$ \& $P$ inlet ramp,  $\Delta t=10^{-7}$};
\node[anchor=center, text=black, font=\scriptsize] at ($(elem11.center)$) {RANS,  $\Delta t=10^{-9}$};
\node[anchor=center, text=black, font=\scriptsize] at ($(elem12.center)$) {DDES,  $\Delta t=10^{-8}$};

\end{ganttchart}
}
\caption{CFD Simulation Workflow - RANS and DDES Phases}
\label{fig:gantt_timeline}
\end{figure*}
\section{Verification and Validation}\label{sec:valid}

\subsection{Verification}

\begin{figure}[ht]
    \centering
    \includegraphics[width=\linewidth]{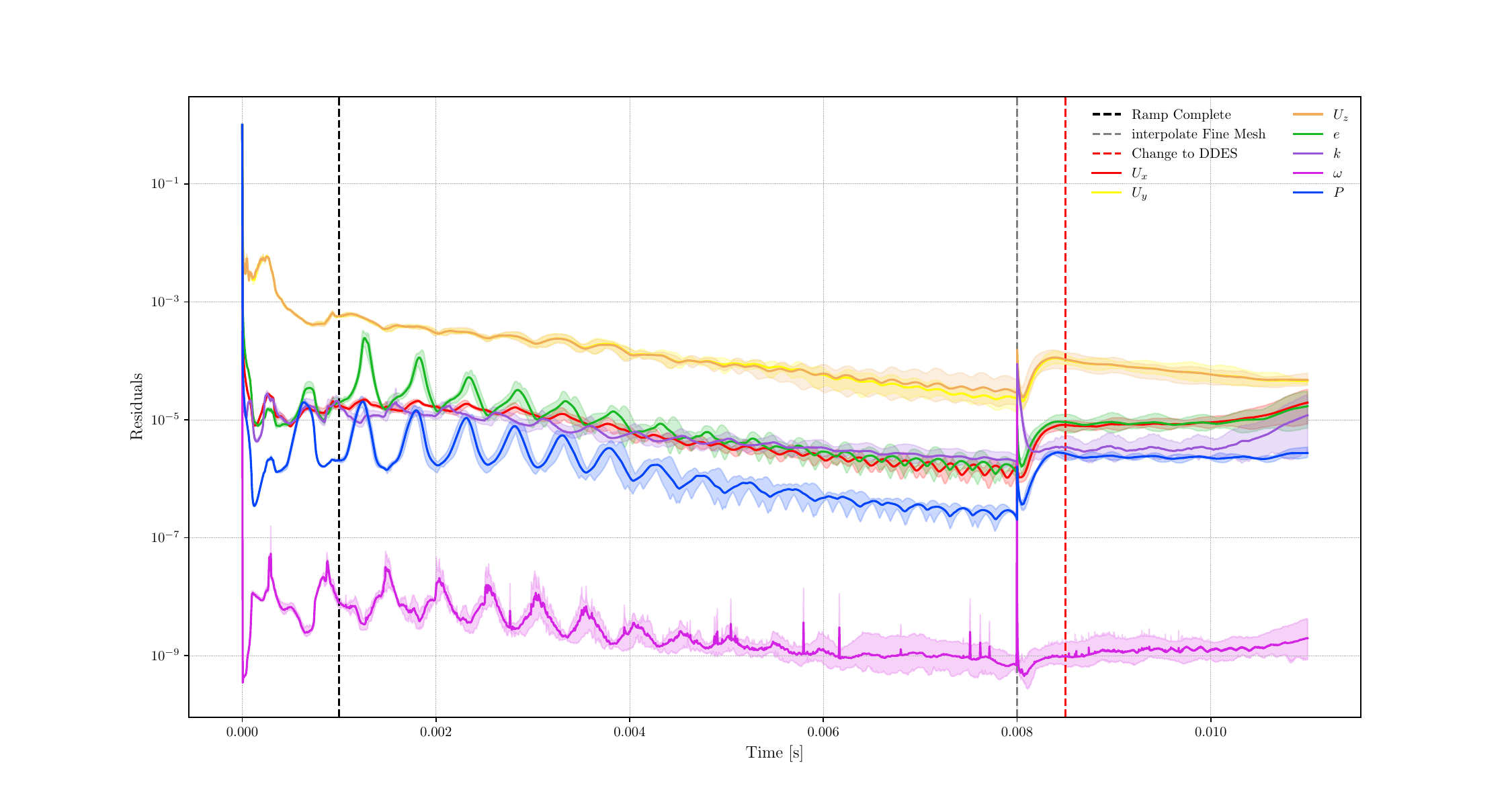}
    \caption{Temporal evolution of the equation residuals during simulation.}
    \label{fig:residuals}
\end{figure}

To verify the model, the residuals depicted in \autoref{fig:residuals} were monitored and ensured that the residuals of all flow variables were continuously below $10^{-2}$.During the first phase, when the temperature and inlet pressure are increasing to the intended operating conditions, the residuals show significant changes. Following this initial phase, the residuals gradually decline. The URANS solution was mapped to a finer mesh at 0.0008 s, which resulted in a brief increase in residuals while the solution stabilized on the new mesh. Nevertheless, after a few hundred iterations, the residuals rapidly decrease. The turbulence model was changed from URANS to DDES after about 2000 iterations. The residuals show consistent oscillations well below the convergence threshold after this transition point.

Additionally, as mass should be conserved in the nozzle, monitoring planes at the inlet, throat, and exit were used to capture the velocities and densities and calculate the mass flux. The particles were released at a point where the mass flux passing through these planes had reached relatively the same value, indicating that the nozzle had passed the start-up phase and achieved pseudo steady-state conditions. Recording of pressure fluctuations started at the same time the particles were injected. Furthermore, as a significant proportion of acoustic noise is generated near the nozzle walls, having adequate mesh resolution near the walls was important. The chosen first cell size was $0.001$ mm for all cases, and throughout all timesteps, the maximum $y^+$ observed was 7.

\begin{figure}
    \centering
    \includegraphics[width=1\linewidth]{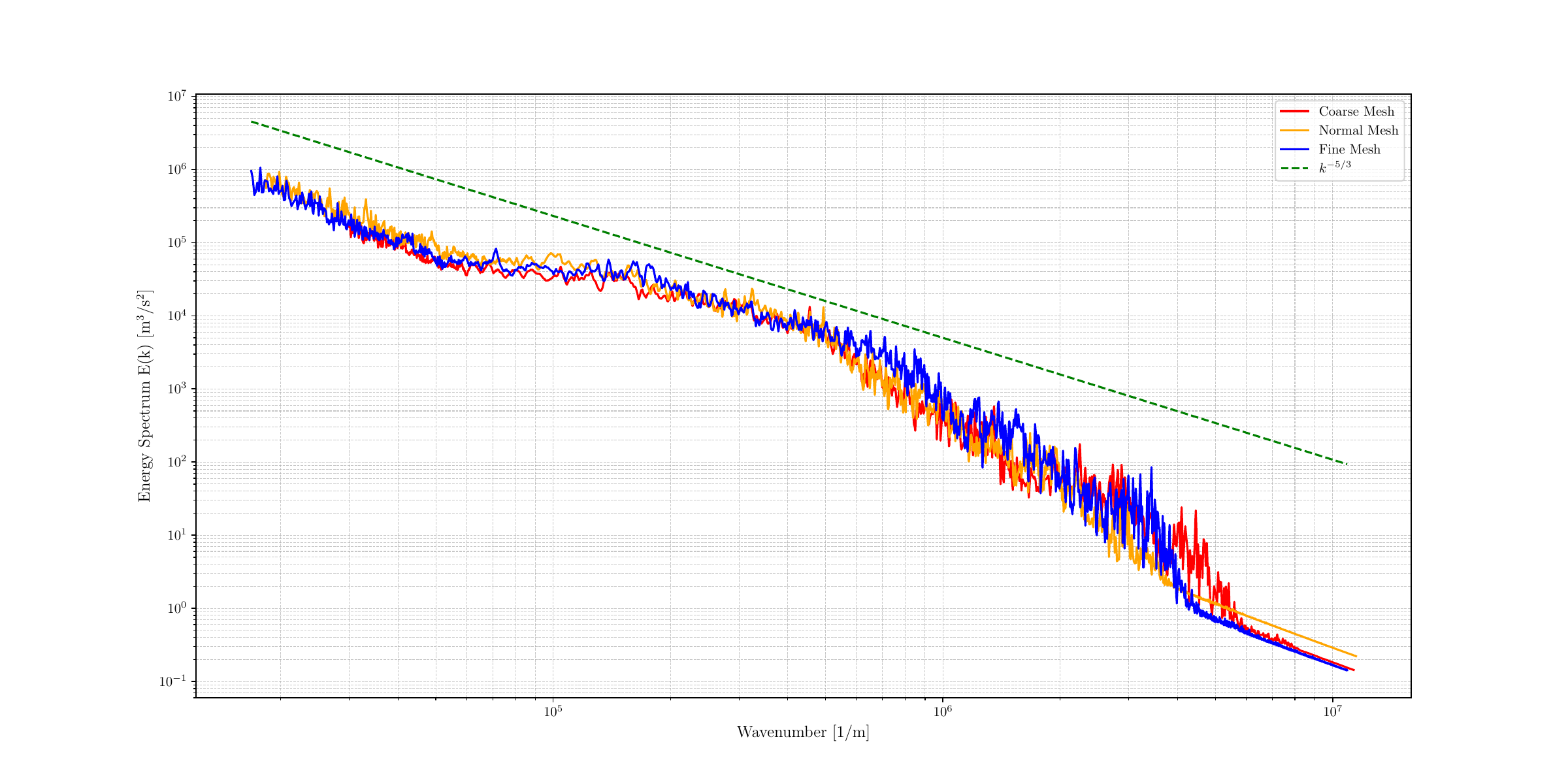}
    
    \caption{Energy spectrum showing velocity fluctuations downstream of the nozzle exit converted to wavenumber vs. energy spectrum.}
    \label{fig:Energy_cascade}
\end{figure}

\autoref{fig:Energy_cascade} demonstrates the energy spectrum computed from velocity fluctuations downstream of the nozzle exit for three different mesh resolutions. Data were collected from multiple velocity probes placed at incremental distances from the nozzle exit, as shown by the blue dots in \autoref{fig:CFD_schematic}. The energy spectrum calculations are described in \autoref{sec:EnergyCascade}

\begin{figure}[htbp]
    \centering
    \begin{subfigure}[b]{0.48\linewidth}
        \centering
        \includegraphics[width=\linewidth, trim=2cm 8cm 2cm 8cm, clip]{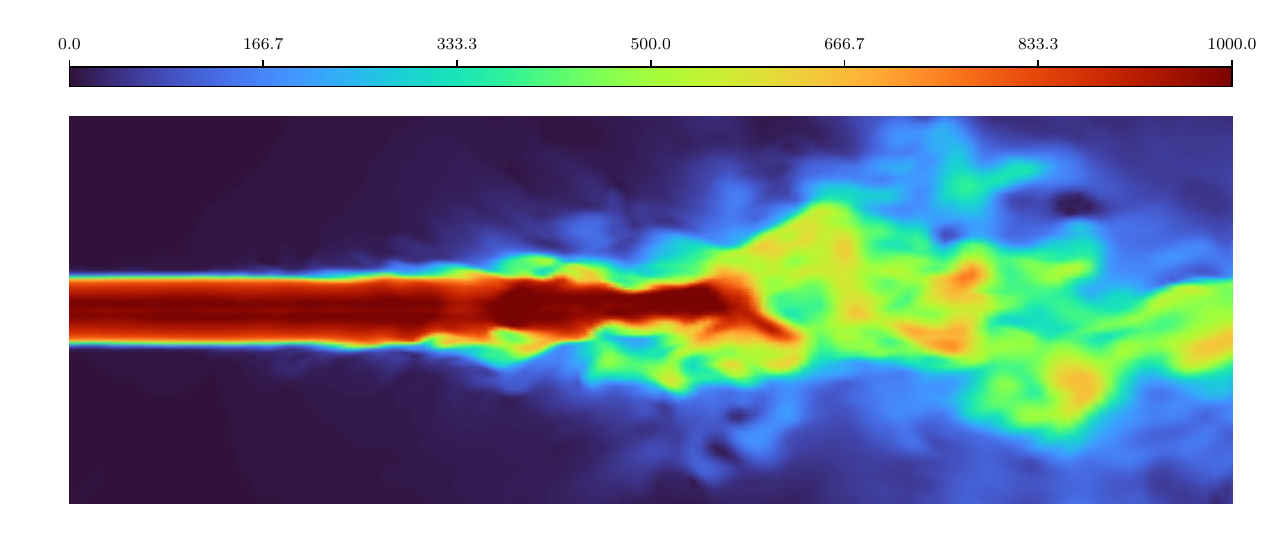}
        \caption{Coarse mesh - Velocity Magnitude}
        \label{fig:mesh9_U}
    \end{subfigure}
    \hfill
    \begin{subfigure}[b]{0.48\linewidth}
        \centering
        \includegraphics[width=\linewidth, trim=2cm 8cm 2cm 8cm,, clip]{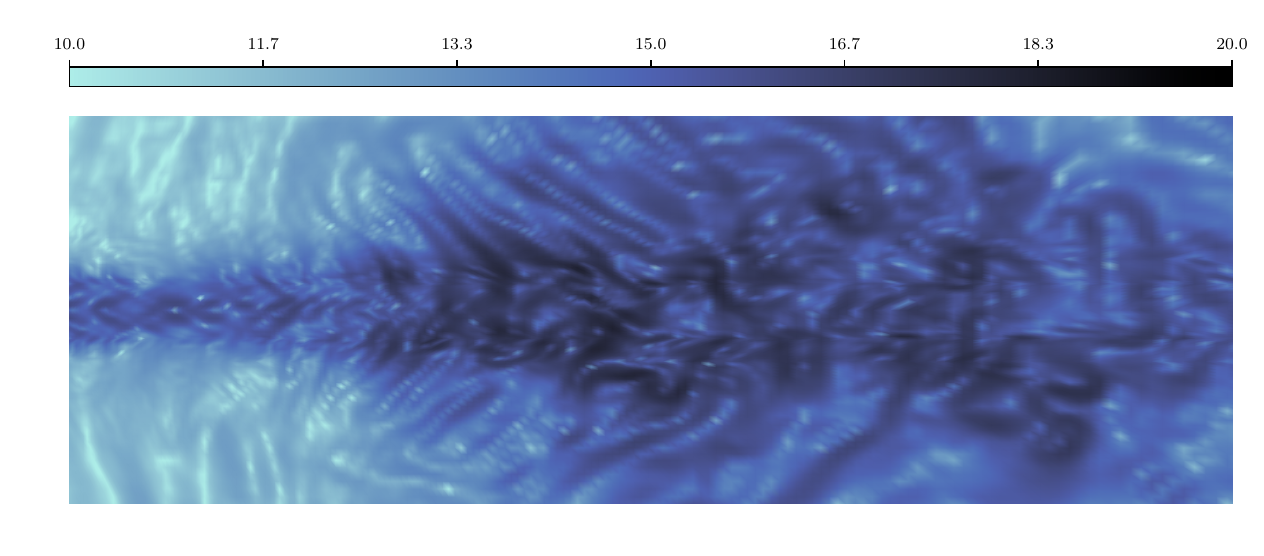}
        \caption{Coarse mesh - Pressure Gradient}
        \label{fig:mesh9_gradP}
    \end{subfigure}
    
    \vspace{0.3cm}
    
    \begin{subfigure}[b]{0.48\linewidth}
        \centering
        \includegraphics[width=\linewidth, trim=2cm 8cm 2cm 8cm, clip]{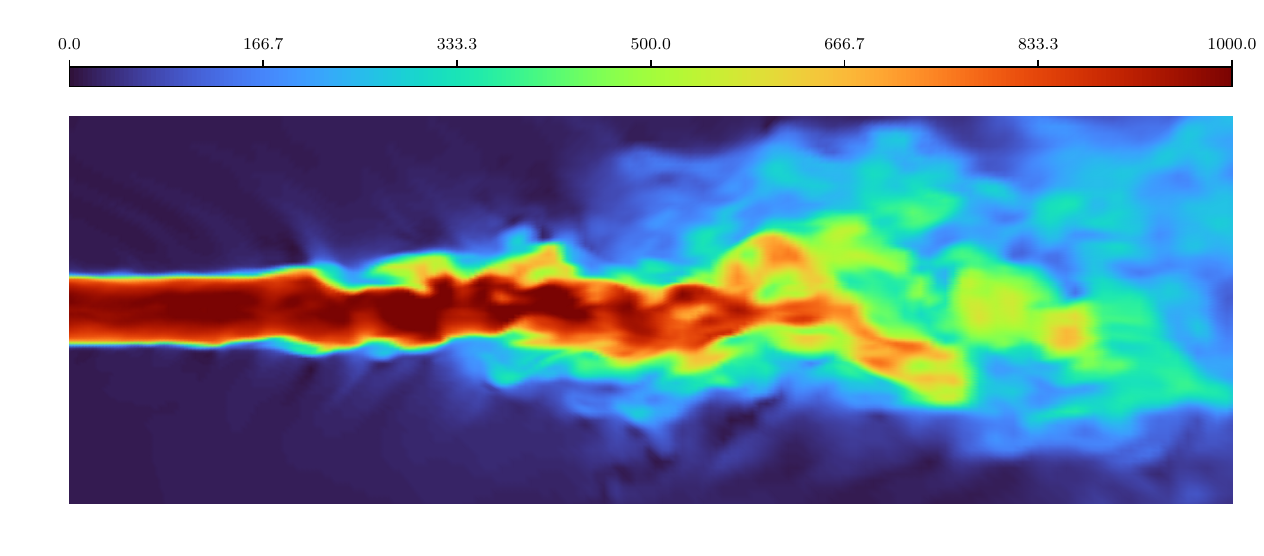}
        \caption{Normal mesh - Velocity Magnitude}
        \label{fig:mesh10_U}
    \end{subfigure}
    \hfill
    \begin{subfigure}[b]{0.48\linewidth}
        \centering
        \includegraphics[width=\linewidth, trim=2cm 8cm 2cm 8cm, clip]{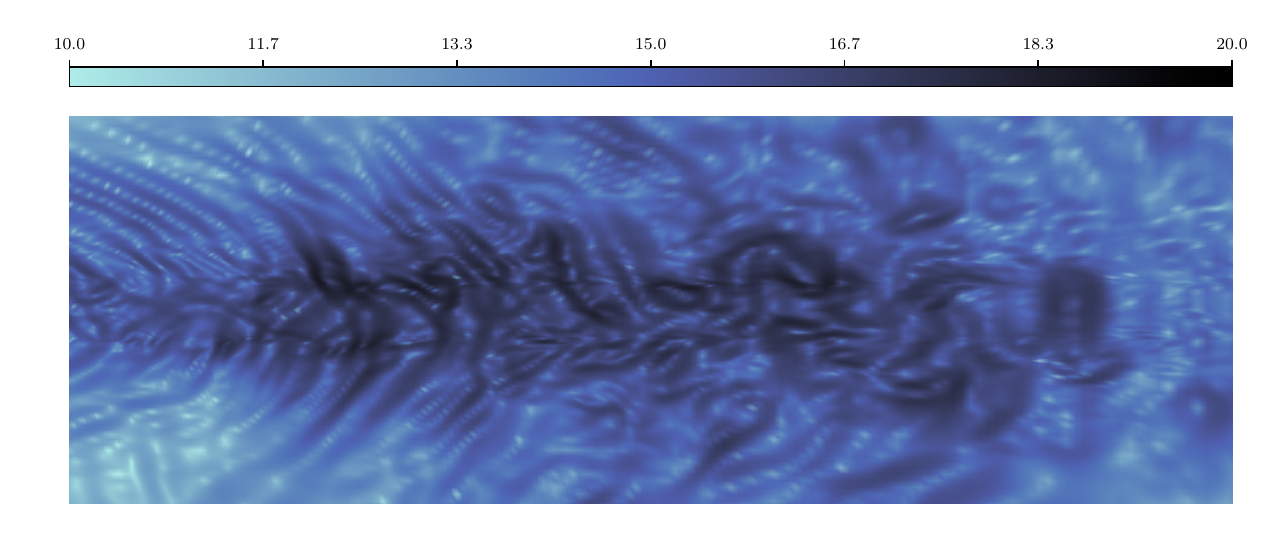}
        \caption{Normal mesh - Pressure Gradient}
        \label{fig:mesh10_gradP}
    \end{subfigure}
    
    \vspace{0.3cm}
    
    \begin{subfigure}[b]{0.48\linewidth}
        \centering
        \includegraphics[width=\linewidth, trim=2cm 8cm 2cm 8cm, clip]{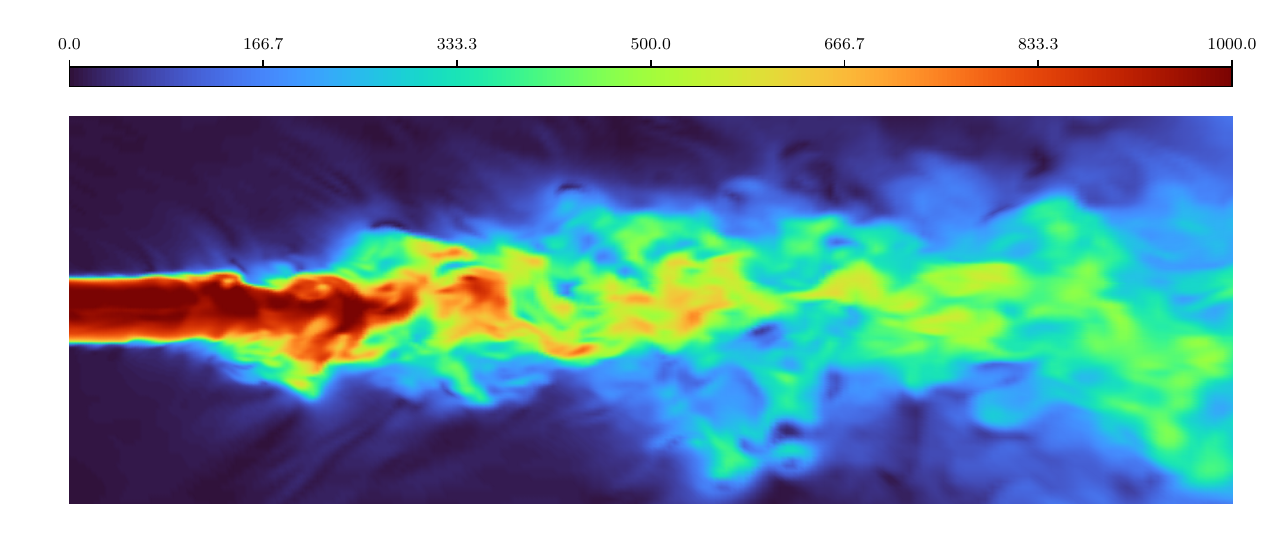}
        \caption{Fine mesh - Velocity Magnitude}
        \label{fig:mesh12_U}
    \end{subfigure}
    \hfill
    \begin{subfigure}[b]{0.48\linewidth}
        \centering
        \includegraphics[width=\linewidth, trim=2cm 8cm 2cm 8cm, clip]{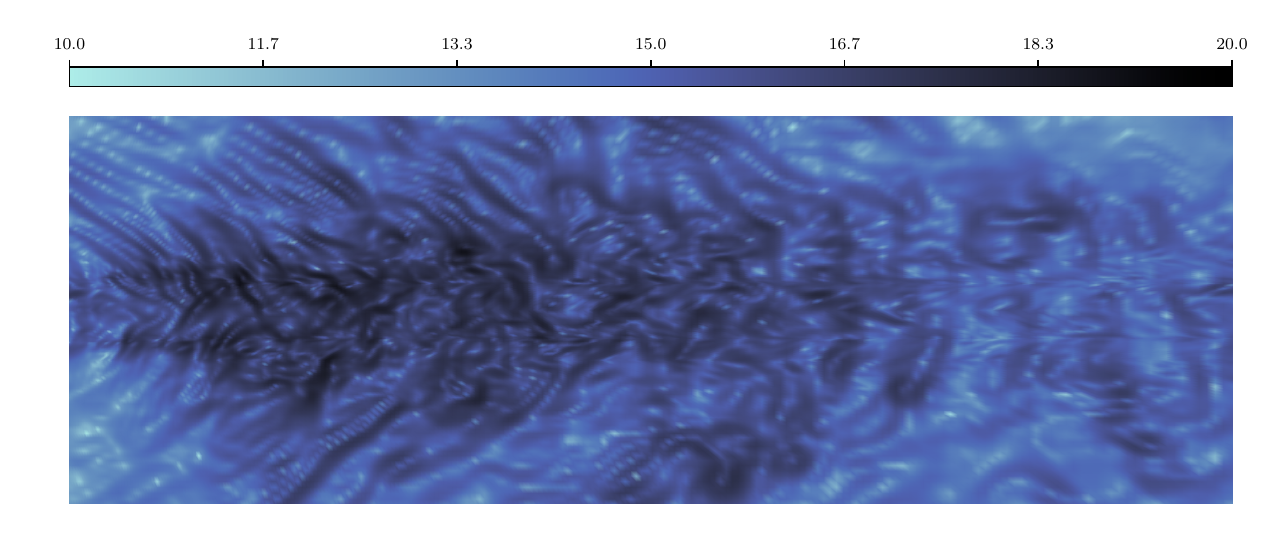}
        \caption{Fine mesh - Pressure Gradient}
        \label{fig:mesh12_gradP}
    \end{subfigure}
    \caption{Comparison of velocity magnitude and pressure gradient for different mesh resolutions at time 0.0015 s}
    \label{fig:mesh_comparison}
\end{figure}

Three different mesh resolutions were compared; details of the mesh configurations are provided in \autoref{tab:mesh_analysis}. The energy spectra for all three meshes follow the $-5/3$ reference line characteristic of the inertial subrange in turbulent flow, indicating that all mesh resolutions adequately capture the turbulent cascade. While the overall trend remains consistent across mesh levels, finer meshes resolve smaller-scale turbulent structures more accurately. This is evident in \autoref{fig:mesh_comparison}, where the velocity magnitude and pressure gradient fields show progressively finer eddy structures as mesh resolution increases.

\begin{table}
\centering
\caption{Mesh resolution comparison}
\label{tab:mesh_analysis}
\begin{tabular}{lrrrr}
\hline
\textbf{Mesh Level} & \textbf{Number of Cells} & \textbf{Min Volume [m$^3$]} & \textbf{Max Volume [m$^3$]} \\
\hline
Coarse & 9,229,340 & $5.38 \times 10^{-16}$ & $9.73 \times 10^{-9}$ \\
Normal & 12,875,088 & $4.33 \times 10^{-16}$ & $7.00 \times 10^{-9}$ \\
Fine & 22,011,652 & $2.97 \times 10^{-16}$ & $4.09 \times 10^{-9}$  \\
\hline
\end{tabular}
\end{table}

\subsection{Validation of Particle Distribution}

\begin{figure}
    \centering
    \includegraphics[width=\linewidth]{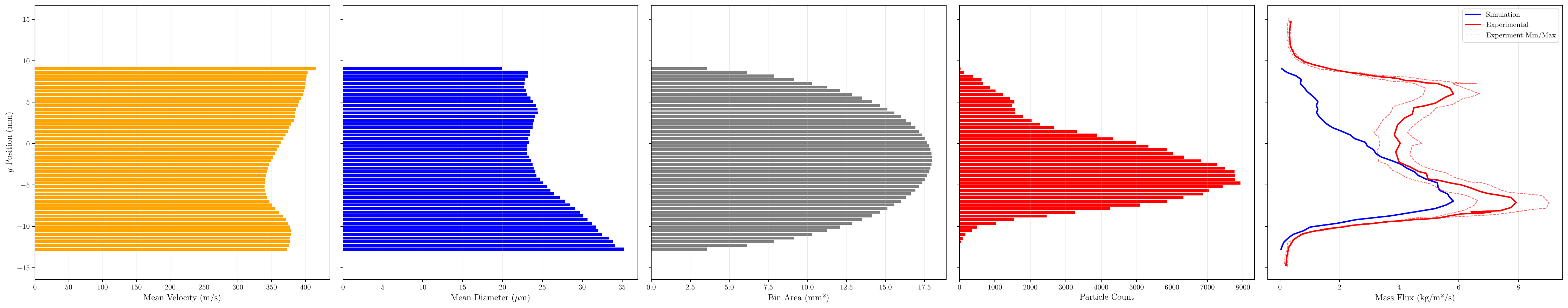}
    \caption{Comparison of particle mass flux distribution at the nozzle exit between DDES-Lagrangian simulation and experimental shadowgraph measurements~\cite{Allofs2022, Allofs2023}.}
    \label{fig:particle_validation}
\end{figure}

To validate the Lagrangian particle tracking model, a simulation with the configuration shown in the second column of \autoref{tab:validation_conditions} was conducted, and the computed particle trajectories were analyzed at the nozzle exit plane and compared with experimental shadowgraph measurements from Allofs et al.~\cite{Allofs2022, Allofs2023}. The details of the validation calculations are described in \autoref{sec:validationCalc}.

As illustrated in \autoref{fig:particle_validation}, the mean axial velocity distribution varies from approximately 250 to 400 m/s, with characteristic variations along the vertical direction. Larger particles tend to concentrate in the lower region of the nozzle exit, and the mean particle diameter is generally consistent at 10–30 $\mu$m across most bins. The particle count distribution exhibits asymmetry, with higher concentrations in the lower half of the nozzle exit. The estimated mass flux and experimental data are compared in the rightmost panel of \autoref{fig:particle_validation}, which demonstrates qualitative agreement in the general distribution pattern. The results show that the particle forces selected for the simulation are sufficiently realistic and the particle distribution model is sufficiently validated, despite some quantitative deviations being seen, especially in the peak positions and magnitudes.
\section{Results}\label{sec:results}

\begin{figure*}[htbp]
    \centering
    \begin{subfigure}[b]{0.8\textwidth}
        \centering
        \includegraphics[width=\textwidth, trim={0cm 4cm 0cm 5cm}, clip]{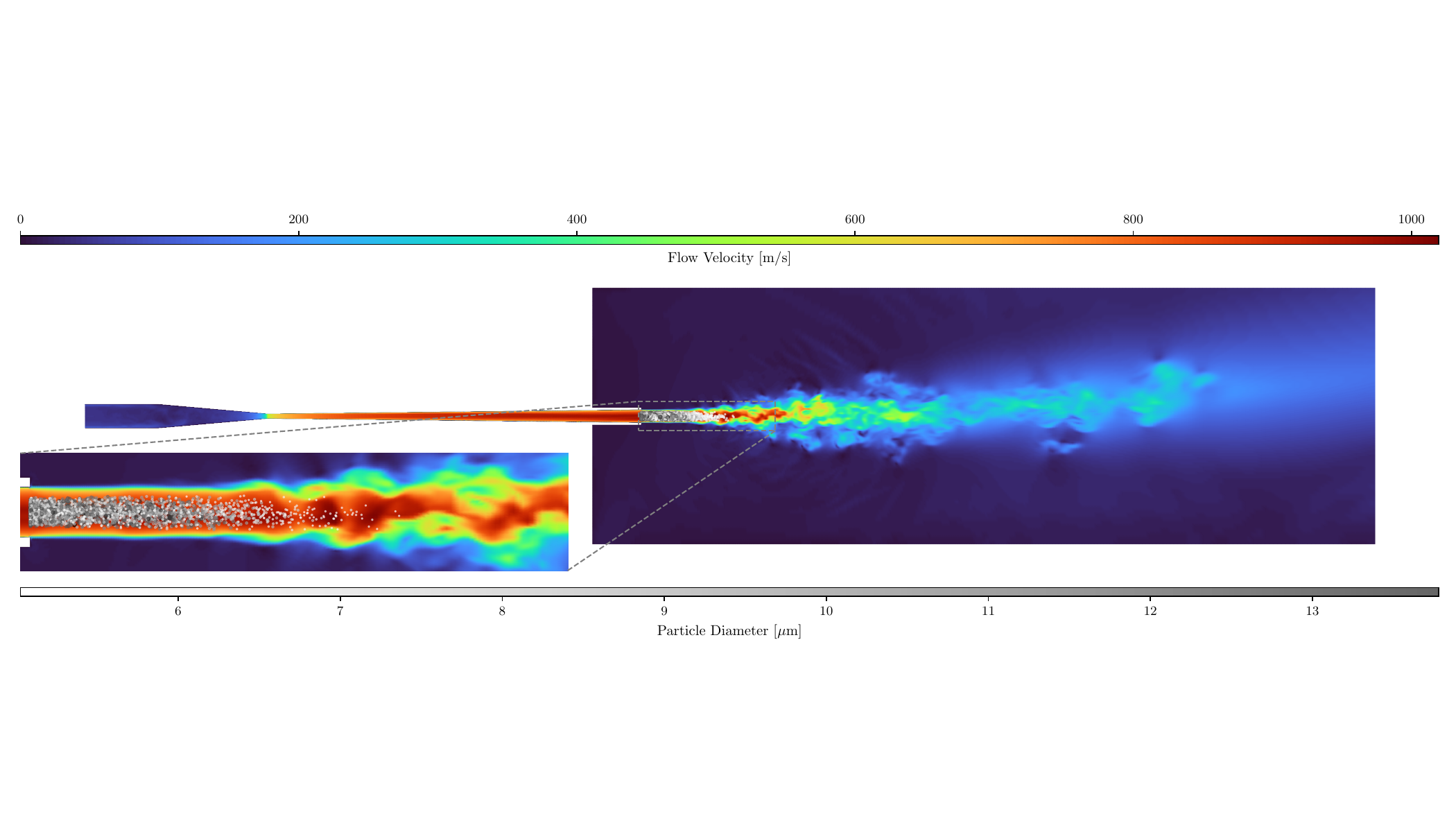}
        \caption{t = 600 $\mu s$}
        \label{fig:ddes_u_0p0006}
    \end{subfigure}
    
    \vspace{0.5cm}
    
    \begin{subfigure}[b]{0.8\textwidth}
        \centering
        \includegraphics[width=\textwidth, trim={0cm 4cm 0cm 5cm},, clip]{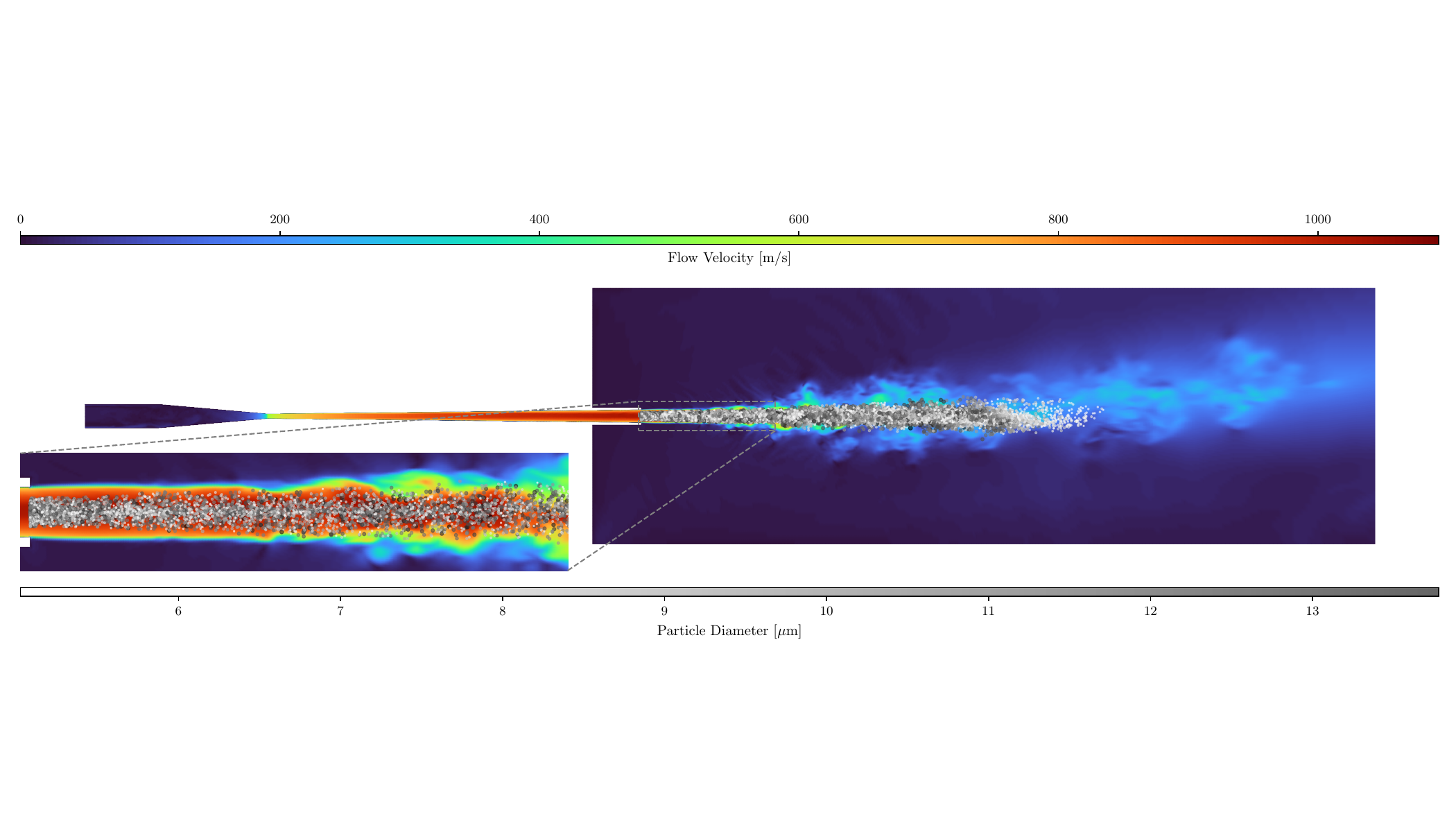}
        \caption{t = 900 $\mu s$}
        \label{fig:ddes_u_0p0015}
    \end{subfigure}
    
    \vspace{0.5cm}
    
    \begin{subfigure}[b]{0.8\textwidth}
        \centering
        \includegraphics[width=\textwidth,trim={0cm 4cm 0cm 5cm},, clip]{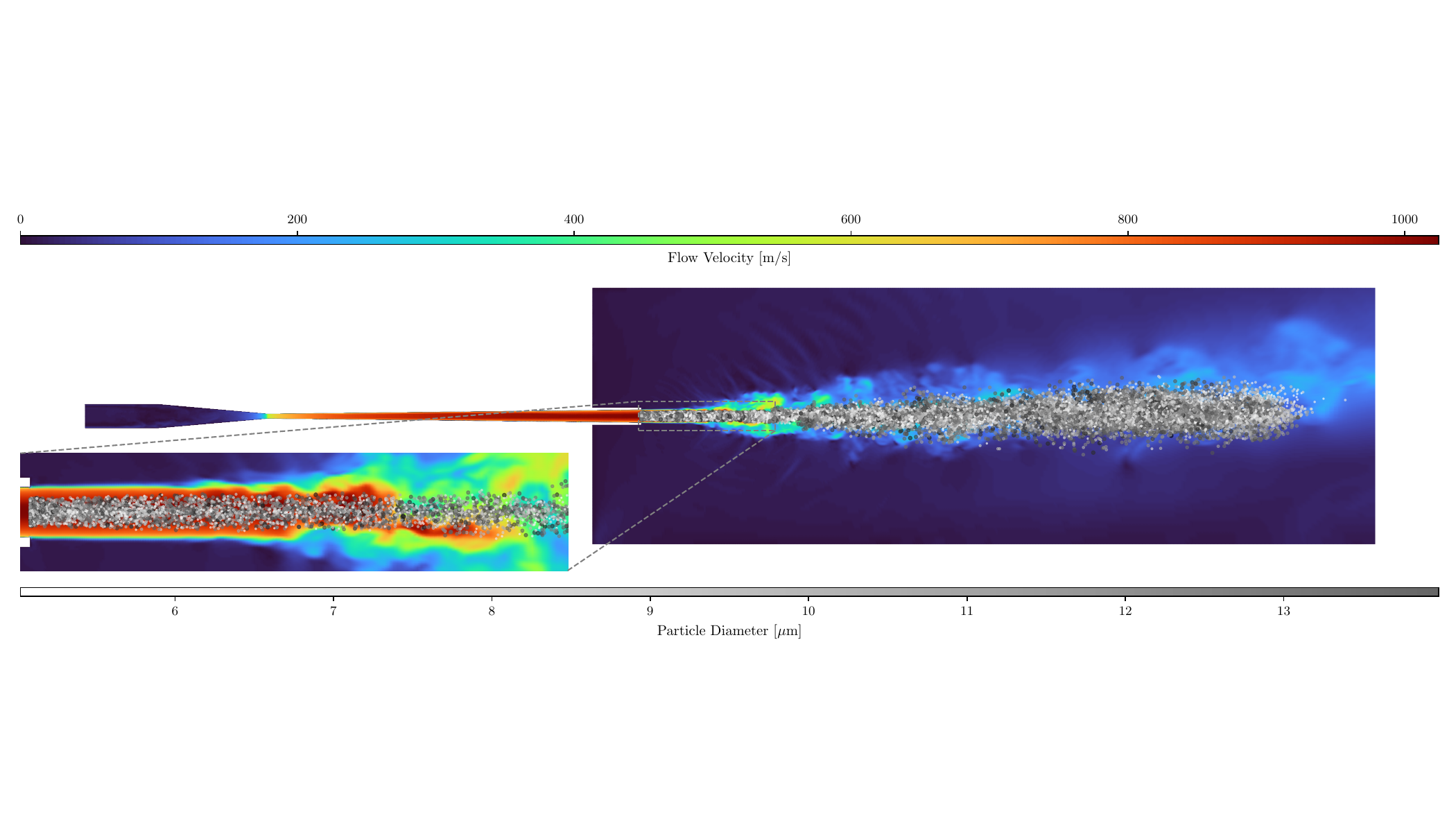}
        \caption{t = 1200 $\mu s$}
        \label{fig:ddes_u_0p0025}
    \end{subfigure}
    
    \vspace{0.5cm}
    
    \begin{subfigure}[b]{0.8\textwidth}
        \centering
        \includegraphics[width=\textwidth,trim={0cm 4cm 0cm 5cm},, clip]{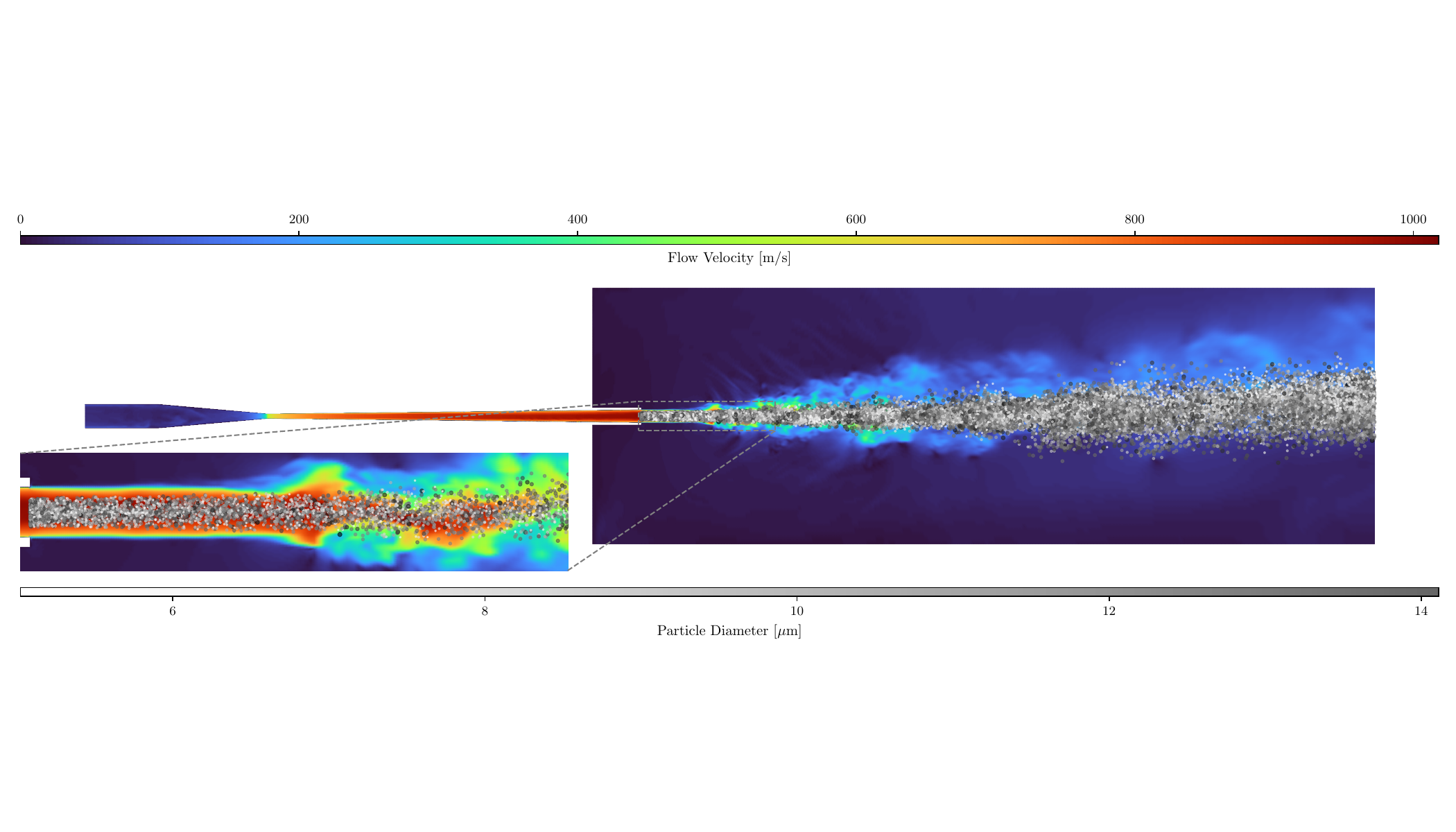}
        \caption{t = 1500 $\mu s$}
        \label{fig:rans_u_0p0035}
    \end{subfigure}
    
    \caption{Velocity Magnitude and particles DDES simulation at different time steps}
    \label{fig:ddes_wave_evolution}
\end{figure*}

The temporal progression of the DDES simulation, depicting particle positions at discrete time intervals, is illustrated in \autoref{fig:ddes_wave_evolution}. This section provides an analysis of the simulation results. Initially, Eulerian flow field data is investigated, including the spatial distribution of Mach number, velocity, and pressure along the jet centerline, with an emphasis on the impacts of chamber temperature and pressure. The principal acoustic source mechanisms within the jet are then identified and described, followed by an analysis of the impacts of operating temperature and pressure on the acoustic field. The investigation then shifts to particle distribution characteristics. Particle trajectories and velocities along the axial centerline are shown first, followed by statistical analysis to determine the relationship between particle percentiles, radial displacement from the centerline, and associated velocity distributions. Radial cross-sectional cuts are then used to characterize particle dispersion and velocity on planes perpendicular to the jet axis. Finally, a two-dimensional histogram is used to depict the relationship between particle diameter and velocity at various distances from the nozzle exit.

\begin{figure*}[t]
    \centering
    \begin{subfigure}[b]{0.49\textwidth}
        \centering
        \includegraphics[width=\textwidth, trim={1cm 6cm 1cm 2cm}, clip]{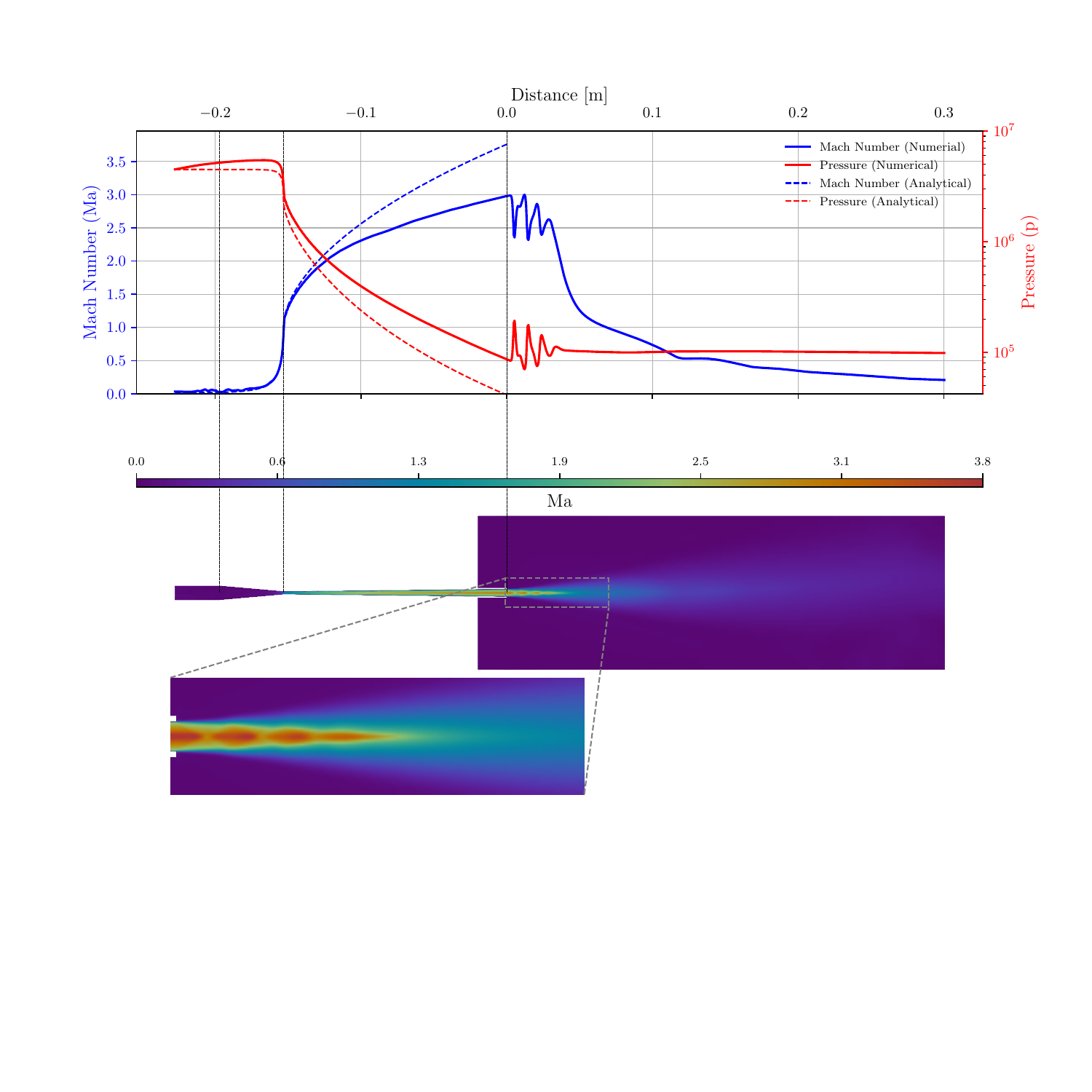}
        \caption{URANS simulation results}
        \label{fig:urans_contour}
    \end{subfigure}
    \hfill
    \begin{subfigure}[b]{0.49\textwidth}
        \centering
        \includegraphics[width=\textwidth, trim={1cm 6cm 1cm 2cm}, clip]{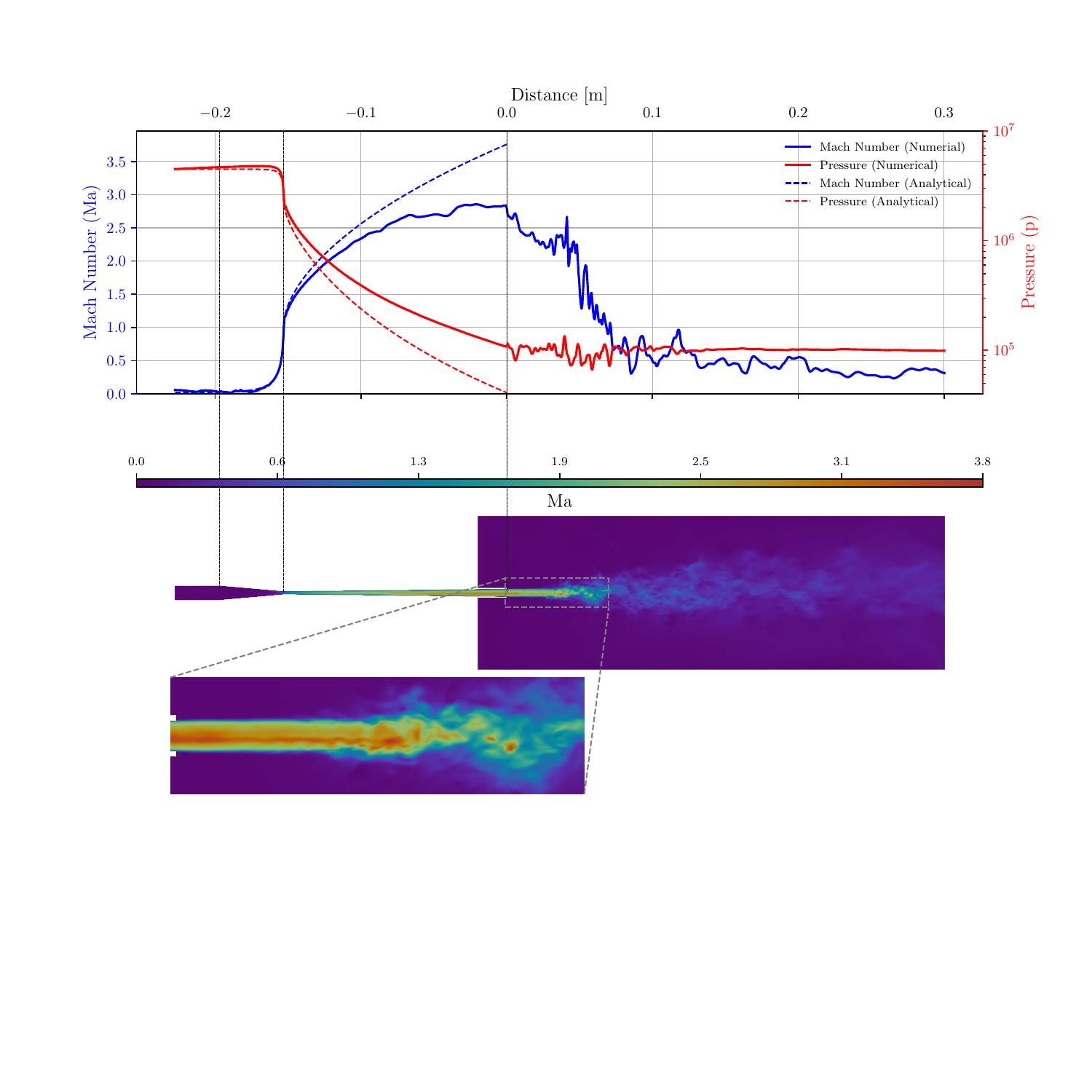}
        \caption{DDES simulation results}
        \label{fig:ddes_contour}
    \end{subfigure}
    \caption{Comparison of Mach number and static pressure distributions along the nozzle centerline between URANS and DDES simulations, with corresponding flow field visualizations.}
    \label{fig:contour_comparison}
\end{figure*}

\subsection{Effect of turbulence model}

\autoref{fig:contour_comparison} compares flow features simulated using URANS and DDES turbulence modelling techniques, with numerical predictions evaluated against analytical isentropic relations in \autoref{fig:urans_contour} and \autoref{fig:ddes_contour}, respectively.  Both URANS and DDES simulations accurately simulated the flow acceleration through the converging section and throat of the nozzle, as evidenced by excellent agreement between numerical predictions and theoretical isentropic curves in this region. The Mach number gradually increases from near-zero values in the upstream chamber to unity at the throat, which is expected for compressible flow in the nozzle. The pressure correspondingly decreases from chamber conditions to the critical pressure ratio at the throat.

The URANS simulation shows a minor pressure rise in the converging region, which is observable as a slight upward departure from the monotone isentropic pressure decrease. The DDES simulations do not show this slight pressure increase.

Significant deviations from isentropic behavior emerge in the diverging section downstream of the throat. The URANS simulation models a Mach number of roughly 3.0 at the nozzle exit, while the analytical isentropic solution indicates a higher exit Mach number of approximately 3.7. The DDES simulation even has more significant different behaviour. The Mach number trend in the diverging region is far more irregular, with fluctuations rather than the smooth monotonic rise indicated by URANS. The DDES exit Mach number reaches only approximately 2.7, representing a more substantial deviation from the isentropic prediction compared to URANS. The lower DDES exit Mach number could result from several factors such as more realistic turbulent dissipation in the boundary layer, resolution of unsteady flow separation or shock-boundary layer interaction in the diverging section, three-dimensional effects captured by the scale-resolving approach, or the instantaneous nature of the DDES where the flow field varies significantly in time. Such behavior is consistent with experimental observations of nozzle flows at high Reynolds numbers, where even nominally steady operating conditions exhibit measurable flow unsteadiness. The URANS approach, while capturing some unsteady features, produces a smoother profile compared to DDES.

Simultaneously, the static pressure predictions show corresponding deviations from isentropic theory. Both turbulence models simulated pressures elevated above theoretical values, with DDES showing slightly higher pressures consistent with its lower Mach numbers. 

Upon exiting the nozzle, both simulations exhibit oscillatory behavior in Mach number and pressure profiles along the centerline, characteristic of the complex shock cell structure in underexpanded supersonic jets. These oscillations represent the periodic compression and expansion of the jet as it adjusts from the nozzle exit pressure to ambient atmospheric conditions through a series of oblique shocks and expansion fans. The pressure profiles eventually stabilize at atmospheric conditions of approximately $101325 \;Pa$ in the far field, while the Mach number gradually decays toward zero as the jet loses momentum through turbulent mixing with ambient air and viscous dissipation. The DDES simulation displays more frequent and sharper oscillations in both Mach number and pressure throughout the downstream region. These rapid fluctuations reflect the instantaneous turbulent structures, unsteady shock motions, and shock-vortex interactions.

In contrast, the URANS simulation exhibits larger amplitude fluctuations immediately downstream of the nozzle exit, corresponding to the initial diamond shock cells where the flow adjustment is most pronounced. The first two to three shock cells show well-defined peaks and troughs in both Mach number and pressure, with the first peak reaching the highest Mach number of approximately 3.0 in the entire flow field. However, these oscillations are subsequently damped more rapidly, resulting in a smoother downstream profile. By approximately 10 to 15 cm downstream, the URANS predictions show reduced oscillatory behavior, approaching a monotonic decay toward ambient conditions. This damping occurs because the enhanced turbulent viscosity used by URANS models promotes faster dissipation of organized flow structures compared to DDES.

In the URANS simulation, distinct diamond-shaped shock patterns form immediately at the nozzle exit, visible as alternating regions of compression and expansion in the Mach number field. The high-velocity core region contracts rapidly with increasing downstream distance. By approximately 5 to 8 cm downstream, the jet cross-section has evolved into a narrow triangular profile, indicating aggressive radial spreading and momentum diffusion. The DDES simulation presents a markedly different flow topology. The jet structure remains coherent and well-defined for several centimeters downstream of the exit, with the high-velocity core maintaining a more cylindrical geometry rather than the triangular URANS profile. Most notably, turbulent mixing appears delayed relative to URANS predictions. Beyond approximately 8 to 10 cm downstream, the DDES flow field exhibits clear signs of large-scale turbulent structures. Small-scale patterns and abnormalities arise in the Mach number field, especially at the jet's edge, indicating instantaneous vortical structures shed from the shear layer.

\subsection{Comparison of velocity and pressure profiles}

\begin{figure}[ht]
    \centering
    \begin{subfigure}[b]{0.48\textwidth}
        \centering
        \includegraphics[width=\textwidth]{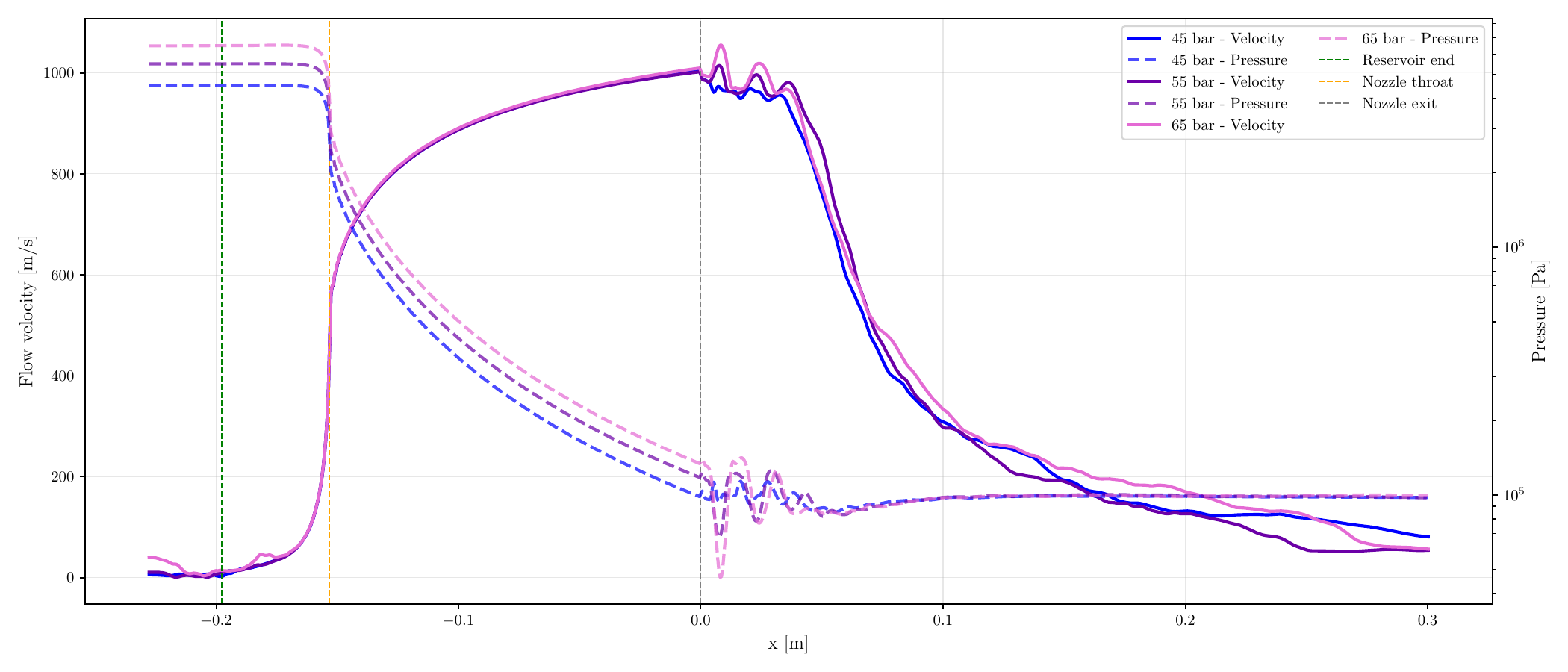}
        \caption{Velocity and pressure profiles for varying chamber pressure (T = 757 K)}
        \label{fig:pcomp_vandp}
    \end{subfigure}
    \hfill
    \begin{subfigure}[b]{0.48\textwidth}
        \centering
        \includegraphics[width=\textwidth]{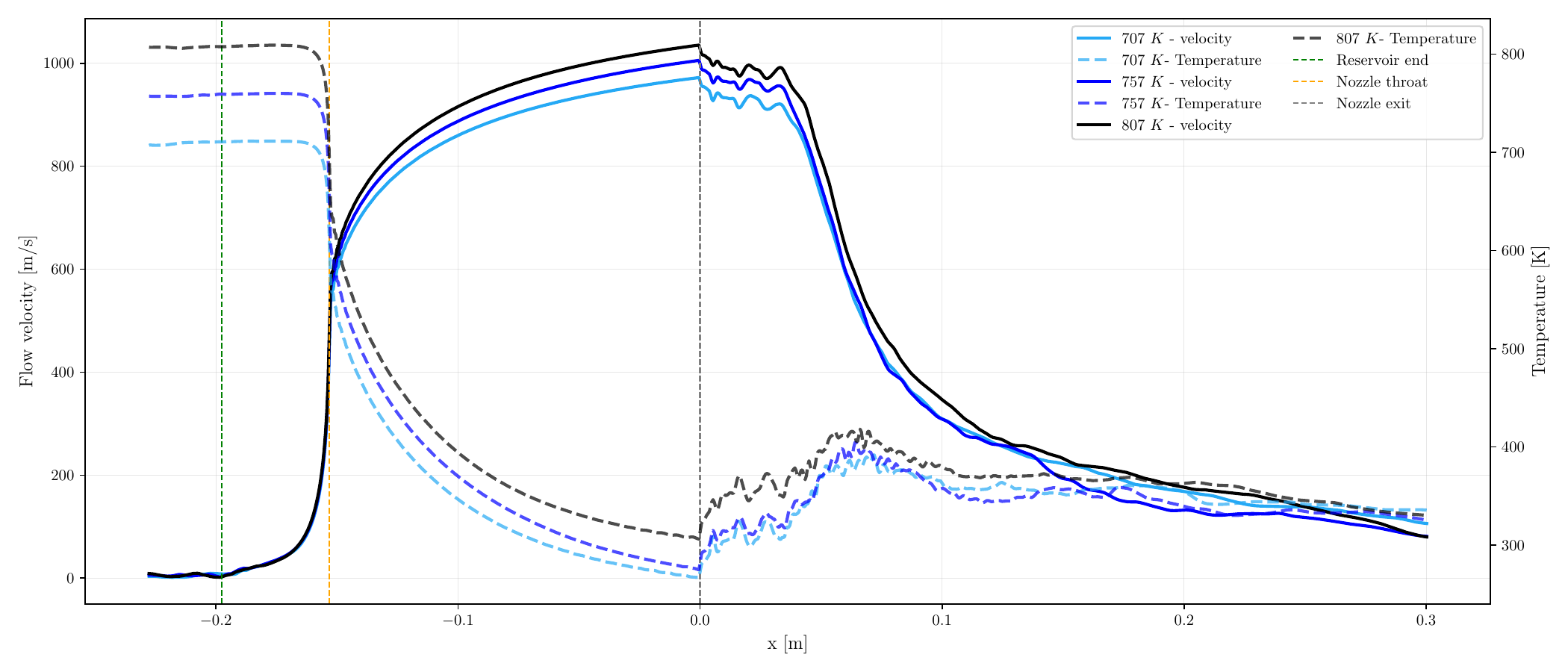}
        \caption{Velocity and temperature profiles for varying gas temperature (p = 45 bar)}
        \label{fig:tcomp_vandt}
    \end{subfigure}
    \caption{Comparison of time-averaged centerline flow properties from DDES simulations under different operating conditions, showing velocity, pressure, and temperature distributions along the axial direction from the nozzle throat to far-field region.}
    \label{fig:flow_comparison}
\end{figure}

\autoref{fig:flow_comparison}  compares time-averaged centerline flow parameters from DDES simulations at different chamber pressure and gas temperature conditions. \autoref{fig:pcomp_vandp} compares the velocity and pressure distributions averaged in time for three chamber pressures (45, 55, and 65 bar) at constant temperature (T = 757 $K$). The mean velocity profiles through the nozzle are virtually identical for all three pressure cases, indicating that the nozzle geometry and gas properties govern the velocity development rather than the chamber pressure. This behavior is consistent with compressible flow theory, where the Mach number distribution in a converging-diverging nozzle depends primarily on the area ratio.

However, significant differences emerge after the nozzle exit. The amplitude of mean velocity fluctuations in the supersonic jet region increases proportionally with chamber pressure. Higher chamber pressures produce stronger shock cells and more intense turbulent mixing, manifesting as larger variations in the time-averaged velocity profile. Additionally, the high-velocity core region extends further downstream for higher chamber pressures. Specifically, the 65 bar case sustains peak velocities over a greater axial distance than the 45 and 55 bar cases, indicating less mean momentum dissipation and delayed jet spreading.

\autoref{fig:tcomp_vandt} illustrates the time-averaged velocity and temperature distributions for three gas temperatures (707, 757, and 807 $K$) at constant chamber pressure (p = 45 bar). In the converging section of the nozzle, the mean velocity profiles are nearly identical regardless of temperature, as the flow remains subsonic and the temperature differences have minimal impact on the acceleration process. However, pronounced differences emerge in the diverging section and become most evident at the nozzle exit.

At the nozzle exit, the velocity differences between the temperature cases reach approximately $100 m/s$. The highest temperature case (807 $K$) achieves mean exit velocities near $1000 m/s$, while the lowest temperature case (707 $K$) reaches approximately $900 m/s$. The temperature-dependent velocity variation results from the relationship between gas temperature and sound speed. As temperature rises, so does the speed of sound (a = $\sqrt{\gamma R T}$). For a given Mach number, this leads to higher absolute gas velocities and, as a result, higher particle velocities due to aerodynamic drag.

The downstream mean velocity decay characteristics also differ significantly between temperature cases. The 807 K case exhibits a more gradual velocity decrease, maintaining higher velocities over longer distances compared to the lower temperature cases. This extended high-velocity region suggests that higher gas temperatures promote better mean momentum preservation in the jet, likely due to altered turbulent mixing rates and viscosity effects. Conversely, the 707 K case shows a steeper mean velocity decay, indicating faster momentum dissipation through enhanced turbulent mixing or viscous losses.

These results demonstrate that chamber pressure and gas temperature influence the mean flow field characteristics through different mechanisms. Pressure changes primarily affect the post-exit jet characteristics—mean shock strength, average turbulence intensity, and potential core length—without significantly altering the in-nozzle velocity development. Temperature changes, however, directly impact the achievable mean particle velocities through their effect on the speed of sound and consequently the absolute exit velocity.

For cold spray optimization, this suggests that temperature is the more effective parameter for controlling average particle impact velocity, while pressure adjustments can be used to tune the mean jet structure and spatial extent of the high-velocity region. The trade-off between maximum mean velocity (favoring high temperature) and spatial uniformity (potentially affected by pressure-driven turbulence) must be considered when selecting optimal operating conditions for specific coating applications.

\subsection{Acoustic Wave propagation}

\begin{figure}[htbp]
    \centering
    \begin{subfigure}[b]{0.49\textwidth}
        \centering
        \includegraphics[width=\textwidth, trim={2cm 7cm 2cm 10cm}, clip]{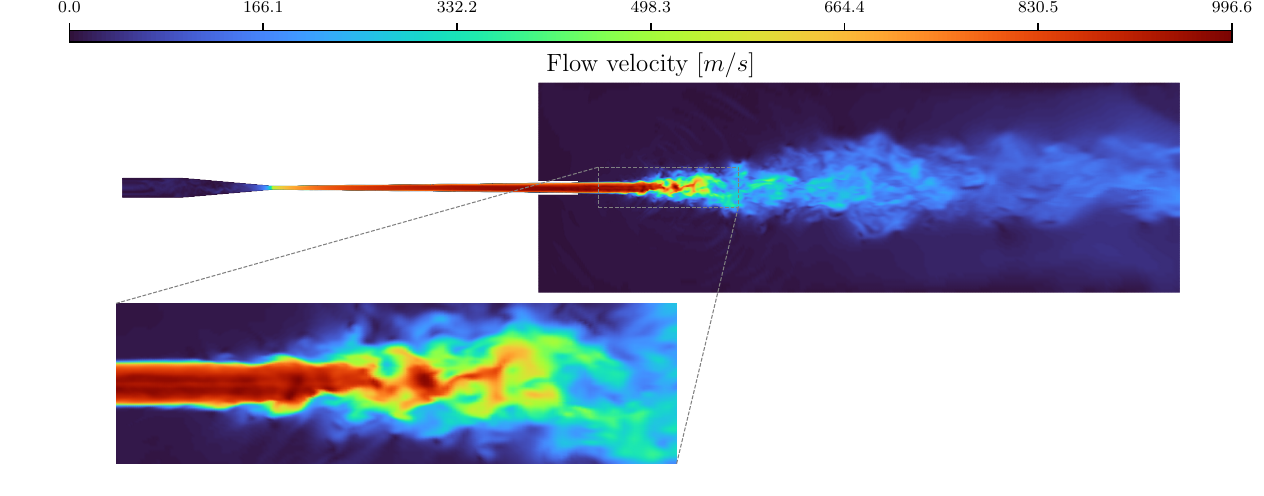}
        \caption{Flow velocity field}
        \label{fig:velocity_instant_0p0005}
    \end{subfigure}
    \hfill
    \begin{subfigure}[b]{0.49\textwidth}
        \centering
        \includegraphics[width=\textwidth, trim={2cm 7cm 2cm 10cm}, clip]{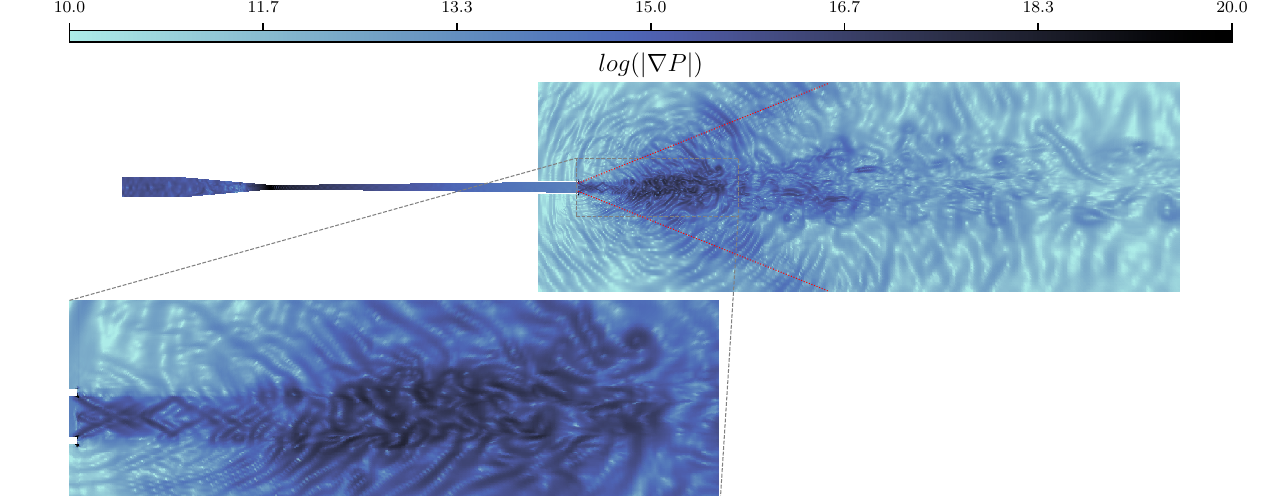}
        \caption{Pressure gradient magnitude}
        \label{fig:pressure_wave_0p0005}
    \end{subfigure}
    \caption{Flow velocity and pressure gradient visualization from DDES simulation at $t = 0.0035\;s$ .}
    \label{fig:instantaneous_flow}
\end{figure}

\autoref{fig:instantaneous_flow} shows snapshots of the flow velocity and pressure waves at $t = 3500 \; \mu s$ from the DDES simulation. Inside the nozzle, the pressure changes smoothly as the gas accelerates through the converging-diverging geometry. However, the largest pressure changes happen a few centimeters after the nozzle exit, where the high-speed jet mixes violently with the surrounding air. This turbulent mixing region is where most of the noise in cold spray systems is generated. When the turbulent flow structures collide with shock waves, they create strong pressure waves that radiate outward, as shown in \autoref{fig:pressure_wave_0p0005}.

The pressure wave image uses a logarithmic scale to make both strong and weak waves visible at the same time. An interesting feature is that waves traveling forward (in the jet direction) are much stronger than waves traveling backward. This happens because the noise sources are moving with the high-speed jet due to the Doppler effect. Two prominent diagonal waves emerge from the nozzle exit at symmetric angles, representing the characteristic wave pattern of supersonic jets. The boundaries of these diagonal waves have been computed using the following equation:
\begin{equation}
    \alpha = \sin^{-1}\left(\frac{1}{Ma}\right)
\end{equation}
where $\alpha$ is the Mach angle and $Ma$ is the Mach number at the exit. For this case, with an exit Mach number of approximately 2.7, the Mach angle is $\alpha = 21.7^\circ$.

Examining the velocity field in \autoref{fig:velocity_instant_0p0005}, it is evident that, at a considerable distance downstream, the hot gas starts to curve upward instead of maintaining a direct trajectory. This upward bending occurs because hot gas is lighter than cold air, so it naturally wants to rise—just like hot air rising from a heater. Near the nozzle exit, the jet's momentum is strong enough to keep the flow moving straight forward. But as the jet slows down due to friction and mixing with ambient air, the upward buoyancy force becomes more important, causing the heated gas to drift upward.

\subsection{Acoustics recorded by observers}
\begin{figure}[ht]
    \centering
    \begin{subfigure}[b]{0.48\textwidth}
        \centering
        \includegraphics[width=\textwidth]{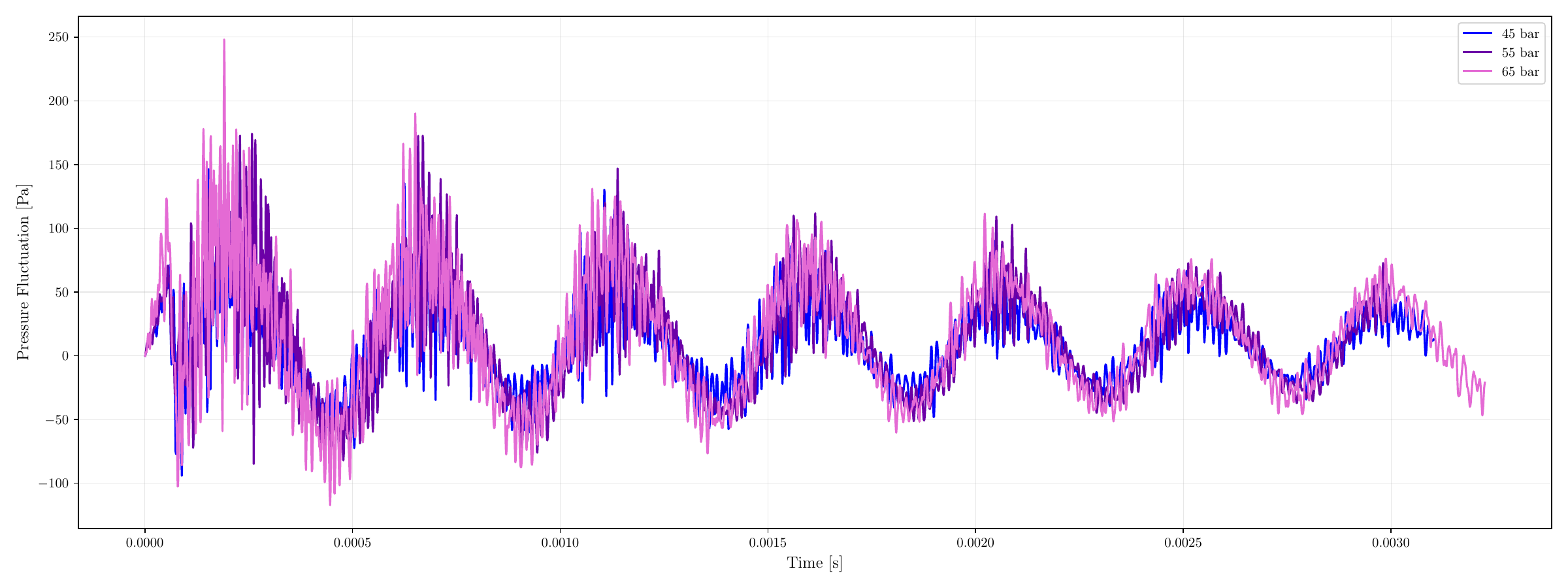}
        \caption{Pressure variation (T = 757 K)}
        \label{fig:pressure_pvar}
    \end{subfigure}
    \hfill
    \begin{subfigure}[b]{0.48\textwidth}
        \centering
        \includegraphics[width=\textwidth]{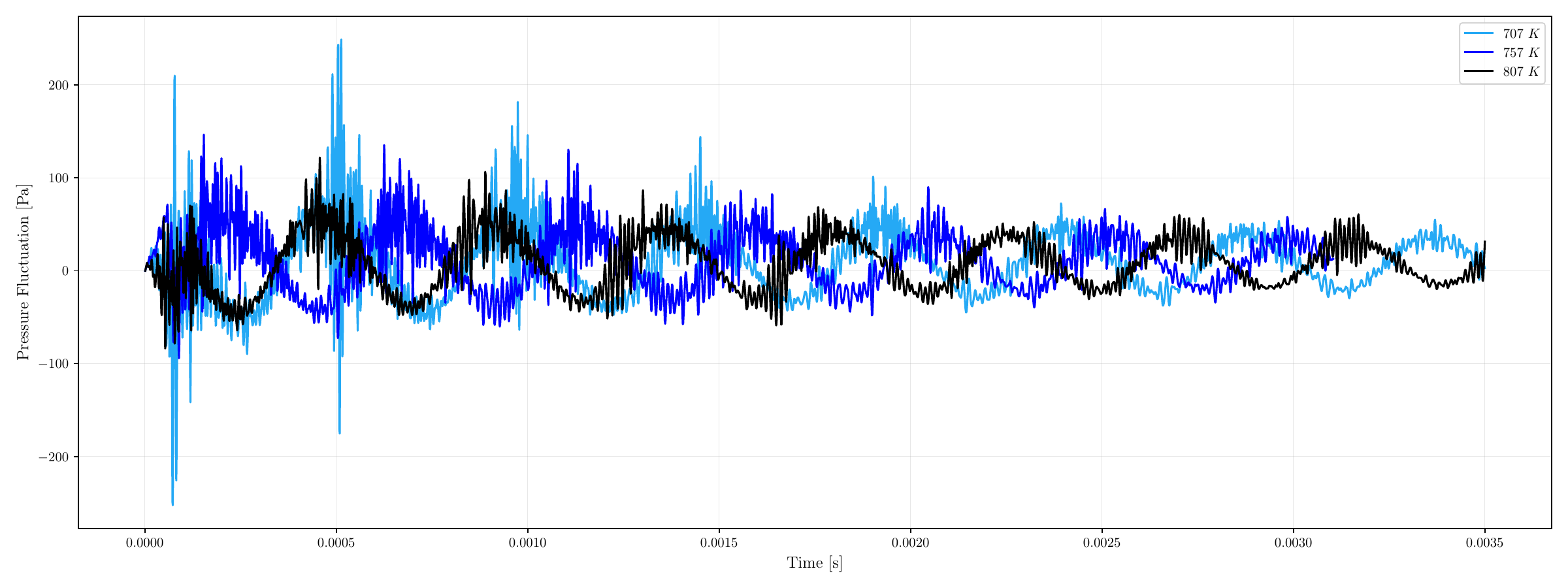}
        \caption{Temperature variation (p = 45 bar)}
        \label{fig:pressure_tvar}
    \end{subfigure}
    \caption{Pressure fluctuations captured at the acoustic observer location under different operating conditions.}
    \label{fig:pressure_fluctuations}
\end{figure}

\autoref{fig:pressure_fluctuations} presents the temporal evolution of pressure fluctuations captured at the acoustic observer location at "Microphone 1" shown in \autoref{fig:CFD_schematic} using the Ffowcs Williams-Hawkings (FW-H) acoustic analogy. \autoref{fig:pressure_pvar} shows pressure fluctuations for three chamber pressure conditions (45, 55, and 65 bar) at constant temperature (T = 757 $K$). The results reveal that changes in chamber pressure do not significantly affect the phase characteristics or the overall envelope of the amplitude of the acoustic signal. However, as the chamber pressure increases, both the amplitude and the sharpness of the fluctuations intensify. This behavior is quantified through the root-mean-square (RMS) pressure, defined as:

\begin{equation}
    P_{\text{rms}} = \sqrt{\frac{1}{N}\sum_{i=1}^{N} {P'_i}^2}
    \label{eq:prms}
\end{equation}

where $P'_i$ represents the instantaneous pressure fluctuation at time step $i$, and $N$ is the total number of samples. The corresponding Overall Sound Pressure Level (OASPL) is calculated as:

\begin{equation}
    \text{OASPL} = 20 \log_{10}\left(\frac{P_{\text{rms}}}{P_{\text{ref}}}\right)
    \label{eq:oaspl}
\end{equation}

where $P_{\text{ref}} = 20$ $\mu$Pa is the reference pressure in air. The quantitative comparison of acoustic metrics for different operating conditions is summarized in \autoref{tab:acoustic_metrics}.

\autoref{fig:pressure_tvar} illustrates the acoustic response to temperature variations (707, 757, and 807 K) at constant chamber pressure (p = 45 bar). The temperature effect is more pronounced than the pressure effect. While the pressure fluctuations for all three cases remain bounded between approximately $-100$ Pa and $+100$ Pa for most of the time history, significant phase shifts are observed between different temperature conditions. The $P_{\text{rms}}$ values remain relatively similar across the three temperature cases, indicating that temperature primarily affects the temporal characteristics rather than the overall energy content of the acoustic field.

An important observation is that lower temperatures produce occasional high-amplitude excursions. For the 707 $K$ case, peak fluctuations reach approximately $\pm 200$ Pa, while the 757 K case exhibits peaks around $\pm 150$ Pa. In contrast, the 807 K case shows more constrained fluctuations, with all recorded pressure deviations remaining within the $\pm 100$ Pa range. This suggests that higher gas temperatures promote more stable acoustic behavior, possibly due to enhanced turbulent mixing and reduced coherence of large-scale vortical structures that drive acoustic radiation.

\begin{table}
    \centering
    \caption{Acoustic metrics comparison between experimental, analytical, and numerical predictions for different cold spray operating conditions}
    \label{tab:acoustic_metrics}
    \setlength{\tabcolsep}{6pt}
    \begin{tabular}{cccccccccccc}
        \toprule
        \multirow{2}{*}{\textbf{$P$ [bar]}} & \multirow{2}{*}{\textbf{$T$ [K]}} & 
        \multicolumn{4}{c}{\textbf{$P_{\text{rms}}$ [Pa]}} & 
        \multicolumn{4}{c}{\textbf{OASPL [dB]}} \\
        \cmidrule(lr){3-6} \cmidrule(lr){7-10}
        & & \textbf{Exp.} & \textbf{Num.} & \textbf{Anal. Orig.} & \textbf{Anal. Cal.} & 
        \textbf{Exp.} & \textbf{Num.} & \textbf{Anal. Orig.} & \textbf{Anal. Cal.} \\
        \midrule
        \multicolumn{10}{l}{\textit{Pressure Variation (T = 757 K)}} \\
        \midrule
        45 & 757 & 34.60 & 34.30 & 73.88 & 36.73 & 124.76 & 124.69 & 131.35 & 125.28 \\
        55 & 757 & 52.97 & 43.78 & 81.66 & 55.06 & 128.46 & 126.80 & 132.22 & 128.80 \\
        65 & 757 & 81.10 & 51.17 & 88.72 & 74.83 & 132.16 & 128.16 & 132.94 & 131.46 \\
        \midrule
        \multicolumn{10}{l}{\textit{Temperature Variation (P = 45 bar)}} \\
        \midrule
        45 & 707 & 31.44 & 41.43 & 72.53 & 33.90 & 123.93 & 126.33 & 131.19 & 124.58 \\
        45 & 757 & 33.31 & 34.30 & 73.80 & 36.55 & 124.43 & 124.69 & 131.34 & 125.24 \\
        45 & 807 & 35.28 & 31.21 & 75.00 & 39.14 & 124.93 & 123.86 & 131.48 & 125.83 \\
        \bottomrule
    \end{tabular}%
\end{table}

As shown in \autoref{tab:acoustic_metrics}, the numerical simulations demonstrate significantly better alignment with experimental data compared to the original analytical model. However, the calibrated analytical model shows the closest agreement with experimental measurements. It is important to note that the calibrated model's reliability is limited, as it was tuned using a small dataset and may not generalize well to all operating conditions beyond the calibration range. For pressure variations at $T = 757$ $K$, the numerical model shows excellent agreement at lower pressures (45 bar: $P_{\text{rms}} = 34.30$ Pa vs. $34.60$ Pa experimental, 0.9\% error) but progressively underpredicts at higher pressures (65 bar: $51.17$ Pa vs. $81.10$ Pa experimental, 36.9\% error). In contrast, the calibrated analytical model maintains close agreement across the entire pressure range with errors within 6--8\%. Regarding temperature variations at $P = 45$ bar, the numerical model shows larger deviations, overpredicting $P_{\text{rms}}$ by 32\% at 707 $K$ and underpredicting by 12\% at 807 $K$, while the calibrated analytical model demonstrates the best performance with errors within 8\%. Overall, while CFD simulations provide reasonable predictions without calibration, the calibrated analytical model demonstrates superior accuracy, though its applicability remains constrained by the limited calibration dataset used for model tuning.

\subsection{Axial Particle Distribution}
\begin{figure*}[htbp]
    \centering
    \includegraphics[width=\linewidth]{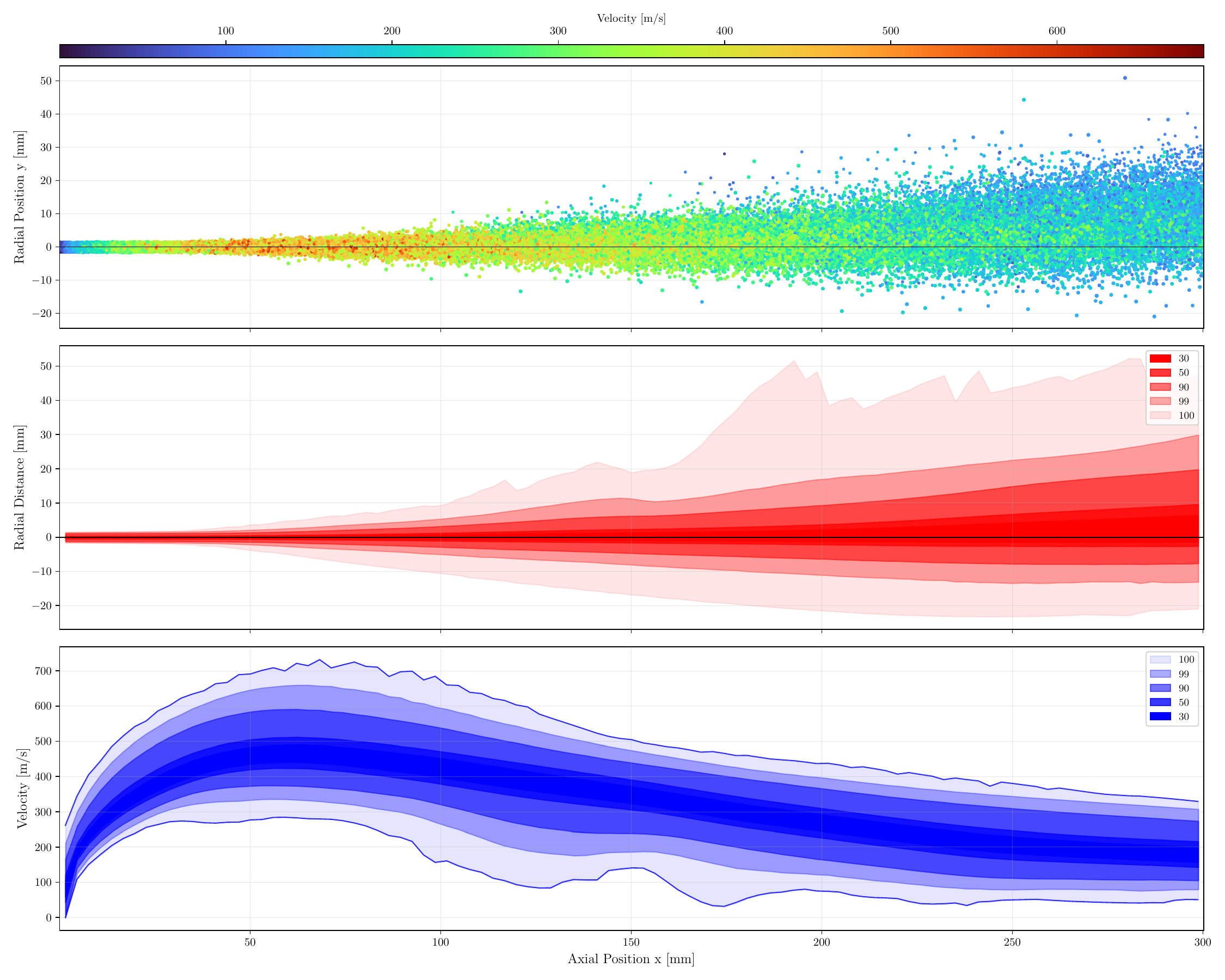}
    \caption{Lagrangian particle tracking analysis showing (top) individual particle trajectories colored by instantaneous velocity, (middle) statistical distribution of radial deviation from centerline with percentile bands, and (bottom) particle velocity evolution along the axial direction. Results obtained for p = 45 bar, T = 757 K operating conditions.}
    \label{fig:particle_analysis}
\end{figure*}

\autoref{fig:particle_analysis} presents a comprehensive Lagrangian particle tracking analysis, illustrating the trajectory, radial dispersion, and velocity evolution of cold spray particles injected into the supersonic gas jet. The upper panel of \autoref{fig:particle_analysis} displays individual particle trajectories colored by velocity magnitude. In the initial 5 cm downstream of injection, particles follow nearly straight trajectories closely aligned with the jet centerline. This behavior reflects the strong axial momentum imparted by the high-velocity gas core in the near-field region, where drag forces rapidly accelerate particles while radial velocity components remain negligible. The particles reach their maximum velocities at approximately this same axial distance (around 5 cm), as shown by the concentration of high-velocity colors in this region.

Beyond 5 cm standoff distance, particle trajectories begin to deviate from the axial direction, with increasing radial dispersion visible as the particles progress downstream. This trajectory spreading results from several physical mechanisms such as turbulent velocity fluctuations in the gas phase that impart transverse momentum to particles, weakening of the axial jet velocity, reducing the stabilizing effect of axial drag forces, and particle-particle interactions in regions of high particle concentration. The progressive velocity decrease reflects the deceleration of particles as the gas jet loses momentum through turbulent mixing with ambient air.

The middle panel of \autoref{fig:particle_analysis} provides a statistical representation of particle radial deviation from the centerline, with shaded bands indicating percentile distributions (50\%, 90\%, 95\% and 99\%). Approximately 50\% of all particles deviate by less than 10 mm from the centerline over a 30 cm standoff distance, indicating that half of the particle population remains tightly focused near the jet axis. Around 90\% of particles remain within a 20 mm radial envelope, which effectively defines the spray diameter for most practical cold spray applications. A small fraction, roughly 1\%, exhibit anomalous trajectories and can deviate by as much as 50 mm after 20 cm of flight.

The gradual expansion of the percentile bands with increasing axial distance quantifies the progressive dispersion of the particle cloud. The relatively narrow 50\% band demonstrates that the core particle population maintains good collimation throughout the analyzed domain. However, the significantly wider outer bands reveal that a small fraction of particles experience substantial radial displacement, likely due to injection into the turbulent shear layer at the jet periphery or interaction with large-scale vortical structures in the flow field.

The outlier particles deviating beyond 50 mm represent approximately 1\% of the total population and warrant special attention. These particles may have been injected at unfavorable radial locations, encountered strong turbulent eddies, or experienced particle-particle collisions that redirected their trajectories. While statistically rare, such outliers can have an impact on coating uniformity in industrial applications, especially when working at long standoff distances, where their radial deviation is most noticeable.

The lower panel of \autoref{fig:particle_analysis} presents the particle velocity distribution as a function of axial position, with color intensity representing particle density at each velocity-position combination. The velocity evolution exhibits several distinct phases: In acceleration phase (0--5 cm), particles rapidly accelerate from their injection velocity from near zero to maximum velocities approaching 600--700 m/s. This acceleration occurs as particles experience intense drag forces within the high-velocity gas jet. The particle velocity increases nearly monotonically with axial distance in this region, with the maximum velocity achieved at approximately 5 cm standoff distance.

In Maximum velocity region (5 cm) The velocity field shows a concentrated band at peak velocities around 5 cm standoff distance, with most particles reaching consistent maximum velocities. This uniformity in peak velocity reflects the relatively homogeneous flow conditions in the jet potential core region, where particles experience similar drag and acceleration histories.

Beyond the peak velocity region, particles gradually decelerate as the gas jet's velocity drops due to turbulent mixing and momentum dissipation. The velocity decline is more gradual than the initial acceleration, and significant velocity dispersion emerges, as evidenced by the enlargement of the distribution band. This dispersion results from particles experiencing varied local gas velocities depending on their radial position inside the expanding jet: particles along the centerline maintain higher velocity for longer periods of time, while those on the jet's periphery decrease more quickly.

The radial deviation statistics directly inform coating uniformity expectations. The 50\% radial deviation envelope (±10 mm at 30 cm) defines the primary coating footprint, while the 90\% envelope (±20 mm) encompasses the effective spray pattern.

Deposition efficiency considerations favor operation in the near-field region. Particles achieving maximum velocity in the 5--10 cm region are most likely to exceed the critical velocity required for successful bonding. The velocity decay beyond this region suggests that deposition efficiency may decrease at larger standoff distances, particularly for materials with high critical velocities.

The small fraction (~1\%) of particles with anomalous trajectories presents challenges for process consistency. These outlier particles may not reach the substrate or may impact at unfavorable angles, reducing deposition efficiency and potentially causing surface defects. 

\subsection{Effect of the conditions of the nozzle chamber on the particles distribution}

\begin{figure}[ht]
    \centering
    \begin{subfigure}[b]{0.48\textwidth}
        \centering
        \includegraphics[width=\textwidth]{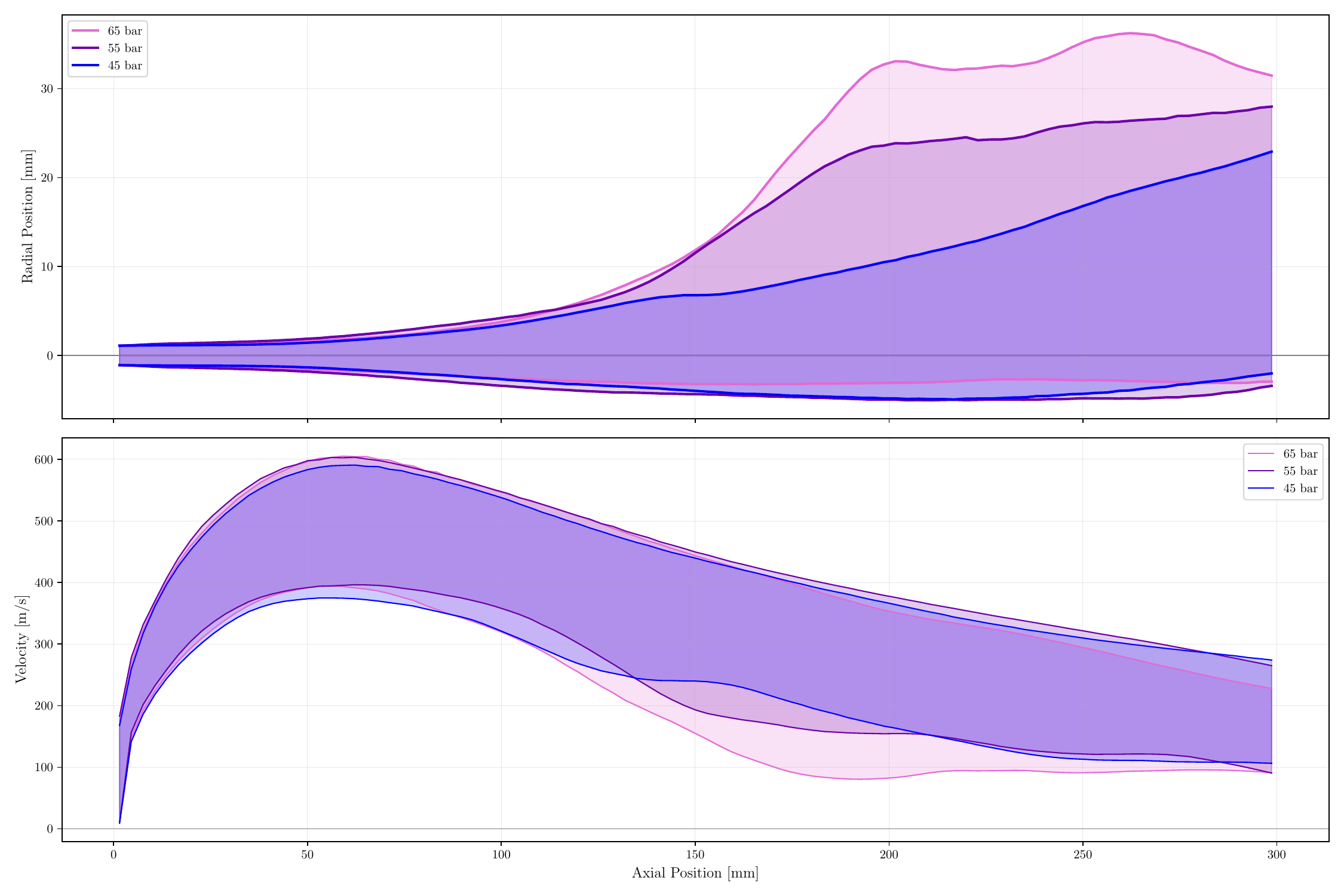}
        \caption{Particle at varying chamber pressures (T = 757 K)}
        \label{fig:particle_pvar}
    \end{subfigure}
    \hfill
    \begin{subfigure}[b]{0.48\textwidth}
        \centering
        \includegraphics[width=\textwidth]{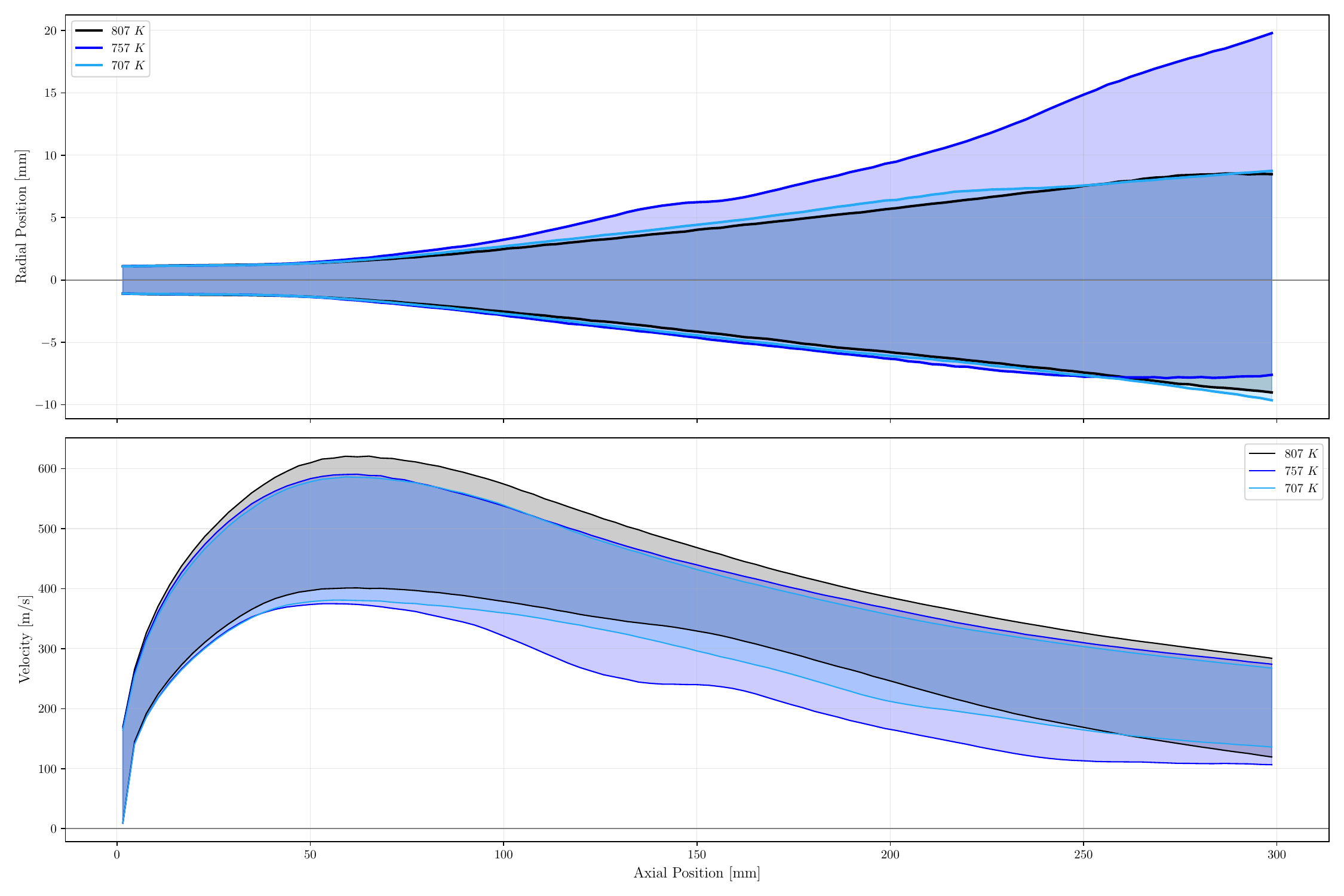}
        \caption{Particle at varying gas temperatures (p = 45 bar)}
        \label{fig:particle_tvar}
    \end{subfigure}
    \caption{Particle percentile distributions for radial position and velocity  under different operating conditions. Shaded regions represent the statistical spread of particle behavior.}
    \label{fig:particle_percentiles}
\end{figure}

\autoref{fig:particle_pvar} illustrates the particle distribution along the axial centerline across different percentile bands for varying chamber pressures. Regarding radial positions, all cases exhibit identical behavior up to 10~cm downstream of the nozzle exit. Beyond this point, the upper radial position percentiles diverge substantially as chamber pressure rises. At a standoff distance of 30 cm, the 65-bar case has a radial spread of around 30 mm, but the 45-bar case has just a 20 mm spread. Notably, the lower percentile bands for all three pressure situations are essentially identical across the majority of the axial domain, implying that the core particle population follows constant radial paths regardless of chamber pressure. For the particle velocity distributions, all three cases follow similar trends until approximately 60~mm downstream of the nozzle exit, where they reach comparable peak velocities with aligned upper and lower percentile bands. Subsequently, the higher pressure case exhibits a more rapid velocity decay compared to the lower pressure conditions. Most significantly, between 12~cm and 23~cm axial distance, the 65~bar case displays a substantially wider gap between upper and lower velocity percentiles than the other cases, indicating enhanced velocity dispersion within the particle field. The upper velocity limit remains relatively consistent across all chamber pressures throughout most of the domain. However, in the final segment (25--30~cm), the 65~bar case undergoes a more pronounced velocity reduction, exiting the computational domain approximately 30--40~m/s slower at its upper percentile bound compared to the two lower pressure configurations. The comparison of different chamber pressures demonstrates that increasing the chamber pressure results in wider particle distributions and more rapid velocity decay at greater distances from the nozzle exit.

For the temperature variation shown in \autoref{fig:particle_tvar}, it can be observed that the 707~K and 807~K cases exhibit nearly identical particle radial distributions. The 757K case deviates significantly from the other two temperature cases, particularly in the top percentile band. The changes in particle velocity distributions between chamber temperatures are more noticeable. The upper bound of the highest temperature case (807~K) consistently exhibits higher velocities than the other two cases across all axial positions. At the peak velocity region, the upper percentile band of the 807~K case is approximately 50~m/s higher than the other temperatures. Additionally, the area between the upper and lower velocity bands for the 807~K case is notably smaller than those of the 757~K and 707~K cases. The lower bound of the 807~K case is predominantly higher than both the 757~K and 707~K cases throughout the domain. Overall, the increase in chamber temperature has less impact on radial particle distribution along the axis but significantly influences particle velocities. Higher temperatures result in elevated upper velocity bounds and reduced dispersion between the upper and lower percentile bands, indicating more uniform particle velocity distributions during deposition.

The parametric analysis of chamber pressure and temperature reveals distinct effects on particle behavior. Increasing chamber pressure mainly causes particles to spread out more radially and creates a wider range of velocities in regions farther from the nozzle exit. In contrast, increasing chamber temperature primarily boosts particle velocities while having little effect on how far particles spread radially. These two parameters work in opposite ways, higher pressure creates more variation in particle velocities (wider velocity distribution), while higher temperature makes particle velocities more uniform (narrower velocity distribution). This difference occurs because pressure changes intensify turbulent mixing and increase the velocity difference between particles and gas, whereas temperature changes provide more energy to accelerate all particles more uniformly. In practical terms, chamber pressure is the key parameter for controlling where particles go spatially, while temperature is more effective for controlling how fast particles travel and ensuring they move at similar speeds during thermal spray deposition.

\subsection{Stand-off distance effect on particles distribution}

\begin{figure*}[ht]
    \centering
    \begin{subfigure}[b]{0.32\textwidth}
        \centering
        \includegraphics[width=\textwidth]{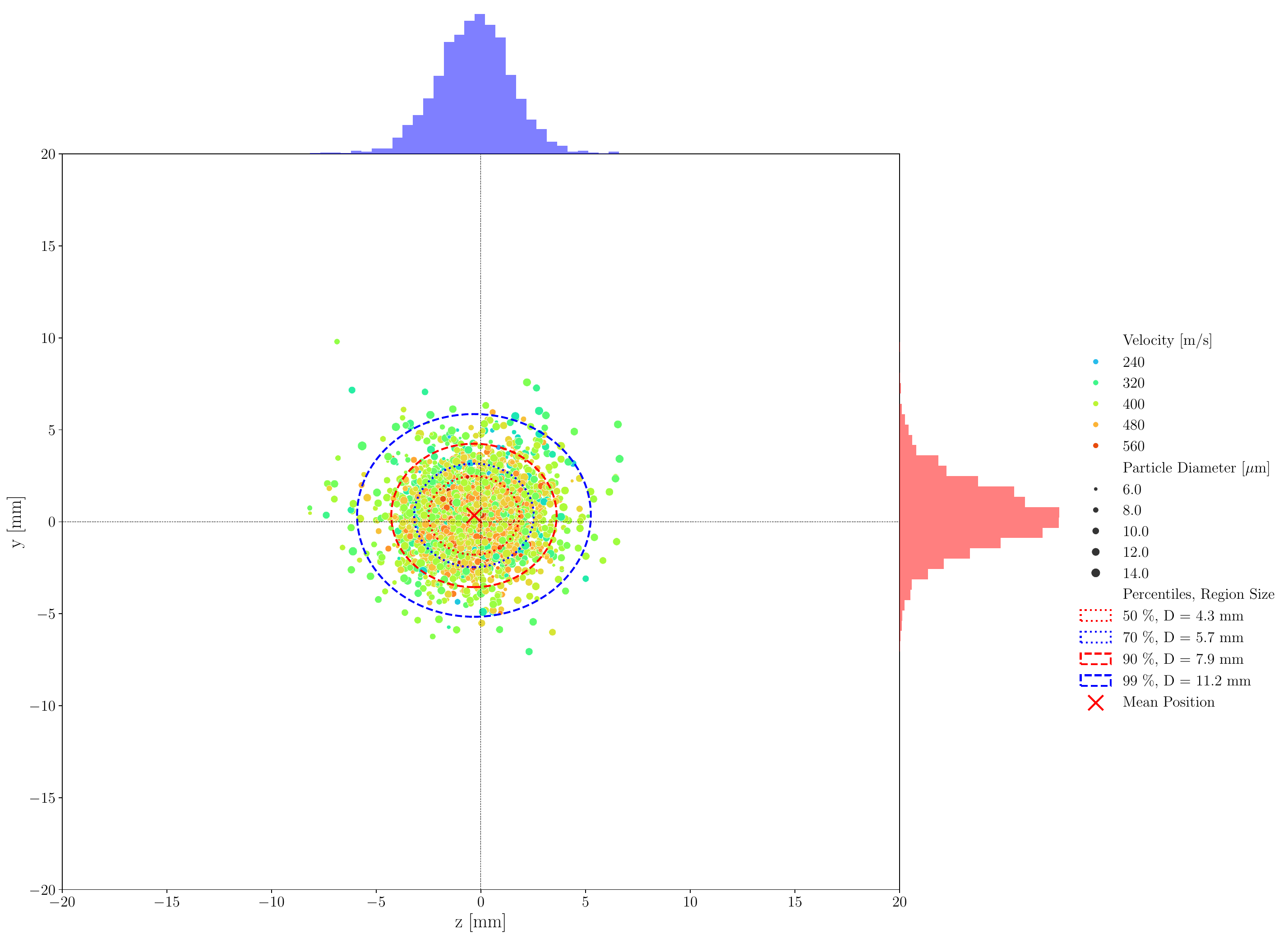}
        \caption{45 bar, 100 mm}
        \label{fig:particles_p45_100}
    \end{subfigure}
    \hfill
    \begin{subfigure}[b]{0.32\textwidth}
        \centering
        \includegraphics[width=\textwidth]{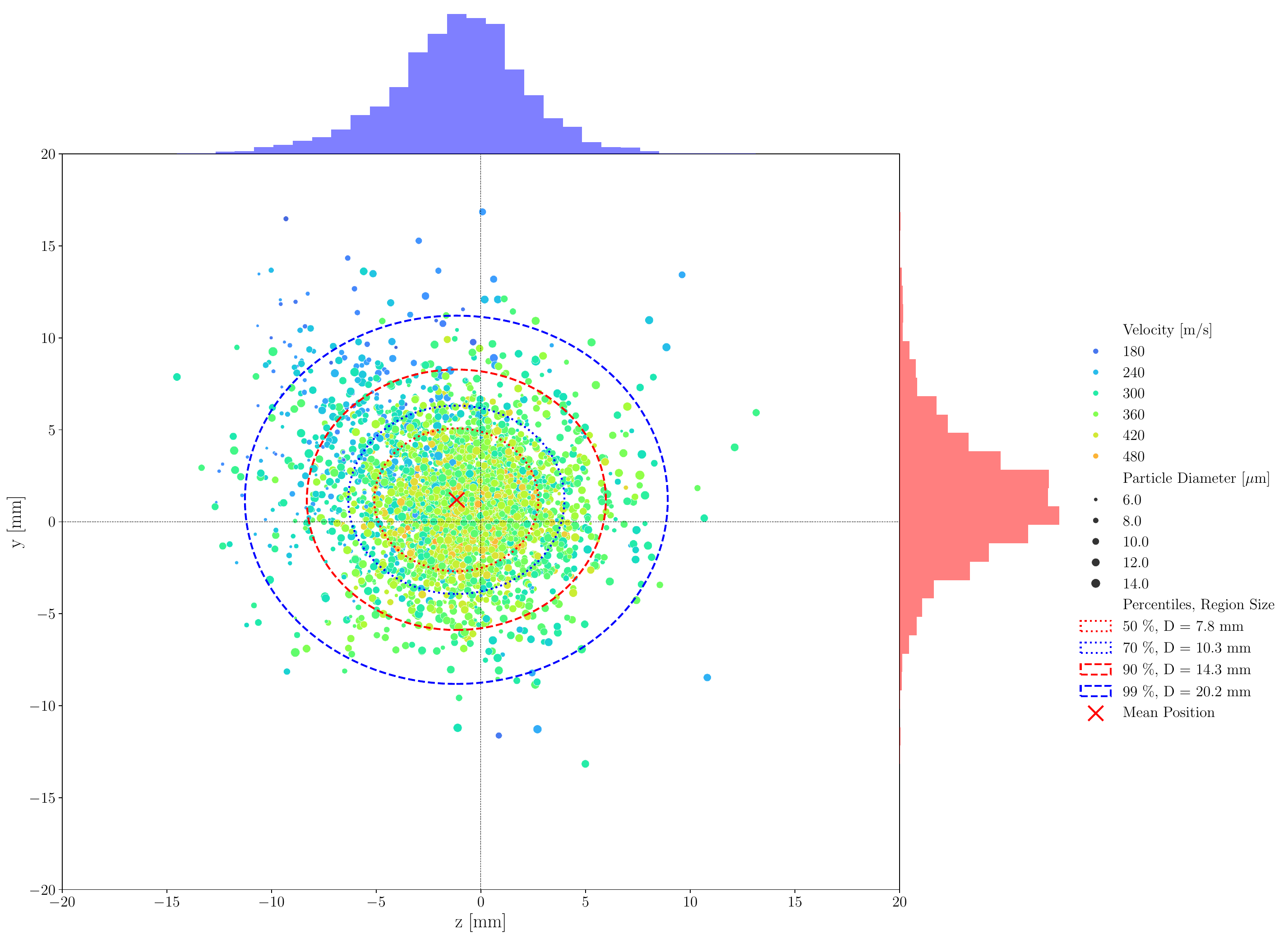}
        \caption{45 bar, 150 mm}
        \label{fig:particles_p45_150}
    \end{subfigure}
    \hfill
    \begin{subfigure}[b]{0.32\textwidth}
        \centering
        \includegraphics[width=\textwidth]{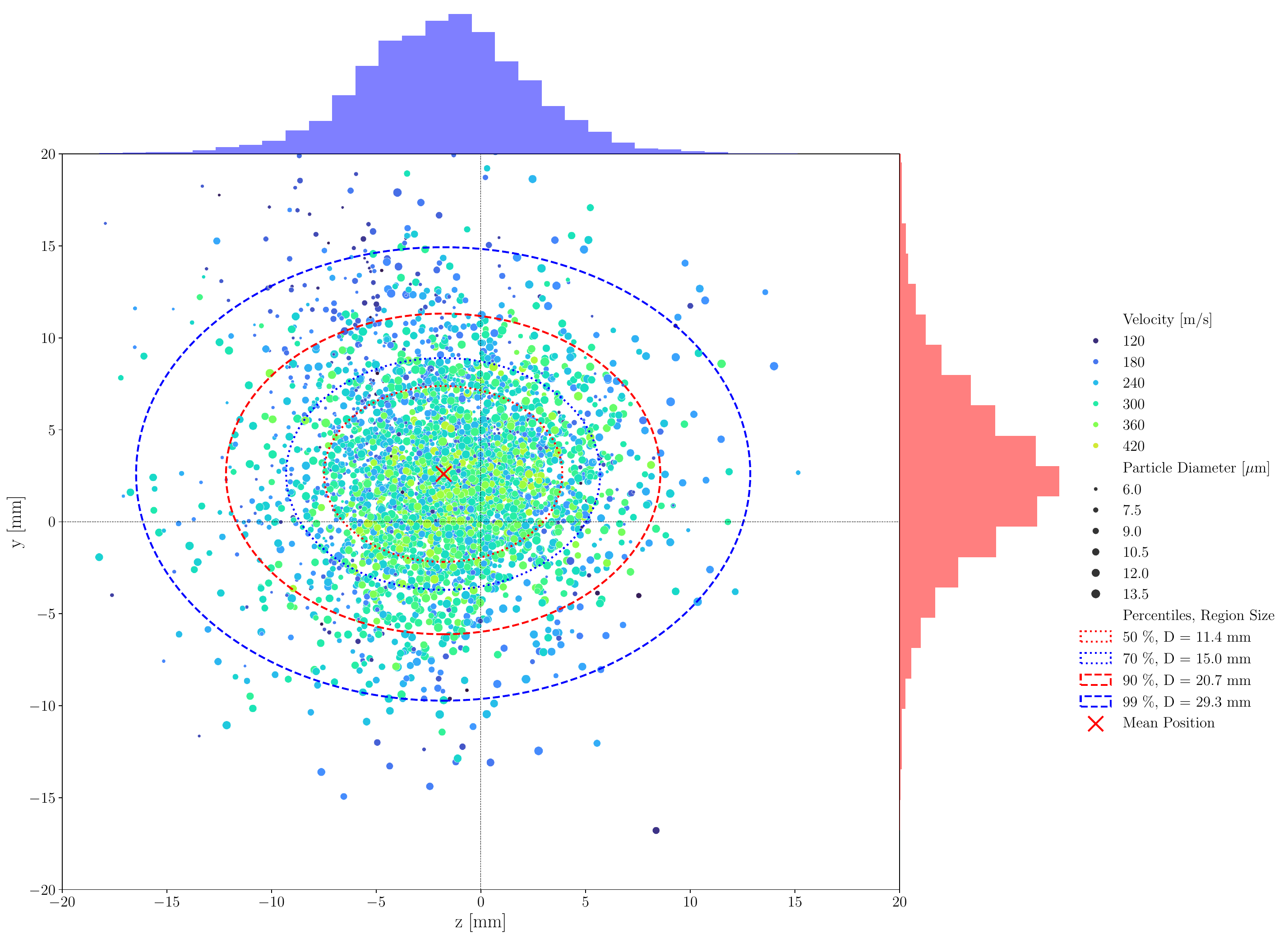}
        \caption{45 bar, 200 mm}
        \label{fig:particles_p45_200}
    \end{subfigure}
    
    \vspace{0.3cm}

    \begin{subfigure}[b]{0.32\textwidth}
        \centering
        \includegraphics[width=\textwidth]{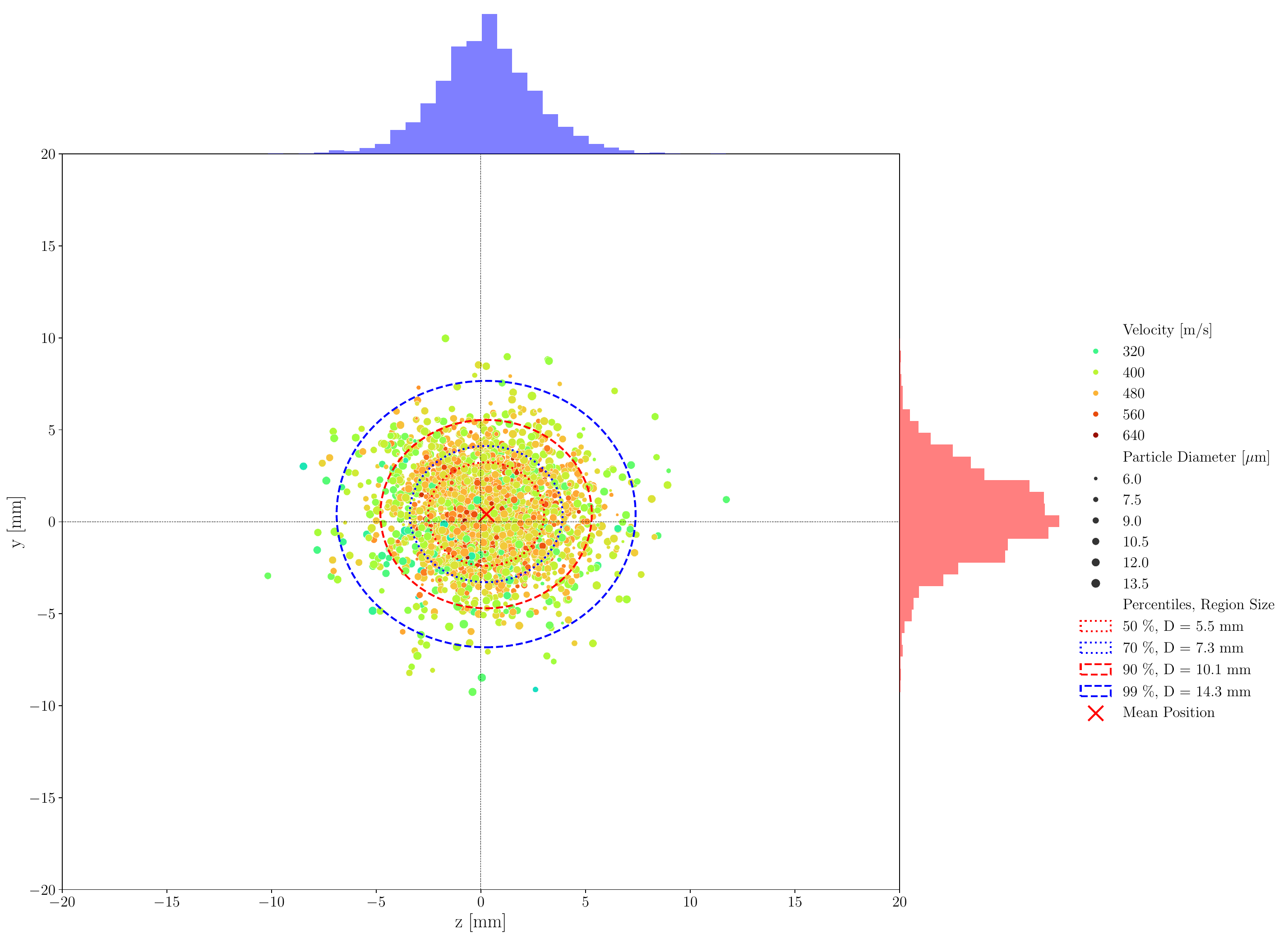}
        \caption{55 bar, 100 mm}
        \label{fig:particles_p65_100}
    \end{subfigure}
    \hfill
    \begin{subfigure}[b]{0.32\textwidth}
        \centering
        \includegraphics[width=\textwidth]{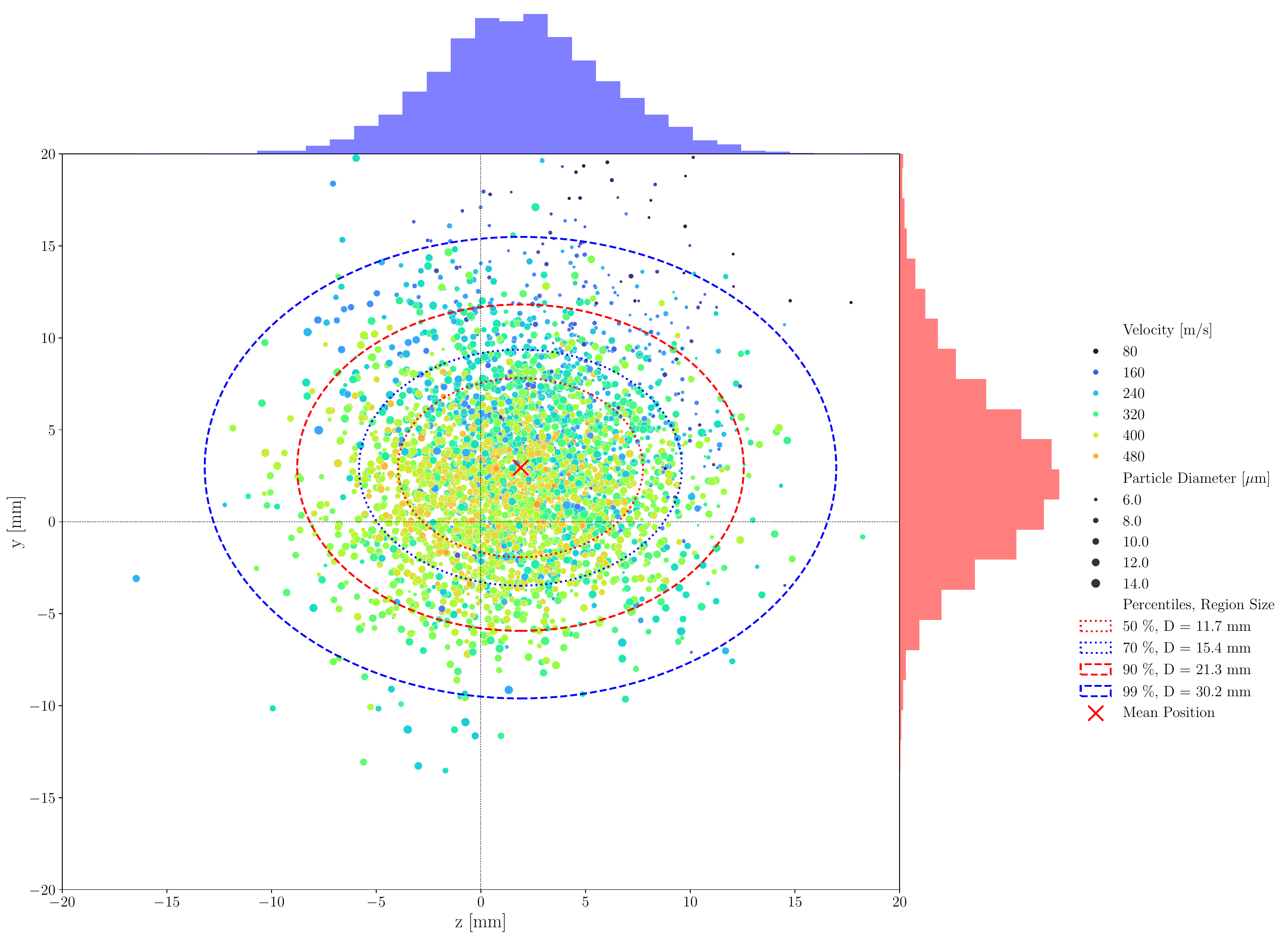}
        \caption{55 bar, 150 mm}
        \label{fig:particles_p65_150}
    \end{subfigure}
    \hfill
    \begin{subfigure}[b]{0.32\textwidth}
        \centering
        \includegraphics[width=\textwidth]{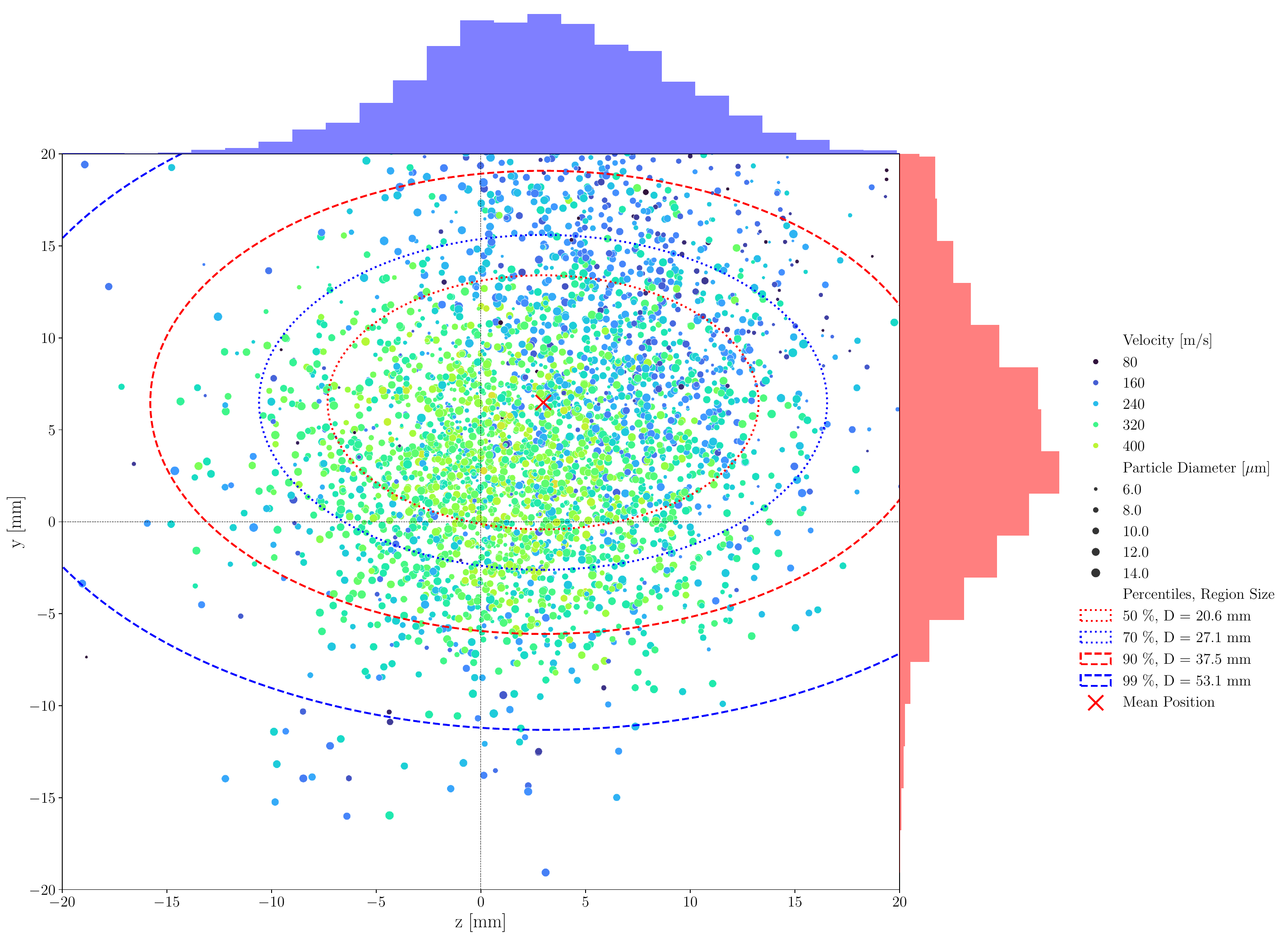}
        \caption{55 bar, 200 mm}
        \label{fig:particles_p65_200}
    \end{subfigure}

    
    \begin{subfigure}[b]{0.32\textwidth}
        \centering
        \includegraphics[width=\textwidth]{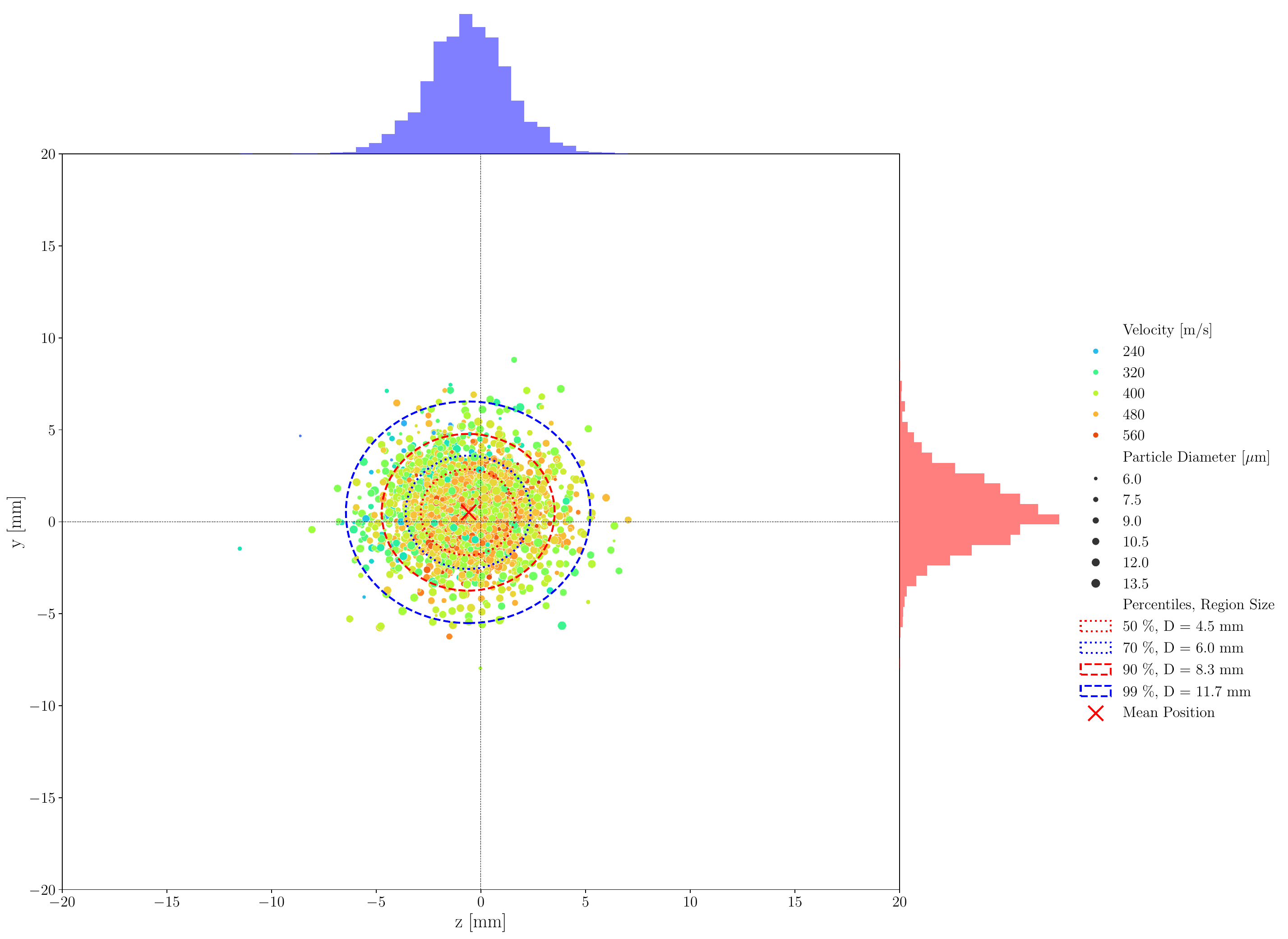}
        \caption{65 bar, 100 mm}
        \label{fig:particles_p55_100}
    \end{subfigure}
    \hfill
    \begin{subfigure}[b]{0.32\textwidth}
        \centering
        \includegraphics[width=\textwidth]{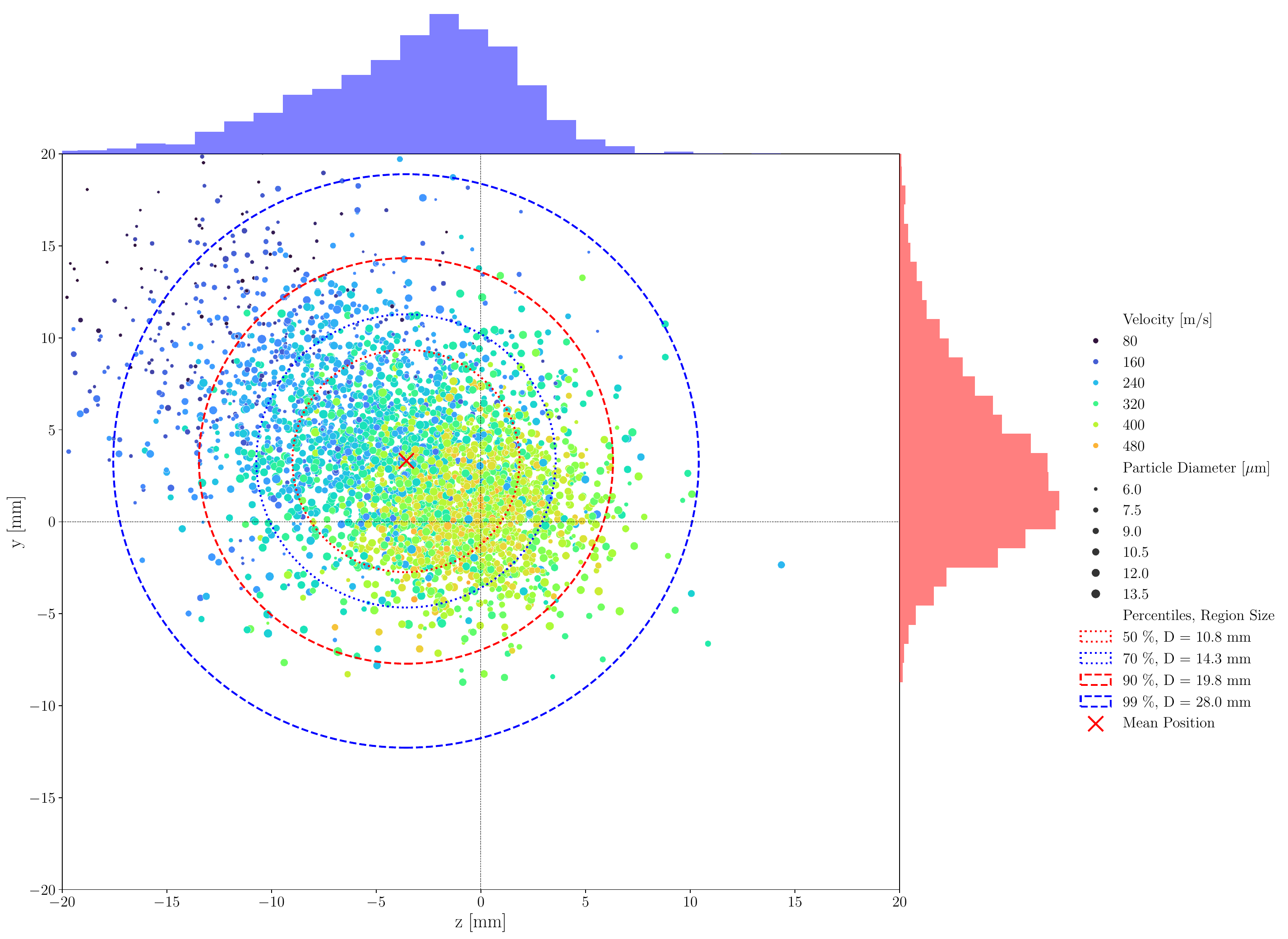}
        \caption{65 bar, 150 mm}
        \label{fig:particles_p55_150}
    \end{subfigure}
    \hfill
    \begin{subfigure}[b]{0.32\textwidth}
        \centering
        \includegraphics[width=\textwidth]{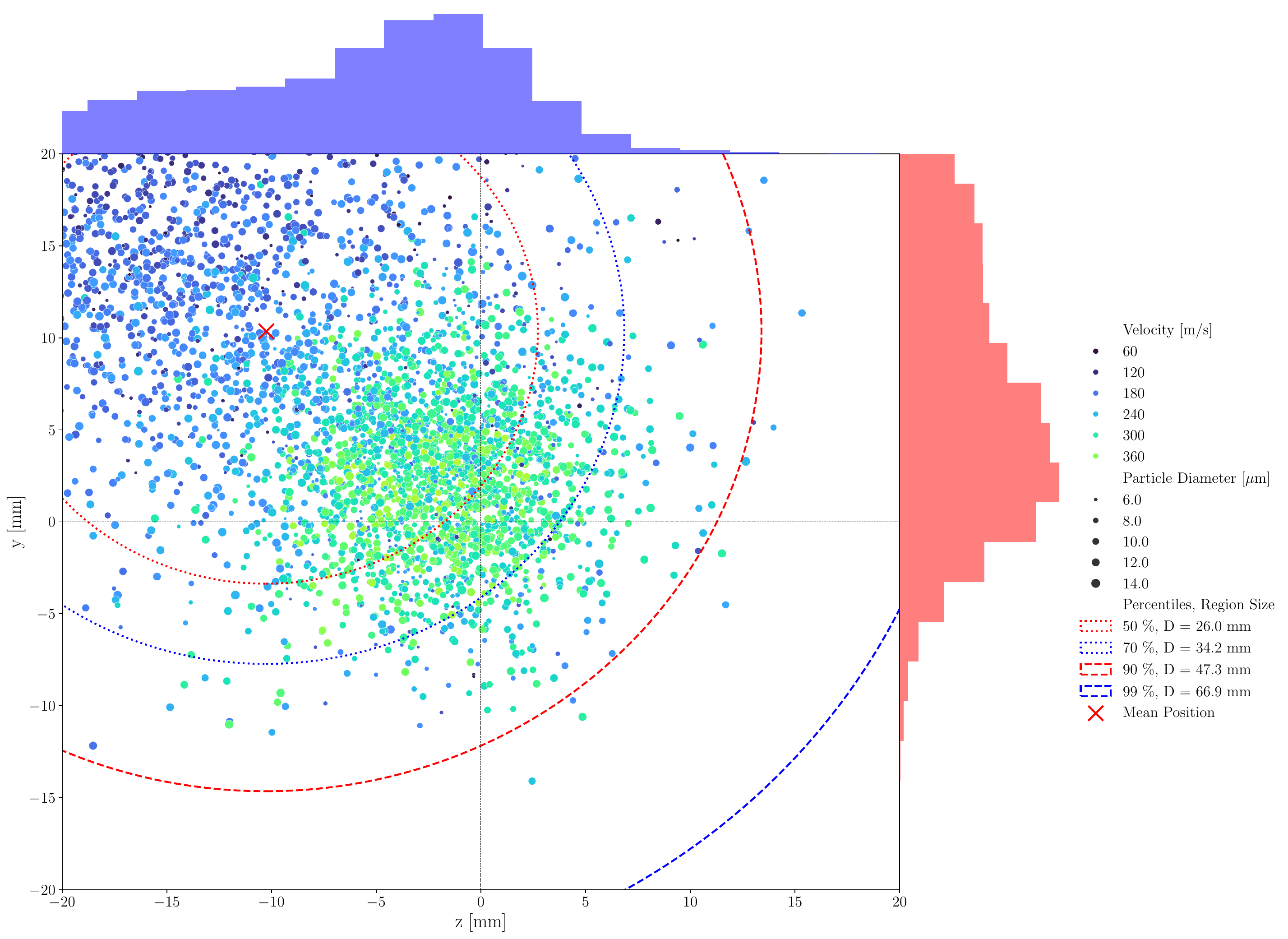}
        \caption{65 bar, 200 mm}
        \label{fig:particles_p55_200}
    \end{subfigure}
    
    \caption{Particle spatial distribution at three standoff distances (100, 150, 200 mm) for chamber pressures of 45, 55, and 65 bar at constant temperature T = 757 K.}
    \label{fig:particles_grid}
\end{figure*}

\autoref{fig:particles_grid} presents a cross-sectional view of the particle distributions shown in \autoref{fig:particle_analysis}. The particle spatial distributions are visualized at three standoff distances (100~mm, 150~mm, and 200~mm) for chamber pressures of 45~bar, 55~bar, and 65~bar at constant temperature ($T = 757$~K). Each panel displays particle positions in a cross-sectional plane perpendicular to the jet axis, with color indicating particle velocity magnitude. Marginal histograms on the top and right edges show the projected particle distributions along the vertical and horizontal axes, respectively. Concentric circles represent radial distances from the centerline.

\begin{table}
    \centering
    \caption{Particle distribution diameters (mm) at different percentiles for varying chamber pressures and standoff distances at constant temperature ($T = 757$~K)}
    \label{tab:particle_diameters}
    \begin{tabular}{ccccccccccccc}
        \toprule
        \textbf{Pressure} & \multicolumn{4}{c}{\textbf{100 mm}} & \multicolumn{4}{c}{\textbf{150 mm}} & \multicolumn{4}{c}{\textbf{200 mm}} \\
        \cmidrule(lr){2-5} \cmidrule(lr){6-9} \cmidrule(lr){10-13}
        \textbf{(bar)} & \textbf{50\%} & \textbf{70\%} & \textbf{90\%} & \textbf{99\%} & \textbf{50\%} & \textbf{70\%} & \textbf{90\%} & \textbf{99\%} & \textbf{50\%} & \textbf{70\%} & \textbf{90\%} & \textbf{99\%} \\
        \midrule
        45 & 4.3 & 5.7 & 7.9 & 11.2 & 7.8 & 10.3 & 14.3 & 20.2 & 11.4 & 15.0 & 20.7 & 29.3 \\
        55 & 5.5 & 7.3 & 10.1 & 14.3 & 11.7 & 15.4 & 21.3 & 30.2 & 20.6 & 27.1 & 37.5 & 53.1 \\
        65 & 4.5 & 6.0 & 8.3 & 11.7 & 10.8 & 14.3 & 19.8 & 28.0 & 26.0 & 34.2 & 47.3 & 66.9 \\
        \bottomrule
    \end{tabular}
\end{table}

Circles representing the regions through which different percentiles of particles have passed are visualized and the corresponding diameters are provided in \autoref{tab:particle_diameters}. The center of all particles passing through each cross-sectional plane has been marked as well. It can be observed that as the standoff distance increases, the particle center shifts toward the positive y-direction. It is worth noting that gravity has been modeled in the negative y-direction, and as previously shown in earlier sections, the hot gas jet deflects toward positive y due to its lower density compared to the ambient surrounding air. For the  z-axis, no clear trend emerges, in two cases, the majority of particles lean to the left, while in one case they lean to the right. A notable observation from this visualization is that particles closer to the centerline have higher velocities due to increased interaction with the high-velocity jet coming from the nozzle, whereas particles farther from the centerline have lower velocities as they move away from the central jet region. Another observation is that as chamber pressure increases, the center of mass of particles passing through the surface tends to move farther from the centerline. At 200~mm, the 65~bar case shows a more diffuse velocity field with a wider transition zone between the high-velocity core and low-velocity periphery compared to the 45~bar case, which maintains a more distinct velocity gradient. This observation aligns with the increased gap between upper and lower velocity percentiles discussed in \autoref{fig:particle_pvar}, confirming that elevated pressure promotes greater velocity stratification within the particle population.

\begin{figure*}[htbp]
    \centering
    \begin{subfigure}[b]{0.32\textwidth}
        \centering
        \includegraphics[width=\textwidth]{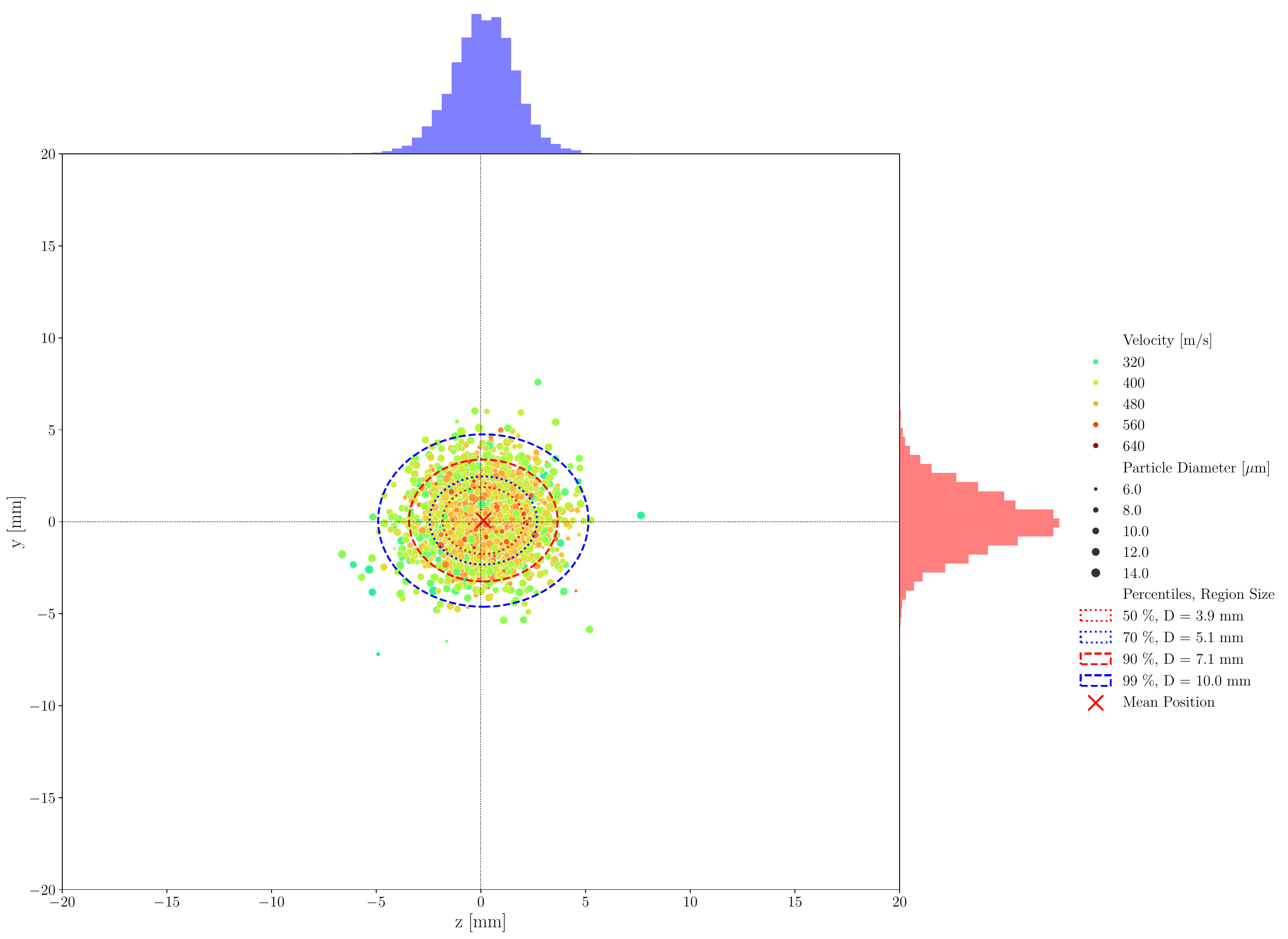}
        \caption{707 K, 100 mm}
        \label{fig:particles_t707_100}
    \end{subfigure}
    \hfill
    \begin{subfigure}[b]{0.32\textwidth}
        \centering
        \includegraphics[width=\textwidth]{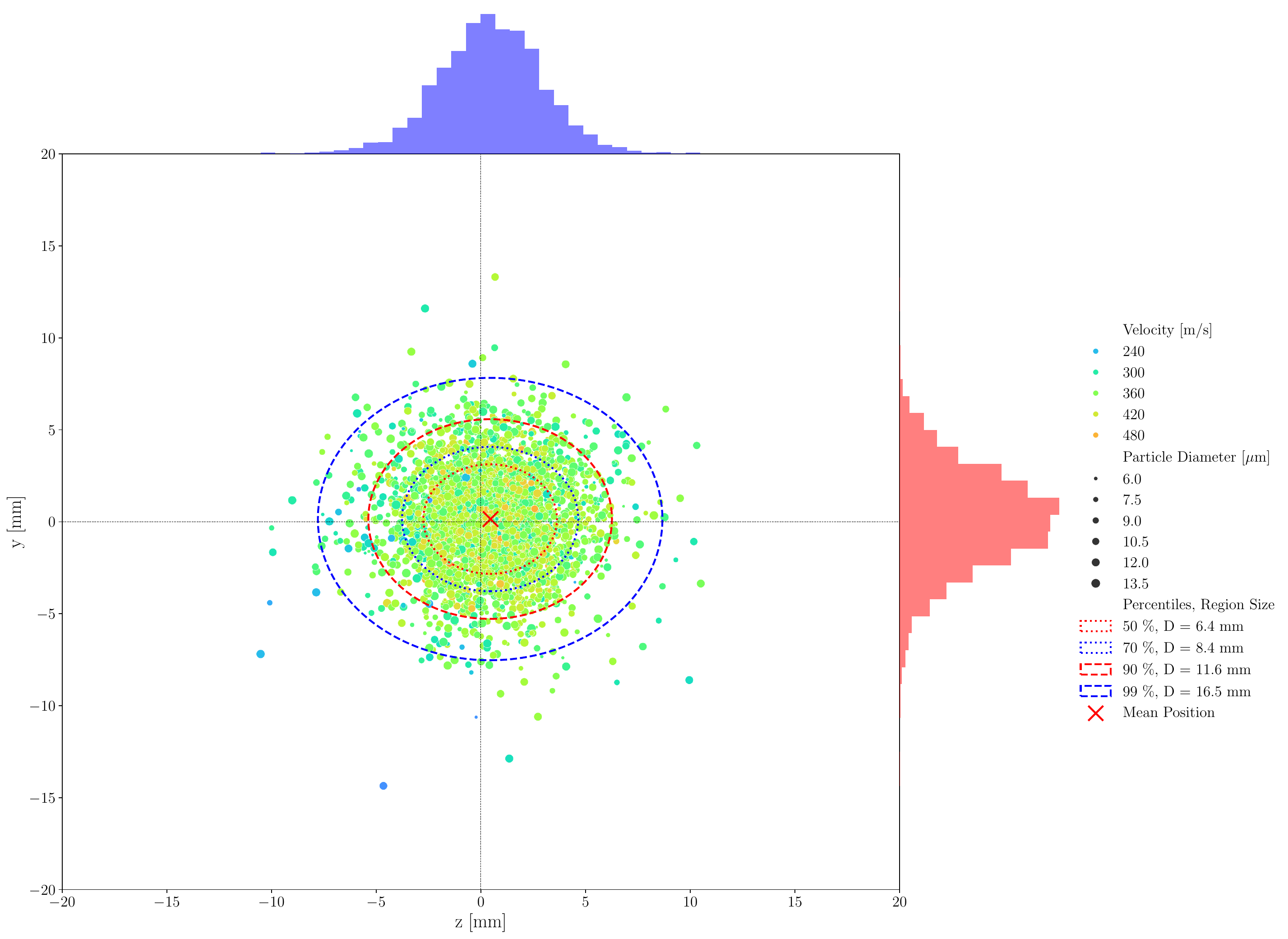}
        \caption{707 K, 150 mm}
        \label{fig:particles_t707_150}
    \end{subfigure}
    \hfill
    \begin{subfigure}[b]{0.32\textwidth}
        \centering
        \includegraphics[width=\textwidth]{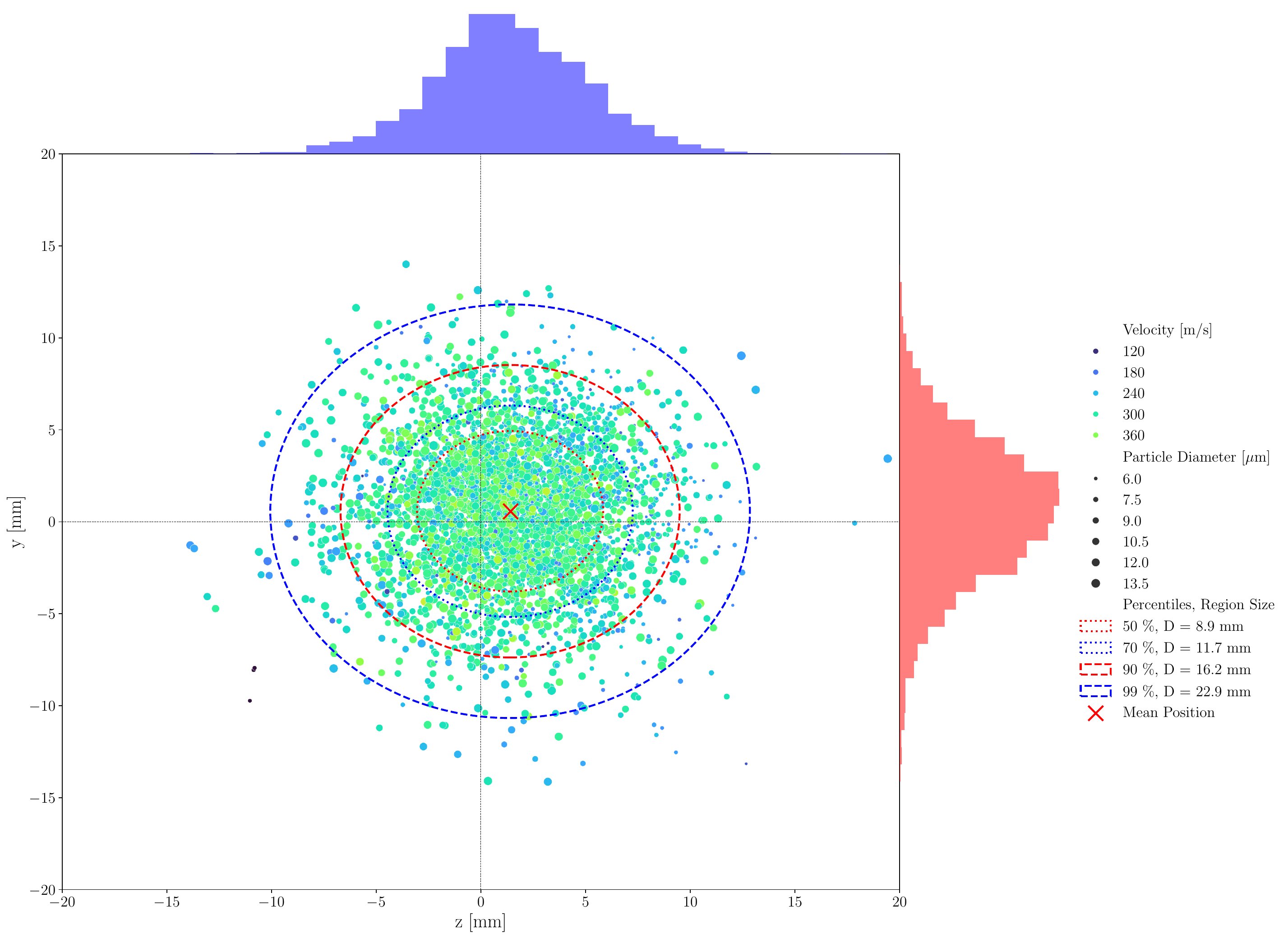}
        \caption{707 K, 200 mm}
        \label{fig:particles_t707_200}
    \end{subfigure}
    
    \vspace{0.3cm}
    
    \begin{subfigure}[b]{0.32\textwidth}
        \centering
        \includegraphics[width=\textwidth]{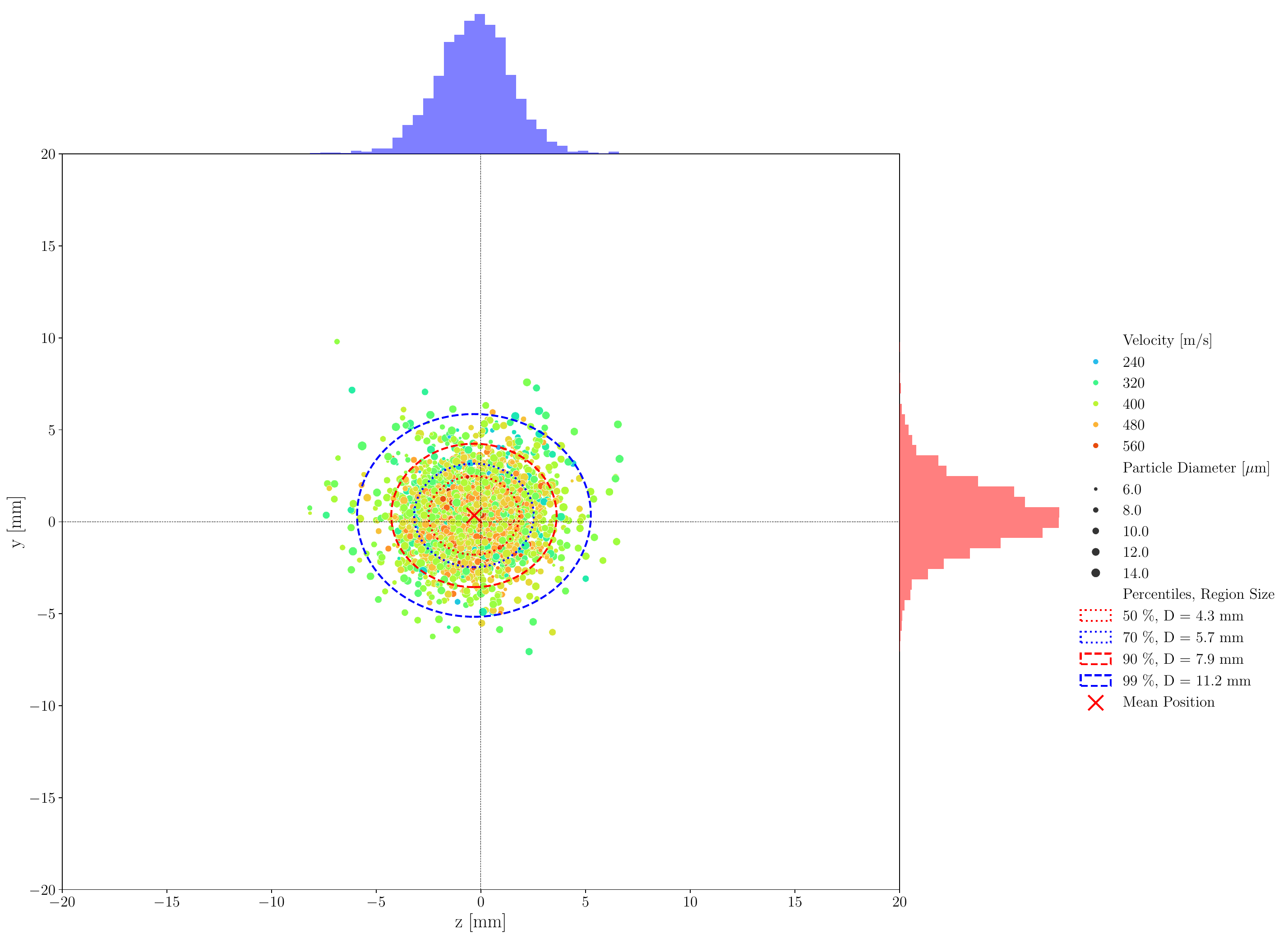}
        \caption{757 K, 100 mm}
        \label{fig:particles_t757_100}
    \end{subfigure}
    \hfill
    \begin{subfigure}[b]{0.32\textwidth}
        \centering
        \includegraphics[width=\textwidth]{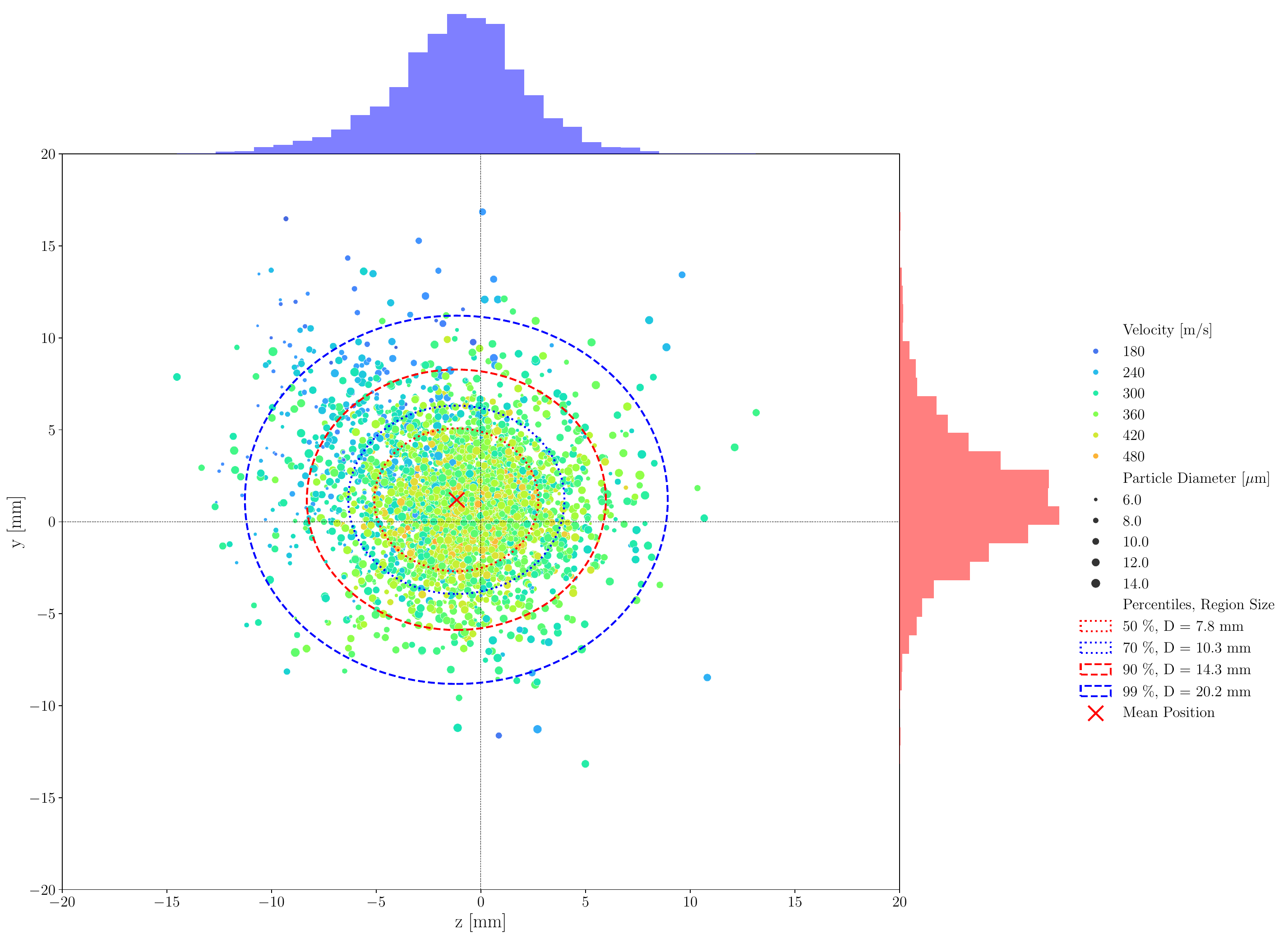}
        \caption{757 K, 150 mm}
        \label{fig:particles_t757_150}
    \end{subfigure}
    \hfill
    \begin{subfigure}[b]{0.32\textwidth}
        \centering
        \includegraphics[width=\textwidth]{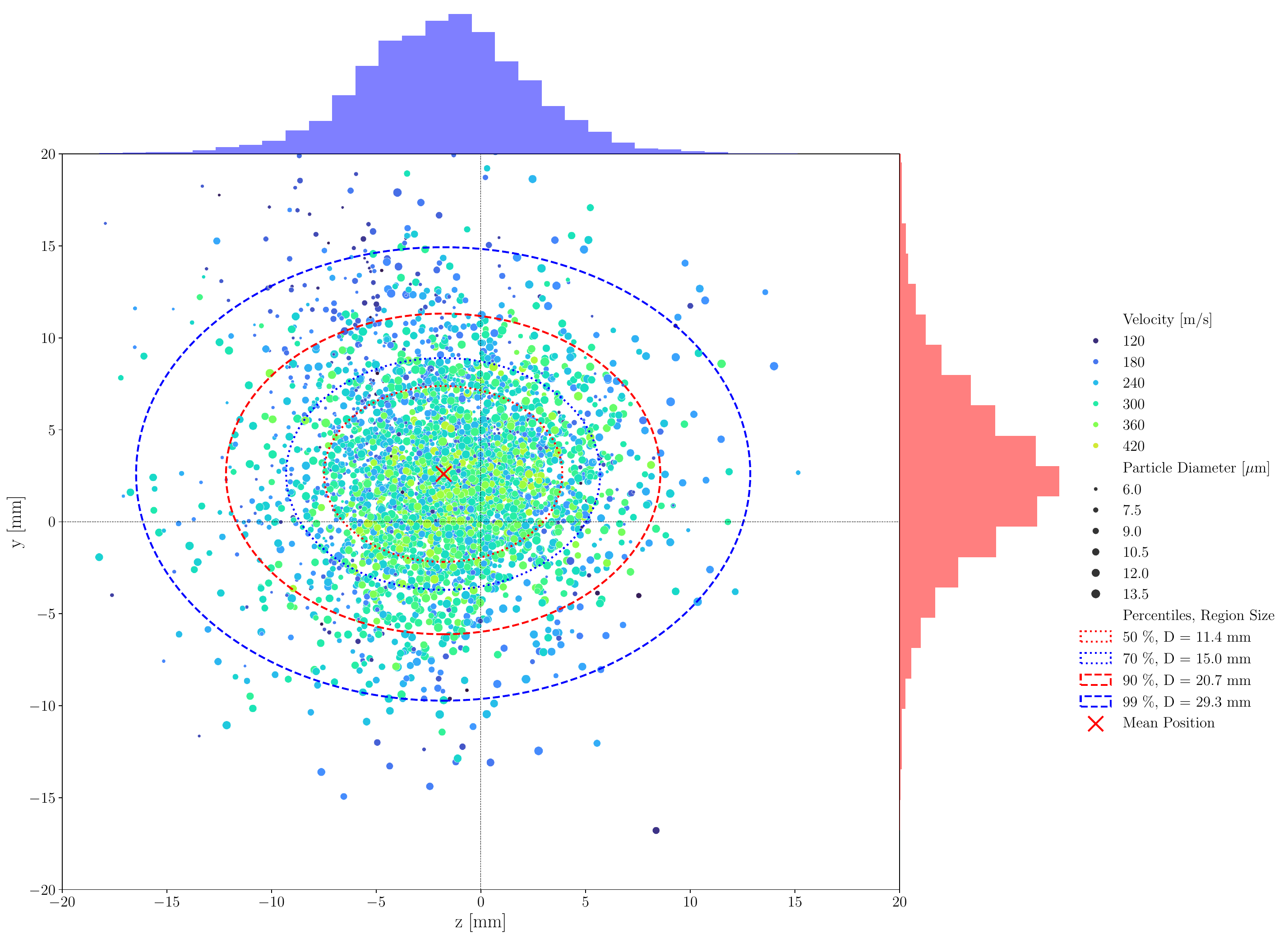}
        \caption{757 K, 200 mm}
        \label{fig:particles_t757_200}
    \end{subfigure}
    
    \vspace{0.3cm}
    
    \begin{subfigure}[b]{0.32\textwidth}
        \centering
        \includegraphics[width=\textwidth]{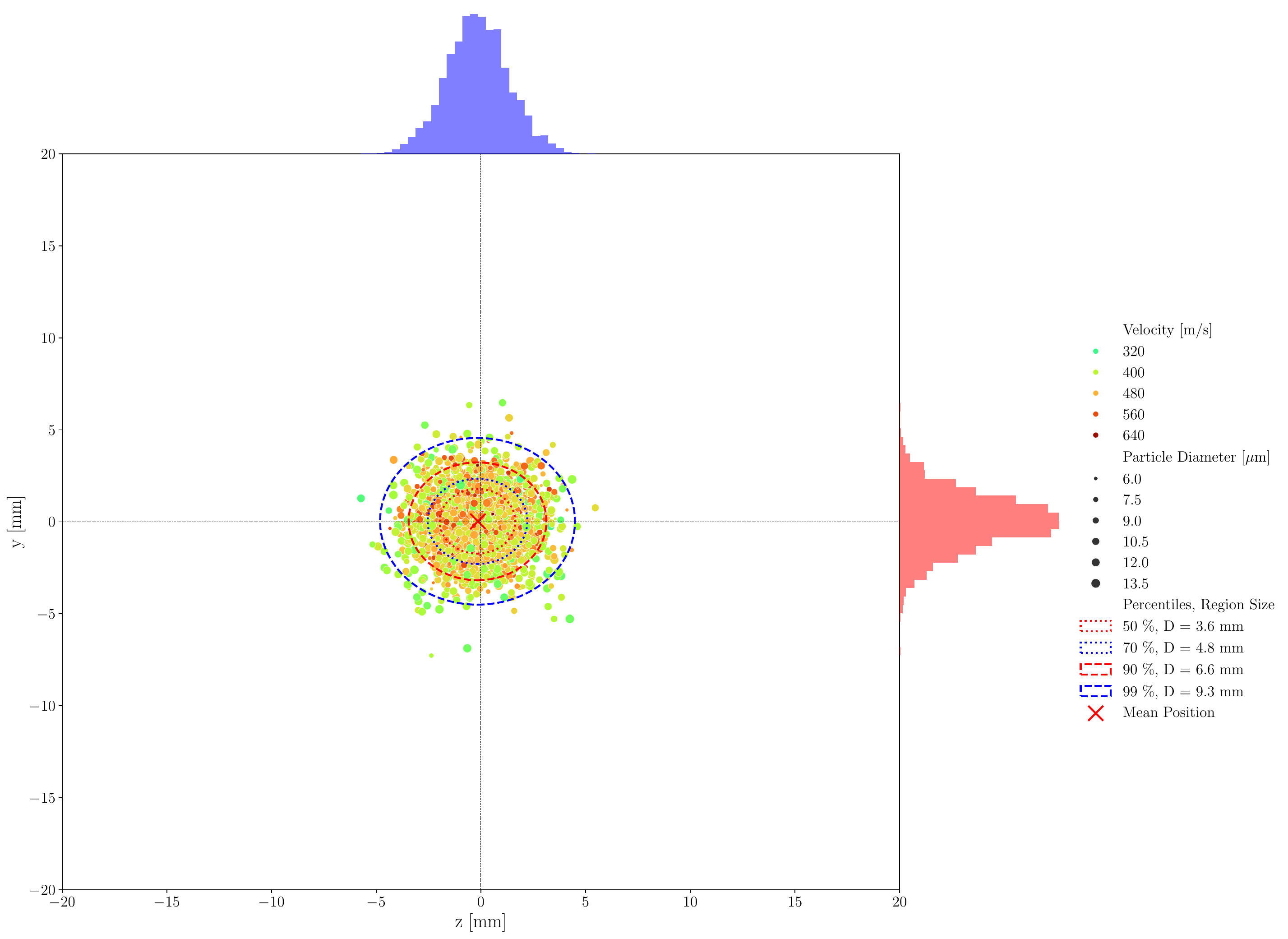}
        \caption{807 K, 100 mm}
        \label{fig:particles_t807_100}
    \end{subfigure}
    \hfill
    \begin{subfigure}[b]{0.32\textwidth}
        \centering
        \includegraphics[width=\textwidth]{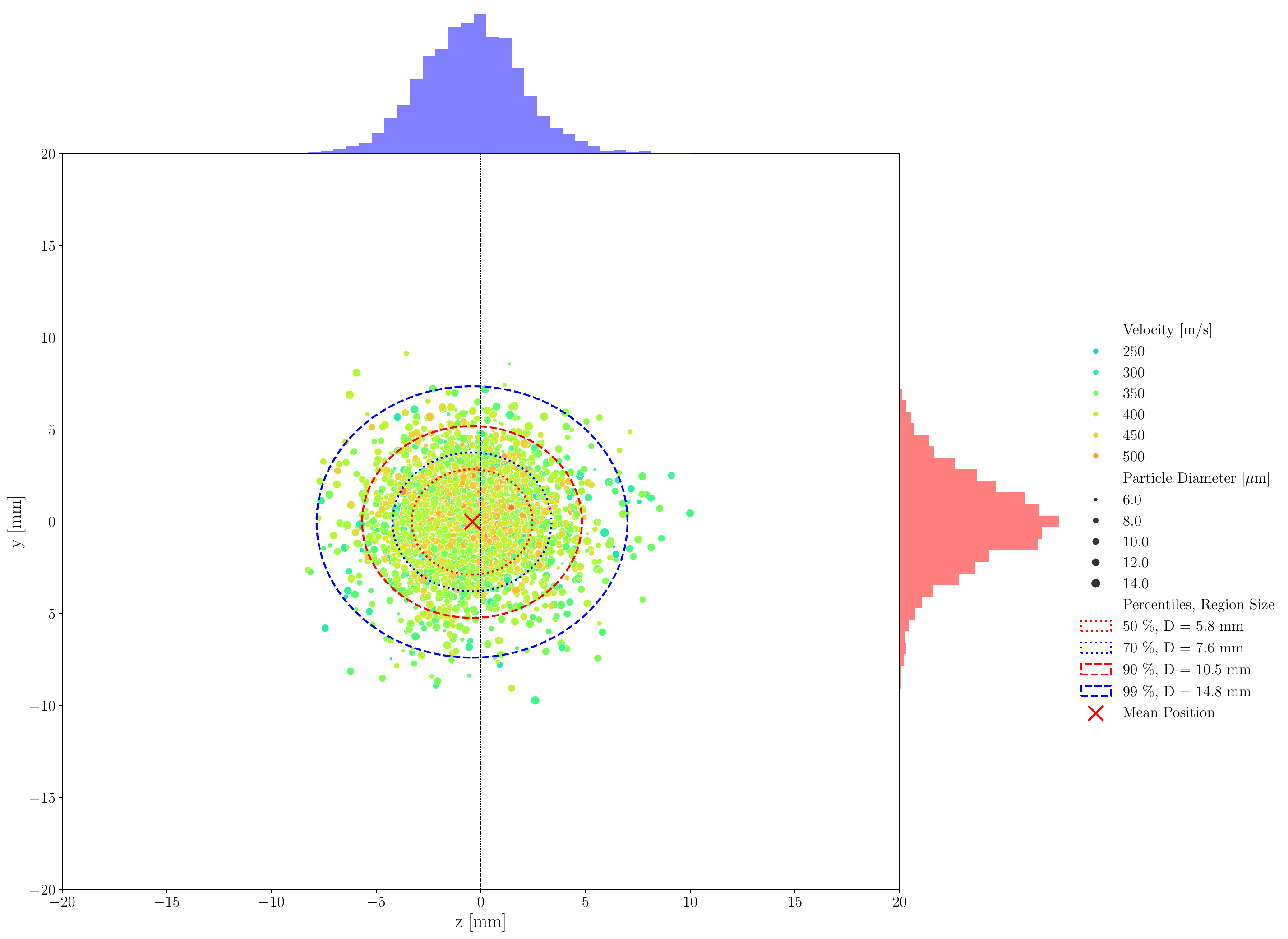}
        \caption{807 K, 150 mm}
        \label{fig:particles_t807_150}
    \end{subfigure}
    \hfill
    \begin{subfigure}[b]{0.32\textwidth}
        \centering
        \includegraphics[width=\textwidth]{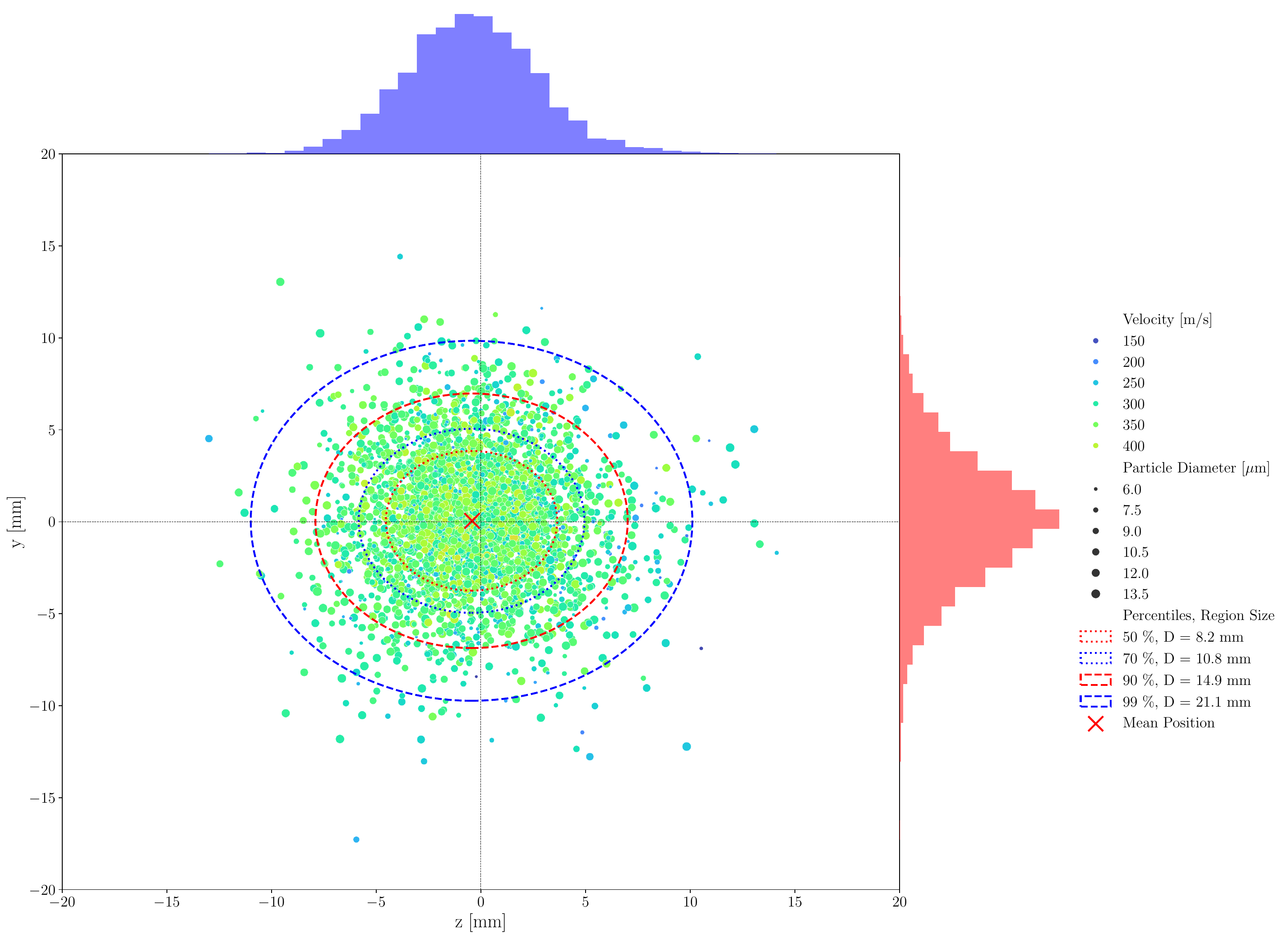}
        \caption{807 K, 200 mm}
        \label{fig:particles_t807_200}
    \end{subfigure}
    
    \caption{Particle spatial distribution at three standoff distances (100, 150, 200 mm) for gas temperatures of 707, 757, and 807 K at constant chamber pressure p = 45 bar.}
    \label{fig:particles_temp_grid}
\end{figure*}

\autoref{fig:particles_temp_grid} shows the same behavior but for different chamber temperatures. The first notable observation is that, compared to the increase in pressure, increasing the temperature does not push the center of particles farther from the centerline. As the majority of particles remain closely confined to the core jet region near the centerline, the particle velocities are generally higher and more uniform. These results confirm the velocity bands demonstrated in \autoref{fig:particle_tvar}. The values noted in \autoref{tab:particle_diameters_temp} show that the increase in temperature, in contrast to pressure increase, spreads the percentiles of the particles much less.

\begin{table}
    \centering
    \caption{Particle distribution diameters (mm) at different percentiles for varying chamber temperatures and standoff distances at constant pressure ($p = 45$~bar)}
    \label{tab:particle_diameters_temp}
    \begin{tabular}{ccccccccccccc}
        \toprule
        \textbf{Temperature} & \multicolumn{4}{c}{\textbf{100 mm}} & \multicolumn{4}{c}{\textbf{150 mm}} & \multicolumn{4}{c}{\textbf{200 mm}} \\
        \cmidrule(lr){2-5} \cmidrule(lr){6-9} \cmidrule(lr){10-13}
        \textbf{(K)} & \textbf{50\%} & \textbf{70\%} & \textbf{90\%} & \textbf{99\%} & \textbf{50\%} & \textbf{70\%} & \textbf{90\%} & \textbf{99\%} & \textbf{50\%} & \textbf{70\%} & \textbf{90\%} & \textbf{99\%} \\
        \midrule
        707 & 3.9 & 5.1 & 7.1 & 10.0 & 6.4 & 8.4 & 11.6 & 16.5 & 8.9 & 11.7 & 16.2 & 22.9 \\
        757 & 4.3 & 5.7 & 7.9 & 11.2 & 7.8 & 10.3 & 14.3 & 20.2 & 11.4 & 15.0 & 20.7 & 29.3 \\
        807 & 3.6 & 4.8 & 6.6 & 9.3 & 5.8 & 7.6 & 10.5 & 14.8 & 8.2 & 10.8 & 14.9 & 21.1 \\
        \bottomrule
    \end{tabular}
\end{table}

Examining the percentile diameters across all three standoff distances, the temperature cases show significantly less variation in particle spreading. At a standoff distance of 200~mm, the 99th percentile diameters range from 21.1~mm (807~K) to 29.3~mm (757~K), with a span of only 8.2~mm, compared to the 37.6~mm spread reported for pressure variations (29.3~mm to 66.9~mm). Notably, the highest temperature (807 K) consistently shows the most compact particle distribution over all standoff lengths, with the smallest percentile diameters at each measurement location. This finding implies that greater thermal energy improves particle acceleration and momentum coupling with the gas phase, keeping particles more firmly bound to the core jet direction rather than increasing radial dispersion.The marginal histograms in \autoref{fig:particles_temp_grid} further support this observation, showing narrower, more peaked distributions for the 807~K case compared to the lower temperature conditions.

The visualization shows that all three temperature conditions maintain well-defined, compact high-velocity cores even at 200~mm standoff distance. This contrasts with the diffuse velocity fields found for high-pressure situations at equal distances. The 807~K case exhibits the most concentrated high-velocity region, with a sharper transition to lower velocities at the periphery, consistent with the reduced gap between upper and lower velocity percentiles discussed in \autoref{fig:particle_tvar}. This tighter velocity distribution indicates more uniform particle acceleration and reduced velocity stratification at elevated temperatures. The combination of compact spatial distribution and uniform velocity field suggests that temperature increase primarily enhances the overall momentum transfer efficiency without significantly amplifying turbulent mixing or radial particle dispersion,.

\subsection{Particle Diameter-Velocity Correlation}

\begin{figure}[h]

        \centering
        \includegraphics[width=\textwidth, trim={5cm 0cm 9cm 0cm}, clip]{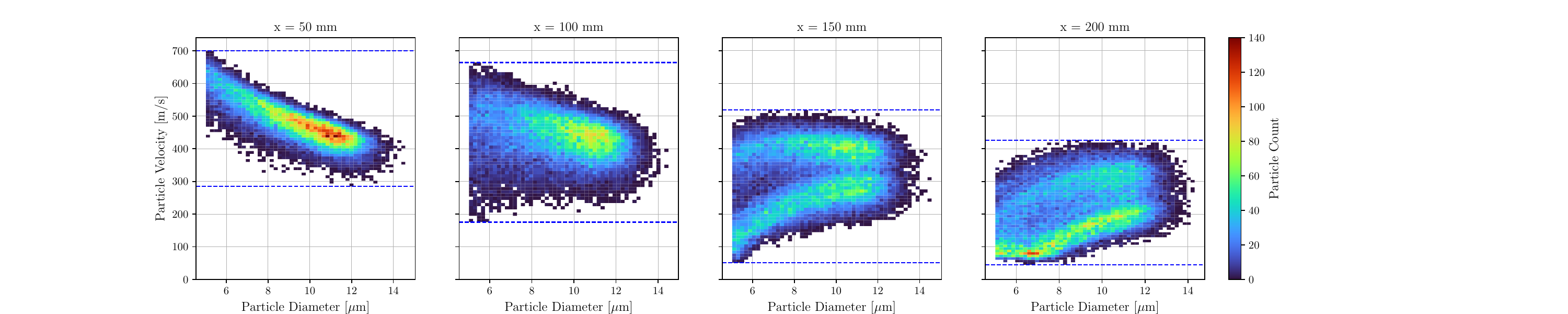}

    \caption{Two-dimensional histograms showing the relationship between particle diameter and particle velocity at three standoff distances (100, 150, 200, 250 mm) for chamber pressures of 55 bar at constant temperature T = 757 K. Color intensity represents particle count density, with purple indicating regions of highest particle concentration. Horizontal dashed lines mark reference velocities.}
    \label{fig:velocity_maps_pvar}
\end{figure}

Finally, in \autoref{fig:velocity_maps_pvar} the 2D histograms of particle diameter versus particle velocity for different standoff distances are visualized. It can be seen that for short distances, the histogram is well-shaped and uniform. The particle velocity upper and lower bounds both follow a steady decrease with increasing particle diameter. At 100~mm standoff distance, the upper bound starts from nearly 700~m/s for particles with 5~$\mu$m diameter and drops to 450~m/s for particles with approximately 14~$\mu$m diameter. The lower bound follows the same trend, decreasing from 450~m/s to 290~m/s from smallest to largest particles. As the standoff distance increases, the upper bound decreases and its slope flattens out. For the lower bound, as the standoff distance increases, the slope first completely flattens at approximately 150~mm from the nozzle exit and then reverses. Additionally, the velocity range from the lower to upper bound for each particle diameter at short distances is relatively identical across particle sizes. However, as the distance increases, smaller particles lose their velocity faster while larger particles maintain their speed over longer distances. This results in a greater velocity gap between the slowest and fastest particles for small diameters compared to the gap between the slowest and fastest particles for large diameters at farther standoff distances.

\section{Conclusion}
In this research, an analytical model was developed and described. It has been demonstrated that the $P_{\text{rms}}$ from a jet can be correlated with various factors, including nozzle design, microphone placement, chamber pressure, and temperature. Although multiple simplifying assumptions have been made, the resulting equation can serve as a useful instrument for preliminary forecasting of aeroacoustic noise produced by a jet. In the subsequent phase, it was demonstrated how the analytical model can be calibrated utilizing experimental data to enhance its accuracy and alignment with real-world conditions.

In the subsequent section, an OpenFOAM instance designed to numerically simulate DDES of a jet was described, documenting the pressure fluctuations and modelling particle flight trajectories. It was demonstrated that employing a coarse mesh with URANS simulations, followed by mapping the results onto a finer mesh and conducting DDES, can markedly expedite the simulation process. The model was validated using two separate sets of experimental data. Finally, the results obtained from the simulations were examined, assessing the influence of chamber pressure and temperature on both the aeroacoustic noise produced by the jet and the particle distribution and velocity at various standoff distances.

The workflow developed in this work enables improved  understanding of particle distribution, aeroacoustic generation of the nozzle, and the effect of geometry and chamber conditions on those properties. It could also provide a non-intrusive approach to monitor particle deposition  and coating quality using the thermal spray acoustic signature.

For the analytical model, many simplifying assumptions were made. For example, turbulence effects, friction effects, heat transfer effects, and observation angle had no influence on the SPL and $P_{\text{rms}}$ captured by the observer. In future continuations of this study, the model should be expanded to incorporate these phenomena. Additionally, for the calibration method, very few experimental data points were used. More testing with different nozzle geometries, chamber pressures, and temperatures should be conducted to create a comprehensive dataset that increases the precision of calibration values.

For the numerical simulation, the turbulence model can be further tuned, and different DES and LES turbulence models can be tested to determine which model performs most accurately. The Lagrangian particle forces were calculated as a whole, but it should be possible to calculate each acting force on the particle and analyze them separately. Additionally, different models such as various collision models and different drag equations can be analyzed to study in-flight particle trajectories. Furthermore, the developed solver 'sonicDPMFoam' is currently unable to handle heat transfer for particles, so particle temperature is not analyzed. This represents another potential development direction. Finally, with this case setup and solver, many more designs can be investigated: particles with different feed rates, densities, and sizes can be simulated; various inlet temperatures, pressures, and nozzle designs can be explored. These simulations would enrich the knowledge base for thermal spray specialists, researchers, and engineers regarding the aeroacoustic behavior of jets, particle behavior at different standoff distances, and overall process optimization. It should be noted that the current simulations were performed for free flight of particles without a substrate located in front of the jet. To achieve more precise particle simulations, a substrate should be added, as it would redirect the jet flow and consequently alter particle trajectories.

\section*{Supplementary Material}
See the supplementary material for five video animations from the DDES simulations. Supplementary Video 1 shows the pressure gradient magnitude visualization. Supplementary Video 2 presents the Mach number and static pressure distributions along the nozzle centerline with corresponding flow field visualizations. Supplementary Video 3 displays the temperature contour visualization. Supplementary Video 4 illustrates the velocity magnitude contour. Supplementary Video 5 shows the velocity magnitude with in-flight particle tracking.

\section*{Acknowledgments}
We would like to acknowledge Natural Resources Canada, Concordia University, and the Gina Cody School of Engineering and Computer Science for funding this project. Additionally, we thank Compute Alliance Canada for providing high-performance computing (HPC) resources to run the computationally intensive simulations for this project. OpenFOAM, an open-source CFD simulation tool, was the primary code package used for the simulations. libAcoustics was the main package coupled with OpenFOAM to record aeroacoustic noise generated by the jet, and we are grateful to the developers of libAcoustics who patiently answered our questions and helped us debug the integration of the module into our code. CMasher~\cite{cmasher} was the primary tool used to obtain colormaps for visualizations, with all colormaps used in this paper being perceptually uniform, accessible to colorblind viewers and suitable for black-and-white printing. Finally, PyVista and Matplotlib were the primary tools used to visualize, plot, and graph the contours, scatter plots, and figures presented in this paper.

\section*{Data availability}
The data that support the findings of this study are available from the corresponding author upon reasonable request. This includes the OpenFOAM case files, post-processed field data, particle trajectories, and Python scripts used for visualization and analysis.

\nomenclature[A]{$A$}{Area [m$^2$]}
\nomenclature[A]{$A^*$}{Sonic throat area [m$^2$]}
\nomenclature[A]{$A_e$}{Nozzle exit area [m$^2$]}
\nomenclature[A]{$A_p$}{Particle projected area [m$^2$]}
\nomenclature[A]{$A_s$}{Sutherland's law coefficient [kg/(m·s·K$^{1/2}$)]}
\nomenclature[A]{$c$}{Speed of sound [m/s]}
\nomenclature[A]{$c_0$}{Ambient speed of sound [m/s]}
\nomenclature[A]{$c_a$}{Speed of sound in ambient medium [m/s]}
\nomenclature[A]{$c_e$}{Speed of sound at nozzle exit [m/s]}
\nomenclature[A]{$c_p$}{Specific heat at constant pressure [J/(kg·K)]}
\nomenclature[A]{$c_{p,p}$}{Particle specific heat [J/(kg·K)]}
\nomenclature[A]{$C_D$}{Drag coefficient [--]}
\nomenclature[A]{$C_{DES}$}{DES model constant [--]}
\nomenclature[A]{$C_{d1}$}{Shielding function coefficient [--]}
\nomenclature[A]{$C_{d2}$}{Shielding function exponent [--]}
\nomenclature[A]{$C_\delta$}{Damping coefficient [--]}
\nomenclature[A]{$CD_{k\omega}$}{Cross-diffusion term [--]}
\nomenclature[A]{$Co$}{Courant number [--]}
\nomenclature[A]{$D_e$}{Nozzle exit diameter [m]}
\nomenclature[A]{$D_t$}{Nozzle throat diameter [m]}
\nomenclature[A]{$d_p$}{Particle diameter [m]}
\nomenclature[A]{$d_w$}{Distance to nearest wall [m]}
\nomenclature[A]{$d_w^+$}{Non-dimensional wall distance [--]}
\nomenclature[A]{$E$}{Wall function constant [--]}
\nomenclature[A]{$e$}{Restitution coefficient [--]}
\nomenclature[A]{$F$}{Force [N]}
\nomenclature[A]{$F_1, F_2$}{Blending functions [--]}
\nomenclature[A]{$f$}{Function, frequency [Hz]}
\nomenclature[A]{$f_d$}{Shielding function [--]}
\nomenclature[A]{$f_s$}{Sampling frequency [Hz]}
\nomenclature[A]{$f_N$}{Nyquist frequency [Hz]}
\nomenclature[A]{$G$}{Green's function [--]}
\nomenclature[A]{$h$}{Specific enthalpy [J/kg], heat transfer coefficient [W/(m$^2$·K)]}
\nomenclature[A]{$I$}{Sound intensity [W/m$^2$]}
\nomenclature[A]{$I_e$}{Sound intensity at nozzle exit [W/m$^2$]}
\nomenclature[A]{$k$}{Turbulent kinetic energy [m$^2$/s$^2$], thermal conductivity [W/(m·K)]}
\nomenclature[A]{$k_{\mathrm{eff}}$}{Effective thermal conductivity [W/(m·K)]}
\nomenclature[A]{$k_t$}{Turbulent thermal conductivity [W/(m·K)]}
\nomenclature[A]{$k_n$}{Normal spring stiffness [N/m$^{3/2}$]}
\nomenclature[A]{$K$}{Kinetic energy [J/kg]}
\nomenclature[A]{$l$}{Length scale [m]}
\nomenclature[A]{$l_i$}{Surface loading component [Pa]}
\nomenclature[A]{$l_{DDES}$}{DDES length scale [m]}
\nomenclature[A]{$l_{LES}$}{LES length scale [m]}
\nomenclature[A]{$l_{RANS}$}{RANS length scale [m]}
\nomenclature[A]{$\mathbf{M}$}{Mach number vector [--]}
\nomenclature[A]{$Ma$}{Mach number [--]}
\nomenclature[A]{$Ma_e$}{Exit Mach number [--]}
\nomenclature[A]{$M_r$}{Radial Mach number component [--]}
\nomenclature[A]{$m_p$}{Particle mass [kg]}
\nomenclature[A]{$m_{\mathrm{eff}}$}{Effective mass [kg]}
\nomenclature[A]{$n$}{Normal direction, iteration number [--]}
\nomenclature[A]{$Nu$}{Nusselt number [--]}
\nomenclature[A]{$p$}{Pressure [Pa]}
\nomenclature[A]{$p'$}{Acoustic pressure fluctuation [Pa]}
\nomenclature[A]{$P_0$}{Stagnation pressure [Pa]}
\nomenclature[A]{$P_a$}{Ambient pressure [Pa]}
\nomenclature[A]{$P_e$}{Pressure at nozzle exit [Pa]}
\nomenclature[A]{$P_k$}{Turbulence production term [kg/(m·s$^3$)]}
\nomenclature[A]{$P_{ref}$}{Reference pressure [Pa]}
\nomenclature[A]{$P_{rms}$}{Root-mean-square pressure [Pa]}
\nomenclature[A]{$Pr$}{Prandtl number [--]}
\nomenclature[A]{$Pr_t$}{Turbulent Prandtl number [--]}
\nomenclature[A]{$\mathbf{q}$}{Heat flux vector [W/m$^2$]}
\nomenclature[A]{$r$}{Distance, radial coordinate [m]}
\nomenclature[A]{$r_d$}{Turbulence ratio [--]}
\nomenclature[A]{$R$}{Gas constant [J/(kg·K)], distance [m]}
\nomenclature[A]{$Re_p$}{Particle Reynolds number [--]}
\nomenclature[A]{$S$}{Strain rate magnitude [s$^{-1}$]}
\nomenclature[A]{$S_{ij}$}{Strain rate tensor component [s$^{-1}$]}
\nomenclature[A]{$t$}{Time [s]}
\nomenclature[A]{$T$}{Temperature [K]}
\nomenclature[A]{$T_0$}{Stagnation temperature [K]}
\nomenclature[A]{$T_a$}{Ambient temperature [K]}
\nomenclature[A]{$T_e$}{Exit temperature [K]}
\nomenclature[A]{$T_s$}{Sutherland's law reference temperature [K]}
\nomenclature[A]{$T_{ij}$}{Lighthill stress tensor [Pa]}
\nomenclature[A]{$\mathbf{U}$}{Velocity vector [m/s]}
\nomenclature[A]{$u_i$}{Velocity component [m/s]}
\nomenclature[A]{$u_\tau$}{Friction velocity [m/s]}
\nomenclature[A]{$\mathbf{u}_g$}{Gas velocity [m/s]}
\nomenclature[A]{$\mathbf{u}_p$}{Particle velocity [m/s]}
\nomenclature[A]{$v_n$}{Normal velocity [m/s]}
\nomenclature[A]{$V$}{Volume [m$^3$]}
\nomenclature[A]{$V_p$}{Particle volume [m$^3$]}
\nomenclature[A]{$W$}{Acoustic power [W]}
\nomenclature[A]{$W_{in}$}{Acoustic power input [W]}
\nomenclature[A]{$W_{out}$}{Acoustic power output [W]}
\nomenclature[A]{$\mathbf{x}$}{Position vector [m]}
\nomenclature[A]{$y^+$}{Dimensionless wall distance [--]}
\nomenclature[A]{$y_w$}{Wall-normal distance [m]}

\nomenclature[G]{$\alpha$}{Damping coefficient [--], limiter parameter [--]}
\nomenclature[G]{$\beta$}{Dissipation coefficient [--]}
\nomenclature[G]{$\beta^*$}{Universal dissipation coefficient [--]}
\nomenclature[G]{$\gamma$}{Specific heat ratio [--]}
\nomenclature[G]{$\Gamma$}{Diffusion coefficient [--]}
\nomenclature[G]{$\delta$}{Overlap distance [m], Dirac delta function [--]}
\nomenclature[G]{$\delta_{ij}$}{Kronecker delta [--]}
\nomenclature[G]{$\Delta$}{Grid length scale [m], difference [--]}
\nomenclature[G]{$\Delta t$}{Time step [s]}
\nomenclature[G]{$\Delta x$}{Grid spacing [m]}
\nomenclature[G]{$\epsilon$}{Turbulent dissipation rate [m$^2$/s$^3$]}
\nomenclature[G]{$\theta$}{Polar angle [rad]}
\nomenclature[G]{$\kappa$}{von Kármán constant [--]}
\nomenclature[G]{$\mu$}{Dynamic viscosity [Pa·s], friction coefficient [--]}
\nomenclature[G]{$\mu_t$}{Turbulent viscosity [Pa·s]}
\nomenclature[G]{$\mu_{\mathrm{eff}}$}{Effective viscosity [Pa·s]}
\nomenclature[G]{$\nu$}{Kinematic viscosity [m$^2$/s]}
\nomenclature[G]{$\nu_t$}{Turbulent kinematic viscosity [m$^2$/s]}
\nomenclature[G]{$\rho$}{Density [kg/m$^3$]}
\nomenclature[G]{$\rho_0$}{Ambient density [kg/m$^3$]}
\nomenclature[G]{$\rho_a$}{Ambient medium density [kg/m$^3$]}
\nomenclature[G]{$\rho_e$}{Density at nozzle exit [kg/m$^3$]}
\nomenclature[G]{$\rho_g$}{Gas density [kg/m$^3$]}
\nomenclature[G]{$\rho_p$}{Particle density [kg/m$^3$]}
\nomenclature[G]{$\sigma_k$}{Turbulent Prandtl number for $k$ [--]}
\nomenclature[G]{$\sigma_\omega$}{Turbulent Prandtl number for $\omega$ [--]}
\nomenclature[G]{$\boldsymbol{\tau}$}{Stress tensor [Pa]}
\nomenclature[G]{$\phi$}{General scalar variable [--], azimuthal angle [rad]}
\nomenclature[G]{$\psi$}{Limiter parameter [--]}
\nomenclature[G]{$\omega$}{Specific dissipation rate [s$^{-1}$]}
\nomenclature[G]{$\Omega$}{Vorticity magnitude [s$^{-1}$]}
\nomenclature[G]{$\Omega_{ij}$}{Vorticity tensor component [s$^{-1}$]}

\nomenclature[S]{$_a$}{Ambient}
\nomenclature[S]{$_e$}{Exit}
\nomenclature[S]{$_g$}{Gas phase}
\nomenclature[S]{$_{in}$}{Input, inlet}
\nomenclature[S]{$_{max}$}{Maximum}
\nomenclature[S]{$_n$}{Normal direction}
\nomenclature[S]{$_{out}$}{Output, outlet}
\nomenclature[S]{$_p$}{Particle}
\nomenclature[S]{$_{ref}$}{Reference}
\nomenclature[S]{$_{rms}$}{Root mean square}
\nomenclature[S]{$_t$}{Tangential direction, turbulent}
\nomenclature[S]{$_w$}{Wall}
\nomenclature[S]{$_0$}{Stagnation condition}

\nomenclature[P]{$^+$}{Dimensionless wall units}
\nomenclature[P]{$^*$}{Sonic condition}
\nomenclature[P]{$^n$}{Time level}
\nomenclature[P]{$'$}{Fluctuation, calibrated value}

\nomenclature[Z]{CFD}{Computational Fluid Dynamics}
\nomenclature[Z]{DDES}{Delayed Detached Eddy Simulation}
\nomenclature[Z]{DES}{Detached Eddy Simulation}
\nomenclature[Z]{DILU}{Diagonal Incomplete LU}
\nomenclature[Z]{DPM}{Discrete Particle Method}
\nomenclature[Z]{FW-H}{Ffowcs Williams-Hawkings}
\nomenclature[Z]{GT}{Garrick Triangle}
\nomenclature[Z]{HPC}{High Performance Computing}
\nomenclature[Z]{LES}{Large Eddy Simulation}
\nomenclature[Z]{PBiCGStab}{Preconditioned Biconjugate Gradient Stabilized}
\nomenclature[Z]{PDF}{Probability Density Function}
\nomenclature[Z]{PIMPLE}{Merged PISO-SIMPLE algorithm}
\nomenclature[Z]{PISO}{Pressure-Implicit with Splitting of Operators}
\nomenclature[Z]{RANS}{Reynolds-Averaged Navier-Stokes}
\nomenclature[Z]{SIMPLE}{Semi-Implicit Method for Pressure-Linked Equations}
\nomenclature[Z]{SPL}{Sound Pressure Level}
\nomenclature[Z]{SST}{Shear Stress Transport}
\nomenclature[Z]{TVD}{Total Variation Diminishing}
\nomenclature[Z]{URANS}{Unsteady Reynolds-Averaged Navier-Stokes}
\printnomenclature

\appendix

\section{Governing equations}

\subsection{Turbulence Modeling}\label{sec:turbulence}

The mathematical framework of the SST DDES model is governed by two transport equations for the turbulent kinetic energy ($k$) and the specific dissipation rate ($\omega$). The governing equations read as follows\cite{gritskevich2012development}:

\begin{equation}
\frac{\partial \rho k}{\partial t} + \nabla \cdot (\rho \vec{U} k) = \nabla \cdot [(\mu + \sigma_k \mu_t) \nabla k] + P_k - \rho \sqrt{k^3}/l_{DDES}
\label{eq:k_transport}
\end{equation}

\begin{equation}
\frac{\partial \rho \omega}{\partial t} + \nabla \cdot (\rho \vec{U} \omega) = \nabla \cdot [(\mu + \sigma_\omega \mu_t) \nabla \omega] + 2(1-F_1) \rho \sigma_{\omega 2} \frac{\nabla k \cdot \nabla \omega}{\omega} + \gamma \frac{\rho}{\mu_t} P_k - \beta \rho \omega^2
\label{eq:omega_transport}
\end{equation}

where the model coefficients $\sigma_k$, $\sigma_\omega$, $\gamma$, and $\beta$ are obtained through blending of the inner (subscript 1) and outer (subscript 2) model constants using the blending function $F_1$:

\begin{equation}
\sigma_k = F_1 \sigma_{k1} + (1-F_1) \sigma_{k2}, \quad \sigma_\omega = F_1 \sigma_{\omega1} + (1-F_1) \sigma_{\omega2}
\label{eq:blending_sigma}
\end{equation}

\begin{equation}
\gamma = F_1 \gamma_1 + (1-F_1) \gamma_2, \quad \beta = F_1 \beta_1 + (1-F_1) \beta_2
\label{eq:blending_gamma_beta}
\end{equation}

The $k$-$\omega$ formulation (subscript 1) is active close to the wall where $F_1 \approx 1$ because of the blending technique, but the $k$-$\varepsilon$-like formulation (subscript 2) predominates in the free stream where $F_1 \approx 0$. The turbulent eddy viscosity is calculated as follows:

\begin{equation}
\mu_t = \rho \frac{a_1 \cdot k}{\max(a_1 \cdot \omega, F_2 \cdot S)}
\label{eq:eddy_viscosity}
\end{equation}

where $S$ is the magnitude of the strain rate tensor. 

\begin{equation}
S=\sqrt{2\sum_{i,j} S_{ij}S_{ij}}, \quad 
S_{ij} = \frac{1}{2}\left(\frac{\partial u_i}{\partial x_j} + \frac{\partial u_j}{\partial x_i}\right)
\label{eq:strain_tensor}
\end{equation}

A smooth transition between near-wall and far-field formulations is achieved by the blending functions $F_1$ and $F_2$:

\begin{equation}
F_1 = \tanh(\arg_1^4), \quad \arg_1 = \min \left[ \max \left( \frac{\sqrt{k}}{\beta^* \omega d_w}, \frac{500\mu}{\rho d_w^2 \omega} \right), \frac{4\rho\sigma_{\omega2}k}{CD_{k\omega}d_w^2} \right]
\label{eq:F1_arg1}
\end{equation}

\begin{equation}
CD_{k\omega} = \max \left( 2\rho\sigma_{\omega2} \frac{\nabla k \cdot \nabla \omega}{\omega}, 10^{-10} \right)
\label{eq:CDkomega}
\end{equation}

\begin{equation}
F_2 = \tanh(\arg_2^2), \quad \arg_2 = \max \left( \frac{2\sqrt{k}}{\beta^* \omega d_w}, \frac{500\mu}{\rho d_w^2 \omega} \right)
\label{eq:F2_arg2}
\end{equation}

where $d_w$ is the distance to the nearest wall. The production term in \autoref{eq:k_transport} includes a limiter for numerical stability:

\begin{equation}
P_k = \min(\mu_t S^2, c_1 \cdot \beta^* \rho k \omega)
\label{eq:production}
\end{equation}

The DDES length scale $l_{DDES}$, which controls the transition between RANS and LES modes, is used to modify the dissipation term in \autoref{eq:k_transport} for the DDES formulation:

\begin{equation}
l_{DDES} = l_{RANS} - f_d \max(0, l_{RANS} - l_{LES})
\label{eq:lDDES}
\end{equation}

where
\begin{equation}
l_{LES} = C_{DES} \Delta, \quad l_{RANS} = \frac{\sqrt{k}}{\beta^* \omega}
\label{eq:length_scales}
\end{equation}

and $\Delta$ is the grid length scale computed using the van Driest damping function for wall-bounded flows:

\begin{equation}
\Delta = \min(d_w, C_\delta \Delta_{\max}[1 - e^{-d_w^+/A^+}])
\label{eq:van_driest_delta}
\end{equation}

where $d_w^+ = \rho d_w u_\tau / \mu$ is the non-dimensional wall distance, $u_\tau$ is the friction velocity, and $\Delta_{\max}$ is the cube root of the cell volume. The shielding function $f_d$ prevents premature RANS-to-LES transition:

\begin{equation}
f_d = 1 - \tanh\left[(C_{d1} r_d)^{C_{d2}}\right], \quad r_d = \frac{\nu_t + \nu}{\kappa^2 d_w^2 \sqrt{0.5(S^2 + \Omega^2)}}
\label{eq:fd_rd}
\end{equation}

where $\Omega$ is the magnitude of the vorticity tensor. 
\begin{equation}
\Omega =  \sqrt{2\sum_{i,j} \Omega_{ij}\Omega_{ij}}, \quad
\Omega_{ij} = \frac{1}{2}\left(\frac{\partial u_i}{\partial x_j} - \frac{\partial u_j}{\partial x_i}\right)
\label{eq:vorticity_tensor}
\end{equation}

This formulation ensures that $l_{RANS}$ is preserved in attached boundary layers ($f_d \approx 0$), while $l_{LES}$ activates in separated regions ($f_d \approx 1$). The complete set of model coefficients is summarized in \autoref{tab:ddes_coefficients}.

\begin{table}
\centering
\caption{$k$–$\omega$ SST DDES model coefficients.}
\label{tab:ddes_coefficients}
\begin{tabular}{lll}
\hline
\textbf{Coefficient} & \textbf{Value} & \textbf{Description} \\
\hline
\multicolumn{3}{l}{\textit{DDES Parameters}} \\
\hline
$C_{DES}$ & 0.61 & DES model constant \\
$C_{d1}$ & 20.0 & Shielding function coefficient \\
$C_{d2}$ & 3.0 & Shielding function exponent \\
\\
\multicolumn{3}{l}{\textit{$k$–$\omega$ Region 1 (Near-wall)}} \\
\hline
$\sigma_{k1}$ & 0.85 & Turbulent Prandtl number for $k$ \\
$\sigma_{\omega 1}$ & 0.5 & Turbulent Prandtl number for $\omega$ \\
$\gamma_1$ & 0.5532 & Production coefficient \\
$\beta_1$ & 0.075 & Dissipation coefficient \\
\\
\multicolumn{3}{l}{\textit{$k$–$\omega$ Region 2 (Far-field)}} \\
\hline
$\sigma_{k2}$ & 1.0 & Turbulent Prandtl number for $k$ \\
$\sigma_{\omega 2}$ & 0.856 & Turbulent Prandtl number for $\omega$ \\
$\gamma_2$ & 0.4403 & Production coefficient \\
$\beta_2$ & 0.0828 & Dissipation coefficient \\
\\
\multicolumn{3}{l}{\textit{Universal Constants}} \\
\hline
$\beta^*$ & 0.09 & Universal dissipation coefficient \\
$a_1$ & 0.31 & Viscosity limiter coefficient \\
$c_1$ & 10.0 & Production limiter coefficient \\
$\kappa$ & 0.41 & von Kármán constant \\
\\
\multicolumn{3}{l}{\textit{van Driest Delta Parameters}} \\
\hline
$A^+$ & 26.0 & van Driest damping constant \\
$C_\delta$ & 0.158 & Damping coefficient \\
\hline
\end{tabular}
\end{table}

\subsection{Boundary Conditions}\label{sec:BC}

The temporal evolution of pressure and temperature boundary conditions is defined as:

\begin{align}
p_{\text{inlet}}(t) &= \begin{cases}
1 \times 10^6 & \text{if } t \leq 0 \\
1 \times 10^6 + \frac{t}{t_{\text{ramp}}} \left( p_{\text{target}} - 1 \times 10^6 \right) & \text{if } 0 < t \leq t_{\text{ramp}} \\
p_{\text{target}} & \text{if } t > t_{\text{ramp}}
\end{cases} \label{eq:pressure_ramp}
\end{align}

\begin{align}
T_{\text{inlet}}(t) &= \begin{cases}
400 & \text{if } t \leq 0 \\
400 + \frac{t}{t_{\text{ramp}}} \left( T_{\text{target}} - 400 \right) & \text{if } 0 < t \leq t_{\text{ramp}} \\
T_{\text{target}} & \text{if } t > t_{\text{ramp}}
\end{cases} \label{eq:temperature_ramp}
\end{align}

where $t_{\text{ramp}}$ is the ramp duration, $p_{\text{target}}$ is the target inlet pressure, and $T_{\text{target}}$ is the target inlet temperature. This approach prevents abrupt pressure and temperature changes that could introduce numerical instabilities in compressible flow simulations. In this study, $t_{\text{ramp}}$ is set to $0.0006\,\text{s}$, which was found to be sufficient for the flow to develop smoothly while avoiding numerical instability and residual divergence.

For the turbulence quantities, the inlet is prescribed with fixed values appropriate for the expected flow conditions. The turbulent kinetic energy ($k$) is set to $1 \times 10^{-3}$~m$^2$/s$^2$, while the specific turbulence dissipation rate ($\omega$) is specified as 1~s$^{-1}$ at the inlet.

For the wall boundaries, wall functions were applied to all turbulent quantities. The turbulent kinetic energy ($k$) uses equilibrium wall functions, the specific dissipation rate ($\omega$) employs logarithmic wall functions, and the turbulent viscosity ($\mu_t$) is modeled using wall functions based on the dimensionless wall distance~\cite{blocken2007cfd}.

\begin{equation}
k_w = \frac{u_\tau^2}{\sqrt{C_\mu}} \qquad \omega_w = \frac{u_\tau}{\kappa y_w \sqrt{C_\mu}}, \qquad \mu_{t,w} = \rho_w u_\tau y_w \left[\frac{\kappa}{\ln(Ey^+)} - \frac{1}{y^+}\right], \qquad y^+ = \frac{\rho_w u_\tau y_w}{\mu_w}
\end{equation}
where $C_\mu = 0.09$, $\kappa = 0.41$ is the von Kármán constant, $E = 9.8$ is the wall function constant, $y_w$ is the wall-normal distance to the first grid point, and the subscript $w$ denotes wall values.

The outlet boundary conditions are designed to allow free outflow into the ambient environment. A nonreflecting boundary condition is applied for pressure to prevent reflections back into the domain, with the far-field pressure set to atmospheric conditions (101,325~Pa).
The outlet velocity condition is pressure-dependent, allowing the flow to exit freely while suppressing unintended inflow. The temperature field uses an inletOutlet formulation, which extrapolates from the interior in outflow regions while enforcing the ambient temperature of $300\,\text{K}$ where inflow occurs. A detailed summary of all boundary conditions, including their types and values, is provided in \autoref{tab:boundary_conditions}.

\begin{table}
\centering
\caption{Summary of boundary conditions}
\label{tab:boundary_conditions}
\begin{tabular}{ccccc}
\hline
\textbf{Parameter} &  & \textbf{Inlet} & \textbf{Nozzle Wall \& Outer Wall} & \textbf{Outlet} \\
\hline
\multirow{2}{*}{\textbf{$k$}} 
& Type  & Fixed Value     & Wall Function  & Freestream \\
& Value & $1 \times 10^{-3}$ & $1 \times 10^{-3}$ & $1 \times 10^{-3}$ \\
\hline
\multirow{2}{*}{\textbf{$\nu_t$}} 
& Type  & Zero Gradient   & Wall Function  & Calculated \\
& Value & --              & $1 \times 10^{-10}$ & $1 \times 10^{-10}$ \\
\hline
\multirow{2}{*}{\textbf{$\omega$}} 
& Type  & Fixed Value     & Wall Function  & Freestream \\
& Value & 1               & $1 \times 10^{-12}$ & 10 \\
\hline
\multirow{2}{*}{\textbf{$p$}} 
& Type  & Ramped Pressure & Zero Gradient  & Wave Transmissive \\
& Value & Time-dependent  & --             & 101{,}325 Pa \\
\hline
\multirow{2}{*}{\textbf{$T$}} 
& Type  & Ramped Temperature & Adiabatic   & Inlet/Outlet \\
& Value & Time-dependent     & --          & 300 K \\
\hline
\multirow{2}{*}{\textbf{$\vec{U}$}} 
& Type  & Pressure-driven  & No-slip       & Pressure-driven \\
& Value & --               & $(0,0,0)$     & $(0,0,0)$ \\
\hline
\end{tabular}
\end{table}

\subsection{Acoustic Analogies}\label{sec:acoustics}

 In its general permeable-surface form, the FW–H equation is:

\begin{equation}
\frac{1}{c_0^2}\frac{\partial^2 p'}{\partial t^2}-\nabla^2 p'
=\frac{\partial}{\partial t}\big[\rho_0 v_n \delta(f)\big]
-\frac{\partial}{\partial x_i}\big[l_i \delta(f)\big]
+\frac{\partial^2}{\partial x_i \partial x_j}\big[T_{ij} H(f)\big]
\end{equation}

where $p'$ is the acoustic pressure fluctuation, $c_0$ and $\rho_0$ are the ambient speed of sound and density, and $f(\mathbf{y},t)=0$ defines the control surface that encloses the acoustic sources. The distributions $\delta(f)$ and $H(f)$ are the Dirac delta and Heaviside functions that localize the surface and volume terms. The quantity $v_n$ is the surface-normal velocity, $l_i$ is the $i$th component of surface loading, and $T_{ij}$ is the Lighthill stress tensor\cite{lighthill1952sound, lighthill1954sound},

\begin{equation}
T_{ij}=\rho u_i u_j + p_{ij} - c_0^2(\rho-\rho_0)\delta_{ij}
\end{equation}

with $\rho$ the instantaneous density, $u_i$ the velocity components, $p_{ij}$ the viscous stress tensor, and $\delta_{ij}$ the Kronecker delta.

For numerical robustness in complex, high-speed flows, the Garrick Triangle (GT) formulation\cite{garrick1953theoretical}, was used which reorganizes the source terms and yields improved time-domain convergence. In GT formulation, the acoustic pressure at observer location $\mathbf{x}$ and time $t$ is decomposed as:

\begin{equation}
p'\left(\boldsymbol{x},t\right) = p'_T\left(\boldsymbol{x},t\right) + p'_L\left(\boldsymbol{x},t\right) + p'_Q\left(\boldsymbol{x},t\right)
\end{equation}
where $p'_T$, $p'_L$, and $p'_Q$ denote, respectively, thickness (displacement), loading (aerodynamic force), and quadrupole (turbulent) contributions.

The thickness contribution, associated with fluid displacement by the moving/permeable surface, is

\begin{equation}
p'_T(\mathbf{x},t)=\frac{1}{4\pi}\int_{f=0}\left[\frac{\rho_0 v_n}{r(1-M_r)}\right]_{\mathrm{ret}}\mathrm{d}S
+\frac{1}{4\pi}\int_{f=0}\left[\frac{\rho_0 v_n\big(r M_r+c_0(\mathbf{M}\cdot\hat{\mathbf{r}}-M_r)\big)}{r^2(1-M_r)^2}\right]_{\mathrm{ret}}\mathrm{d}S
\end{equation}

Here $\mathbf{y}$ is a source point on the control surface, $r=\|\mathbf{x}-\mathbf{y}\|$ is the observer–source distance, $\hat{\mathbf{r}}=(\mathbf{x}-\mathbf{y})/r$ the radiation unit vector, $\mathbf{M}=\mathbf{U}/c_0$ the local Mach-number vector based on surface velocity $\mathbf{U}$, and $M_r=\mathbf{M}\cdot\hat{\mathbf{r}}$ its projection along $\hat{\mathbf{r}}$. The subscript $ret$ indicates evaluation at retarded time .

The loading contribution, due to unsteady aerodynamic forces on the surface, is:

\begin{equation}
p'_L(\mathbf{x},t)=\frac{1}{4\pi c_0}\int_{f=0}\left[\frac{\dot{l}_r}{r(1-M_r)^2}\right]_{\mathrm{ret}}\mathrm{d}S
+\frac{1}{4\pi c_0}\int_{f=0}\left[\frac{l_r-l_M}{r^2(1-M_r)^2}\right]_{\mathrm{ret}}\mathrm{d}S
\end{equation}

where $\mathbf{l}$ is the surface loading vector, $l_r=\mathbf{l}\cdot\hat{\mathbf{r}}$ its radiation-direction component, $l_M=\mathbf{l}\cdot\mathbf{M}$ its projection along $\mathbf{M}$, and $\dot{l}_r=\partial l_r/\partial t$.

The quadrupole contribution from turbulent stresses inside the control volume $V$ is

\begin{equation}
p'_Q(\mathbf{x},t)=\frac{1}{4\pi c_0^2}\int_{V}\left[\frac{\ddot{T}_{ij}}{r(1-M_r)}\,\frac{\partial^2 G}{\partial x_i \partial x_j}\right]_{\mathrm{ret}}\mathrm{d}V
\end{equation}

where $G$ is the free-space Green’s function of the wave operator and overdots denote time derivatives at the source\cite{crighton1992modern}.

\subsection{Numerical Schemes}

The numerical discretization schemes used for the simulation are summerized in \autoref{tab:numerical_schemes}.

\begin{table}
\centering
\caption{Numerical discretization schemes}
\label{tab:numerical_schemes}
\begin{tabular}{llll}
\hline
\textbf{Term} & \textbf{Variable} & \textbf{Scheme} & \textbf{Parameters} \\
\hline
\multicolumn{4}{l}{\textit{Temporal Discretization}} \\
\hline
$\frac{\partial \phi}{\partial t}$ & All variables & Backward Euler & 2nd order implicit \\
\\
\multicolumn{4}{l}{\textit{Gradient Terms}} \\
\hline
$\nabla \phi$ & General & Gauss linear & -- \\
$\nabla \vec{U}$ & Velocity & Gauss linear limited & $\alpha = 0.5$ \\
\\
\multicolumn{4}{l}{\textit{Convective Terms}} \\
\hline
$\nabla \cdot (\rho \vec{U} \phi)$ & $\vec{U}$, $h$, $e$, $K$ & Limited linear & $\psi = 0.3$ \\
$\nabla \cdot (\rho \vec{U} \phi)$ & Pressure terms & Limited linear & $\psi = 0.3$ \\
$\nabla \cdot (\rho \vec{U} \phi)$ & $k$, $\omega$, $\epsilon$ & Bounded upwind & TVD limiter \\
$\nabla \cdot (\rho \vec{U} \phi)$ & $\tilde{\nu}$ & Limited linear & $\psi = 0.3$ \\
\\
\multicolumn{4}{l}{\textit{Diffusive Terms}} \\
\hline
$\nabla \cdot (\Gamma \nabla \phi)$ & All variables & Limited corrected & $\alpha = 0.5$ \\
\\
\multicolumn{4}{l}{\textit{Interpolation Schemes}} \\
\hline
Face interpolation & General & Linear & Central difference \\
Reconstruction & $\vec{U}$ & van Albada & Vector limiter \\
Reconstruction & $\rho$, $T$ & van Albada & Scalar limiter \\
\\
\multicolumn{4}{l}{\textit{Surface Normal Gradient}} \\
\hline
$\frac{\partial \phi}{\partial n}$ & All variables & Corrected & Non-orthogonal correction \\
\hline
\end{tabular}
\end{table}

Additionally, the solution procedure parameters are mentioned in \autoref{tab:numerical_parameters}.

\begin{table}
\centering
\caption{Numerical solution parameters.}
\label{tab:numerical_parameters}
\begin{tabular}{ll}
\hline
\textbf{Parameter} & \textbf{Value} \\
\hline
\multicolumn{2}{l}{\textit{Linear Solver Configuration}} \\
\hline
Solver & PBiCGStab \\
Preconditioner & Diagonal \\
Tolerance (pressure) & $10^{-12}$ \\
Tolerance (transport equations) & $10^{-15}$ \\
Minimum iterations & 2 \\
\\
\multicolumn{2}{l}{\textit{Relaxation Factors}} \\
\hline
Pressure field & 0.3 \\
Temperature field & 0.4 \\
Transport equations & 0.5 \\
\\
\multicolumn{2}{l}{\textit{PIMPLE Algorithm}} \\
\hline
Outer correctors & 2 \\
Pressure correctors & 2 \\
Non-orthogonal correctors & 1 \\
Minimum pressure factor & 0.5 \\
Maximum pressure factor & 2.0 \\
\\
\multicolumn{2}{l}{\textit{Run configuration}} \\
\hline
Courant Number & ~$0.1$ \\
Timestep & $10^{-8}$ \\
\hline
\end{tabular}
\end{table}

\subsection{Energy Spectrum calculations}\label{sec:EnergyCascade}

The energy spectrum is computed from the discrete velocity time series $\mathbf{u}(t) = (u_x(t), u_y(t), u_z(t))$ recorded at each probe with sampling interval $\Delta t = 10^{-8}$. First, velocity fluctuations are obtained by subtracting the temporal mean:

\begin{equation}
u'_i(t) = u_i(t) - \langle u_i \rangle, \quad \text{where} \quad \langle u_i \rangle = \frac{1}{N}\sum_{t} u_i(t)
\end{equation}

where $u_i(t)$ represents the velocity component ($i = x, y, z$) at time $t$, $u'_i$ is the fluctuating velocity component, and $N$ is the total number of time samples. The discrete Fourier transform is then applied to each component:

\begin{equation}
\hat{u}_i(f) = \sum_{t} u'_i(t) e^{-2\pi i ft/T}
\end{equation}

where $\hat{u}_i(f)$ is the Fourier coefficient at frequency $f$ and $T$ is the total sampling time. The kinetic energy spectrum in frequency space is calculated as:

\begin{equation}
E(f) = \frac{1}{2N}(|\hat{u}_x(f)|^2 + |\hat{u}_y(f)|^2 + |\hat{u}_z(f)|^2)
\end{equation}

where $E(f)$ is the energy at frequency $f$. Using Taylor's frozen turbulence hypothesis, the frequency spectrum is converted to wavenumber space:

\begin{equation}
k = \frac{2\pi f}{U_{rms}}, \quad \text{where} \quad U_{rms} = \sqrt{\frac{1}{3}(\langle u_x'^2 \rangle + \langle u_y'^2 \rangle + \langle u_z'^2 \rangle)}
\end{equation}

where $k$ is the spatial wavenumber and $U_{rms}$ is the root-mean-square velocity magnitude. The final spectrum $\langle E(k) \rangle$ represents the ensemble average across all probe locations.

\subsection{Particla Validation calculations}\label{sec:validationCalc}

Particles crossing the exit plane within a $\pm 1~\text{mm}$ axial window were extracted and binned spatially along the vertical direction to construct mass flux profiles. For each bin, the particle mass flux per unit area was computed from the simulated particle trajectories.

For each particle $i$ with diameter $d_i$, the volume and mass are:
\begin{equation}
V_{p,i} = \frac{\pi}{6} d_i^3, \quad m_{p,i} = \rho_p V_{p,i}
\end{equation}
where $\rho_p = 3950~\text{kg/m}^3$ is the alumina particle density. The axial momentum of each particle is:
\begin{equation}
p_i = m_{p,i} \cdot U_{x,i}
\end{equation}
where $U_{x,i}$ is the axial velocity component. The nozzle exit plane was divided into $n_{\text{bins}} = 50$ horizontal bins along the vertical coordinate. For each bin $j$, the total momentum is:
\begin{equation}
P_j = \sum_{i \in \text{bin } j} p_i = \sum_{i \in \text{bin } j} m_{p,i} \cdot U_{x,i}
\end{equation}

For a circular nozzle exit with radius $R = 15~\text{mm}$, the cross-sectional area of each horizontal bin is computed as:
\begin{equation}
A_{\text{bin},j} = 2 \sqrt{R^2 - z_j^2} \cdot \Delta z
\end{equation}
where $z_j$ is the vertical position of the bin center relative to the nozzle centerline and $\Delta z$ is the bin width. Finally, the mass flux per unit area for each bin is:
\begin{equation}
\Phi_j = \frac{P_j}{A_{\text{bin},j} \cdot \Delta x}
\end{equation}
$\Phi$ with units of kg/(m$^2{\cdot}$s), representing the rate of mass transport through the cross-sectional area of bin $j$. Here, $\Delta x = 2~\text{mm}$ is the axial sampling window at the nozzle exit plane used to capture particles. This finite window width was necessary because, due to numerical discretization, it was not possible to capture particles passing exactly at the exit plane ($x = 0$).

\bibliographystyle{ieeetran}

\bibliography{00_bib}

\end{document}